\documentclass[twocolumn]{aastex63}

\usepackage{threeparttablex}
\usepackage{longtable,booktabs}

\newcommand{\gr}{$\gamma$-ray}
\newcommand\fermi{{\it Fermi}}

\begin{document}

\title{Gamma-ray spectral properties of the Galactic globular clusters: constraint on the numbers of millisecond pulsars}

\correspondingauthor{Zhongxiang Wang}
\email{wangzx20@ynu.edu.cn}

\author{Wei Wu}
\affiliation{Department of Astronomy, School of Physics and Astronomy, 
Key Laboratory of Astroparticle Physics of Yunnan Province, Yunnan University, 
Kunming 650091, China}

\author{Zhongxiang Wang}
\affiliation{Department of Astronomy, School of Physics and Astronomy, 
Key Laboratory of Astroparticle Physics of Yunnan Province, Yunnan University, 
Kunming 650091, China}
\affiliation{Shanghai Astronomical Observatory, Chinese Academy of Sciences, 
80 Nandan Road, Shanghai 200030, China}

\author{Yi Xing}
\affiliation{Shanghai Astronomical Observatory, Chinese Academy of Sciences,
80 Nandan Road, Shanghai 200030, China}

\author{Pengfei Zhang}
\affiliation{Department of Astronomy, School of Physics and Astronomy,
Key Laboratory of Astroparticle Physics of Yunnan Province, Yunnan University,
Kunming 650091, China}

\begin{abstract}
We study the \gr\ spectra of 30 globular clusters (GCs) thus far detected 
with {\it the Fermi Gamma-ray Space Telescope}. Presuming that \gr\ emission 
of a GC comes from millisecond pulsars (MSPs) contained in, a model that 
generates spectra for the GCs is built based on the \gr\ properties of 
the detected MSP sample. We fit the GCs' spectra with the model, and for 27 
of them, their emission can be explained with arising from MSPs. The spectra 
of the other three, NGC~7078, 2MS-GC01, and Terzan~1, can not be fit with our 
model, indicating that MSPs' emission should not be the dominant one in the 
first two and the third one has a unique hard spectrum.  We also investigate 
six nearby GCs that have relatively high encounter rates as the comparison 
cases. The candidate spectrum of NGC~6656 can be fit with that of one MSP, 
supporting its possible association with the \gr\ source at its position. 
The five others do not have detectable \gr\ emission. Their spectral upper 
limits set limits of $\leq 1$ MSPs in them, consistent with the numbers 
of radio MSPs found in them.  The estimated numbers of MSPs in the \gr\ GCs 
generally match well those reported for radio pulsars.
Our studies of the \gr\ GCs and 
the comparison nearby GCs indicate that the encounter rate should not be 
the only factor determining the number of \gr\ MSPs a GC contains.
\end{abstract}


\section{Introduction}

From high-energy observations with the Large Area Telescope (LAT) onboard 
{\it the Fermi Gamma-ray Space Telescope (Fermi)}, globular clusters (GCs) in 
our Milky Way have been identified as a class of \gr\ sources in the 
sky \citep{abd+09, 4fgl20}. Thus far, approximately 30 GCs have been detected
at $\gamma$-rays 
(\citealt{abd+09, abd+10, khc10, tam+11,zho+15,zha+16,lcb18,dcn19,son+21};
see also Table~\ref{tab:gc}).
Their \gr\ emission presumably arises from millisecond pulsars 
(MSPs) that are contained in. One side is because that pulsars have been 
established by \fermi\ LAT as the dominant \gr\ sources in the Milky 
Way \citep{2fpsr13,4fgl20}, and on the other side GCs, due to their 
$\sim$10\,Gyr old ages and high star densities (e.g., \citealt{har96}),  
naturally contain compact binaries (see, e.g., \citealt{hen+18,oh+20,zha+20}) 
that enable MSP formation.
Direct observational evidence has also been found showing \gr\ MSPs in the GCs
(\citealt{fre+11,wu+13,joh+13}).
\begin{figure}
\begin{center}
\includegraphics[width=0.95\linewidth]{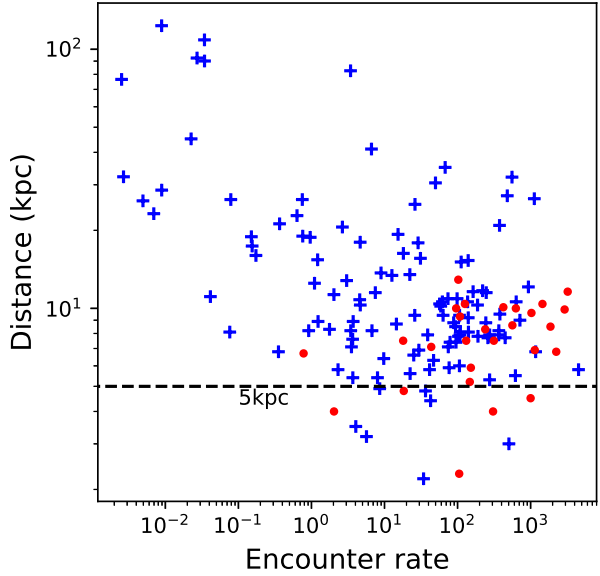}
\end{center}
	\caption{Distances and encounter rates of the GCs in the milky way, 
	where the values are (or calculated) from the catalog given 
	by \citet{har96}. A normalization factor of $\Gamma_e=1000$ is assigned
	to the GC NGC~104 (47\,Tuc). The red dots are the \gr\ GCs reported 
	(Table~\ref{tab:gc}), except
	GLIMPSE-C01, GLIMPSE-C02, and 2MS-GC01 whose structural parameters have
	not been reported.
	}
\label{fig:gc}
\end{figure}

Using the \gr\ detections, the numbers of \gr\ MSPs and the related properties
of the GCs have been studied. Assuming a typical \gr\ luminosity value,
the numbers for MSPs in each GCs can be estimated \citep{abd+10}. Moreover, 
the numbers should be correlated with the stellar encounter
rates $\Gamma_e$ (as well as with the metallicities), 
$\Gamma_e \sim \rho_0^{1.5}r_c^2$, where $\rho_0$ is the central cluster 
density and $r_c$ the cluster core radius. Studies have indicated the possible 
presence of the correlation \citep{abd+10,hui+11,dcn19}. In
Figure~\ref{fig:gc}, we show the GCs from the catalog given by \citet{har96},
among which the \gr-detected GCs are marked. The latter do appear to have 
large $\Gamma_e$, although it can be noted that several 
high-$\Gamma_e$ ones at comparably close distances 
(e.g., $\leq$5\,kpc) have not been reported to have detectable \gr\ emission.
Also there is a possible trend showing farther GCs have smaller $\Gamma_e$,
the reason for which is not clear, but we may speculate that $\rho_0$ 
and/or
$r_c$ of the distant GCs (i.e., the left top ones in Figure~\ref{fig:gc})
could be underestimated. As the simple estimation for $\Gamma_e$ 
given above could suffer large uncertainties (e.g., \citealt{bah+13}), we 
consider the $\Gamma_e$ values as a very approximate indicator.

The numbers of \gr\ MSPs in each GCs may be further investigated by using
the spectral information obtained with \fermi\ LAT. \gr\ emission of pulsars 
generally has a form of a power law with an exponential cutoff 
(PLEC; \citealt{2fpsr13}),
while specifically for MSPs, \citet{xw16} have determined the spectral parameter
ranges using the spectra of 39 \gr\ MSPs known at the time. The \gr\ spectra
of the GCs have a similar PLEC form (e.g., \citealt{abd+10}), and there are the
particular cases, NGC~6624 and NGC~6626 (M28), which were found to have 
detectable \gr\ emission coming from a single bright 
MSP \citep{fre+11,wu+13,joh+13}.
Given the relatively well determined
spectral shape for MSPs, \gr\ emission of the GCs can be studied by fitting
their spectra with that of MSPs and thus the numbers of MSPs in the GCs
can be estimated. Previously, \gr\ spectra of the GCs have been studied in 
detail by \citet{lcb18} and \citep{son+21} through spectral fitting using 
different models. Spectral types different from PLEC or additional spectral
components have been claimed to be present, resulting in suggestions that
other types of high-energy sources in the GCs may also contribute to
the observed \gr\ emission.
A study with consideration 
of MSPs as the main sources in the GCs would potentially be able to help 
clarify the picture, checking if other components are needed in addition to
the MSPs' PLEC component.
\begin{figure}
\begin{center}
\includegraphics[width=0.95\linewidth]{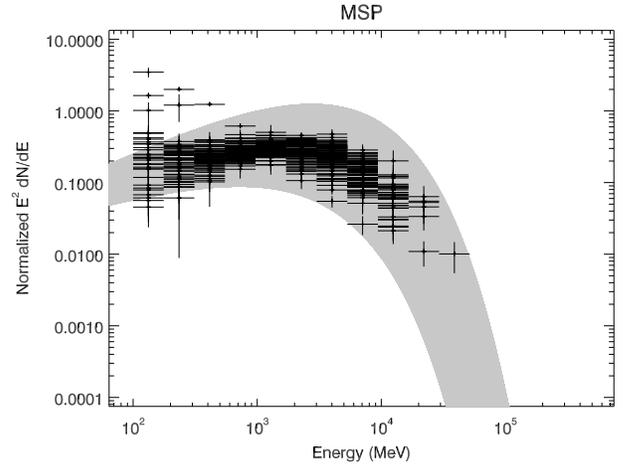}
\end{center}
	\caption{Normalized \gr\ spectra of 104 MSPs. A PLSEC model was used
	to fit the data points. The determined parameters are 
	$\Gamma=1.35^{+0.15}_{-0.16}$ and $E_c=1.45^{+0.60}_{-0.35}$\,GeV,
	while the 3$\sigma$ region is indicated by the gray area.
	}
\label{fig:mspec}
\end{figure}

\begin{figure*}
\begin{center}
\includegraphics[width=0.45\linewidth]{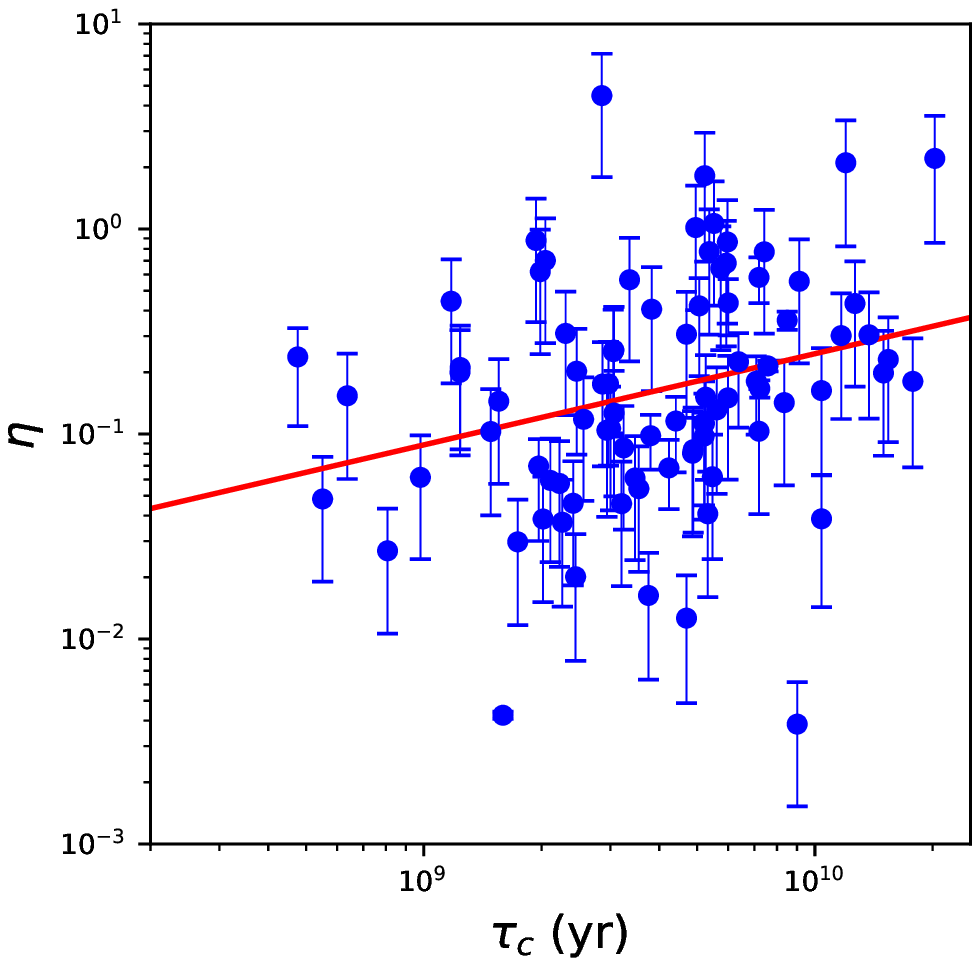}
\includegraphics[width=0.45\linewidth]{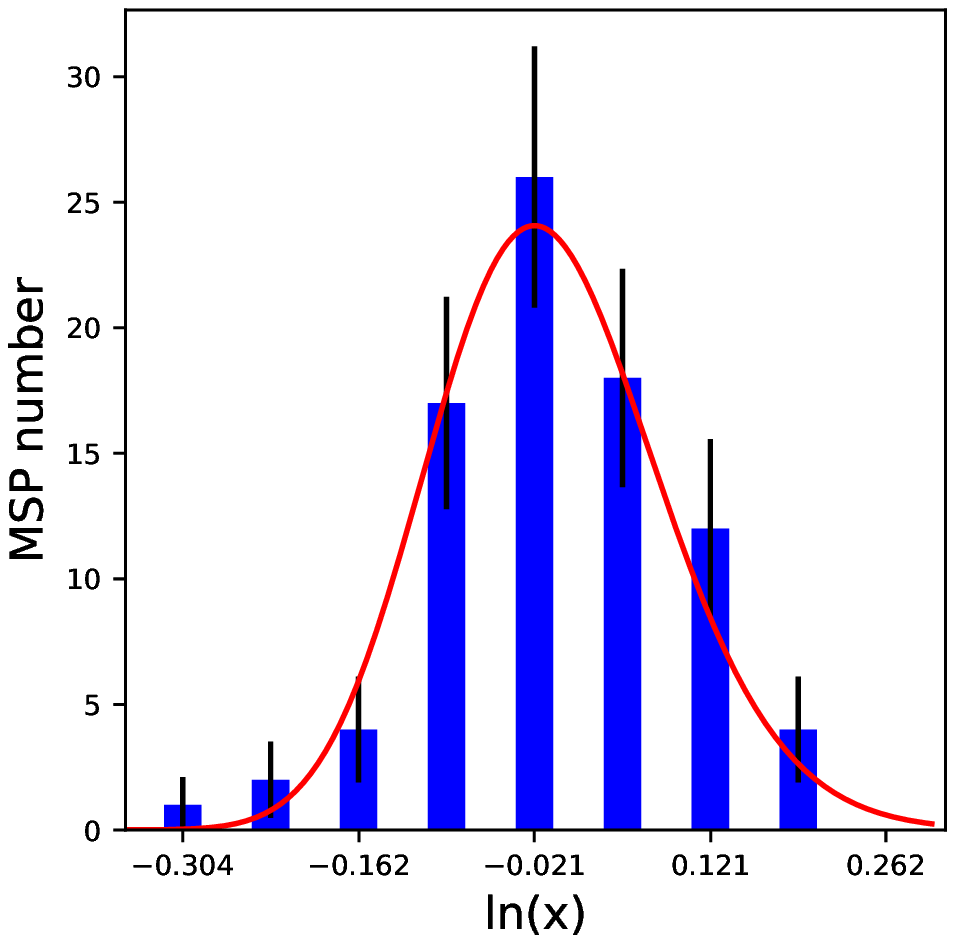}
\end{center}
	\caption{{\it Left panel:} \gr\ efficiency $\eta$ and characteristic 
	age $\tau_c$ of
	85 MSPs. A relationship of $\log\eta = 0.45\log\tau_c - 5.1$ (red line)
	was 
	obtained by fitting the data points. {\it Right panel:} distribution
	of the \gr\ MSPs along the direction perpendicular to the 
	$\log\eta$--$\log\tau_c$ relationship, where $x$ is the perpendicular
	distance of each MSPs from the relationship (cf., the left panel) and
	the uncertainties are the square root of the numbers of the MSPs.
	A log-normal function (red curve) is used
	to describe the distribution.
	}
\label{fig:et}
\end{figure*}
We thus conducted studies of the \gr\ spectra of the GCs, by fitting them
with those generated from the typical MSP \gr\ spectral parameters. In
the studies, we first
obtained the \gr\ spectra of 104 MSPs reported\footnote{https://confluence.slac.
stanford.edu/display/GLAMCOG/Public+List+of+LAT-Detected+Gamma-Ray+Pulsars},
and following \citet{xw16} determined the spectral parameter ranges for 
their PLEC form. Among the MSPs, 85 were found to have estimated distance and 
age values, providing a relationship between \gr\ efficiency $\eta$ and 
characteristic
age $\tau_c$ for \gr-emitting MSPs. We then obtained the \gr\ spectra of 
30 GCs (Table~\ref{tab:gc}; not including NGC~6624 and NGC~6626 
since their emission is dominated by a bright \gr\ MSP). For each of the GCs,
we generated a number of MSPs whose spin period $P$ 
values were randomly given based on the distribution of the 85 \gr\ MSPs 
and $\tau_c$ was assumed to be that of the GC. The spectra of 
the generated MSPs were randomly produced from the spectral parameter ranges.
As the distance of a GC is relatively well known, the spectrum, added from those
of the generated MSPs, was obtained, where the $\eta$--$\tau_c$ relationship 
was used to assign \gr\ luminosities to the MSPs . The model spectrum was 
compared with
the observed \gr\ spectrum of the GC, allowing to determine
the number of MSPs and check the MSP scenario for \gr\ emission of GCs. 

Using this method, the \gr\ spectra of the 30 GCs were studied. In addition,
six GCs (cf., Figure~\ref{fig:gc}) with relatively high $\Gamma_e$ 
and $<$5\,kpc distances, but no \gr\ emission were also studied as 
the comparison cases.
Below in Section~\ref{sec:ana}, we describe the analysis of the \fermi\ LAT
data to obtain the spectra of 104 \gr\ MSPs and those of the GCs. 
The detailed procedure for modeling and spectral fitting is given 
in Section~\ref{sec:mod}.
The results are presented in Section~\ref{sec:res}, and discussed
in Section~\ref{sec:dis}.

\section{\fermi\ LAT Data analysis}
\label{sec:ana}

\subsection{Spectral analysis for \gr\ MSPs}

In the analysis, 108 MSPs listed in the released \fermi\ LAT 10-year source 
catalog (4FGL-DR2; \citealt{bal+20}) were taken as our targets. We selected 
the 0.1--500 GeV LAT events within $20\arcdeg\times 20\arcdeg$ region 
centered at the position of each targets. The time duration of the data was 
from 2008-08-04 15:43:36 (UTC) to 2020-11-25 12:26:35 (UTC). Following 
the recommendations of the LAT team\footnote{\footnotesize http://fermi.gsfc.nasa.gov/ssc/data/analysis/scitools/}, we excluded the events with zenith angles 
larger than 90 degrees (to prevent the contamination from the Earth's limb)
and the events with quality flags of `bad'.

We constructed a source model for each of the targets based on 4FGL-DR2, 
which include the catalog sources within 20-degree radius centered at a
target. Their spectral forms are provided in the catalog. In our analysis,
the spectral normalization parameters of the sources within 5\,degree from a
target were set free, and all other parameters of the sources in the source
model were fixed to their catalog values. 
An MSP target was set as a point source having power-law emission 
with photon spectral index $\Gamma$ fixed at 2.0.
The background Galactic and extragalactic diffuse 
spectral models (gll\_iem\_v07.fits and iso\_P8R3\_SOURCE\_V2\_v1.txt
respectively) were also included in the source model, with their
normalizations set as free parameters in the analysis.

We extracted the \gr\ spectra of the MSP targets by performing the maximum 
likelihood analysis of the LAT data in 15 evenly divided energy band 
from 0.1 to 500 GeV in logarithmic scale. For the spectra, we kept spectral
data points with test statistic (TS) values greater than 16. Among the 108 
MSPs, four were
found to have $<2$ spectral data points and thus they were excluded from
the sample. A total of 670 data points were obtained for 104 MSPs.
\begin{figure}
\begin{center}
\includegraphics[width=0.95\linewidth]{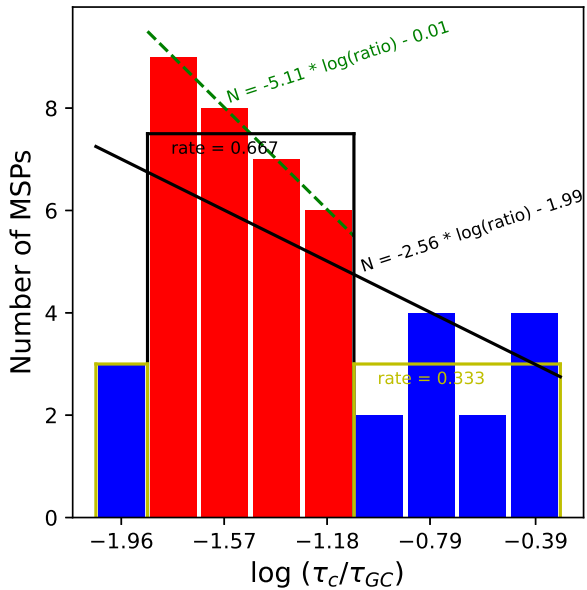}
\end{center}
	\caption{Distribution of the charateristic ages of the MSPs as 
compared to those of their host GCs. A relationship of 
$-2.56\log(\tau_c/\tau_{\rm GC})-1.99$ may be used to represent 
the distribution, but it would over- and under-predict the numbers of the MSPs.
Instead we used a relationship of $-5.11\log(\tau_c/\tau_{\rm GC})-0.01$ to
represent the red-bar MSPs and a constant to represent the blue-bar MSPs, the
fractions of the two groups being 66.7\% and 33.3\% respectively.
	}
\label{fig:tauc}
\end{figure}

\subsection{Spectral Analysis for the \gr\ GCs}
\label{sec:sag}

In the analysis, the 30 \gr\ GCs listed in 4FGL-DR2 were taken as the targets
(Table~\ref{tab:gc}). 
The time period of the LAT events we used was from 2008-08-04 15:43:36  
to 2021-07-06 04:59:55 (UTC), and the updated extragalactic diffuse 
spectral model iso\_P8R3\_SOURCE\_V3\_v1.txt was used in the source models. 
Except these, the data selection and source model construction were the same as 
those for the \gr\ MSPs. 

We extracted the \gr\ spectra of the GCs by performing the maximum likelihood 
analysis in 10 evenly divided energy band from 0.1 to 500 GeV in logarithmic 
scale. Because the GCs are relatively faint due to their large distances and
we would like to have as many as possible data points for their spectra,
we kept spectral data points with TS values greater than 4 
($>$2$\sigma$ significance) and calculated the 95\% flux upper 
limits otherwise.
\begin{figure}
\begin{center}
\includegraphics[width=0.95\linewidth]{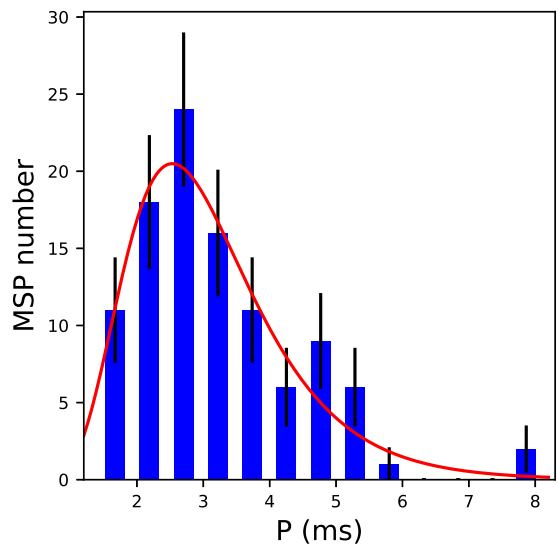}
\end{center}
	\caption{$P$ distribution of the \gr\ MSPs, where the uncertainties
	are the square root of the numbers. A log-normal function 
	(red curve) is used to describe the distribution, which has the
	peak at $P=2.5$\,ms.
	}
\label{fig:pdis}
\end{figure}

\subsection{Analysis for the six nearby GCs}
\label{sec:ngc}

We also searched for \gr\ emission from six nearby GCs, which are NGC~3201, 
NGC~6121, NGC~6254, NGC~6366, NGC~6544, and NGC~6656. Only NGC~6656 was found
to have \gr\ emission at its position (for its properties, see 
Table~\ref{tab:gc}), which had a TS value of 157 in 0.1--500\,GeV.
However this possible \gr\ counterpart
was marked with ASSOC2 in 4FGL-DR2, indicating that the association
probability is not sufficiently high.

We obtained the spectral upper limits for the five none-detected GCs and
spectral fluxes and upper limits for NGC~6656 following the procedure 
given above in Section~\ref{sec:sag}.

\begin{figure*}
\begin{center}
\includegraphics[width=0.41\linewidth]{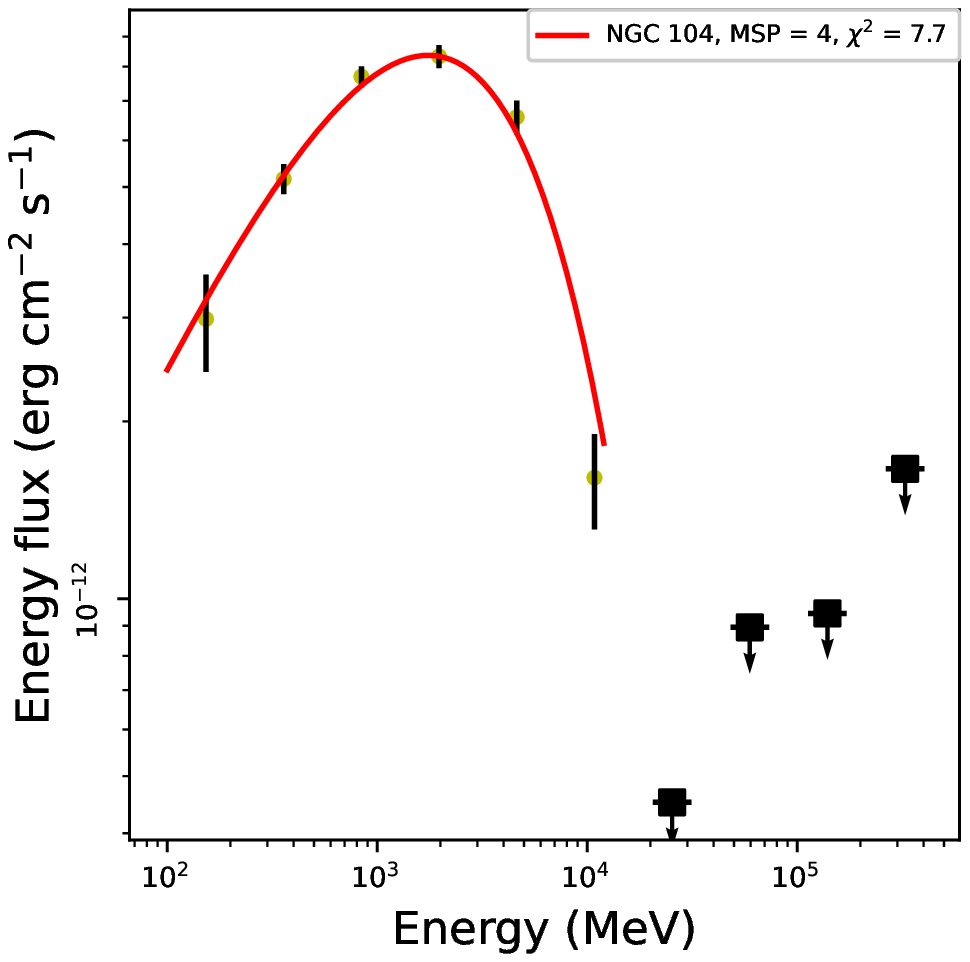}
\includegraphics[width=0.41\linewidth]{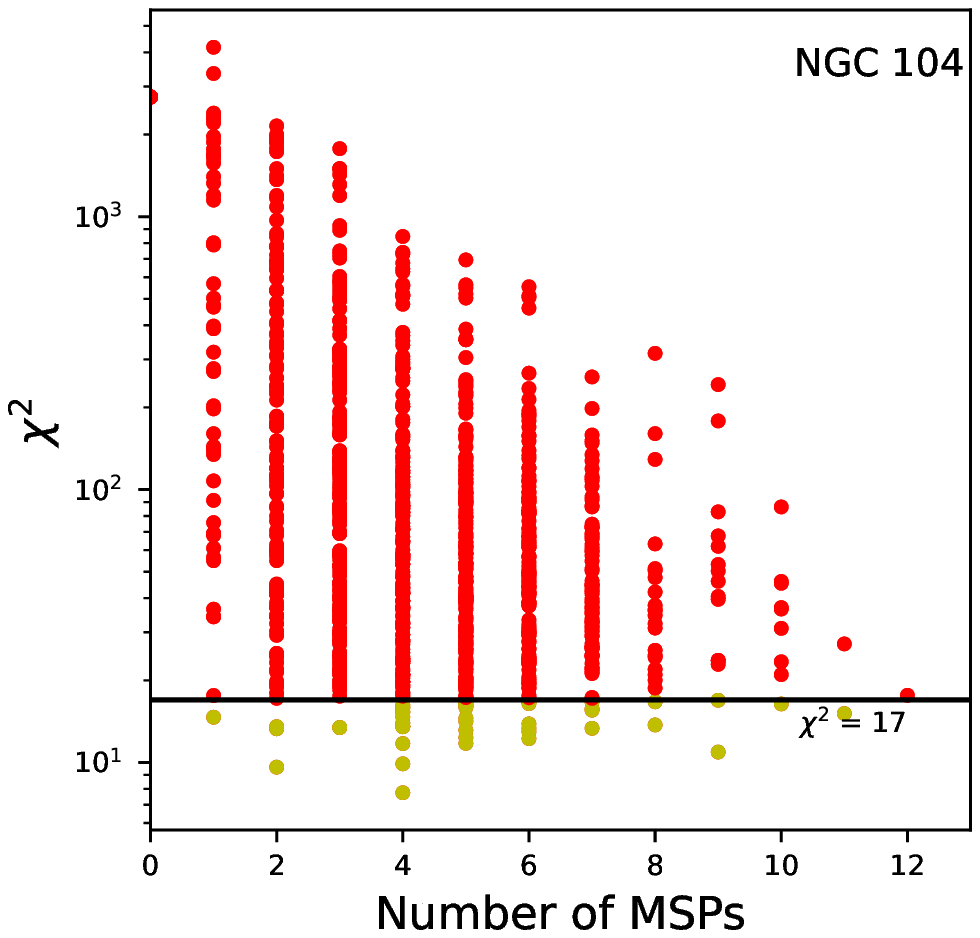}
\end{center}
	\caption{{\it Left} panel: \gr\ spectrum of NGC~104 (47~Tuc). 
	Four MSPs provide the best fit (red curve) to the spectrum.
	{\it Right} panel: $\chi^2$ values from 1000 runs, among which the 5\%
	smallest values are marked as golden data points. This 5\% limit 
	(black line) gives a range of 1--11 for the number of MSPs in this GC.
	}
\label{fig:tuc}
\end{figure*}
\section{Modeling and spectral fitting}
\label{sec:mod}
\subsection{Spectral parameter range determination for the \gr\ MSPs}
\label{sec:ssd}

Following the procedure described in \citet{xw16}, we determined the spectral 
shape based on the spectra of the 104 \gr\ MSPs. The spectral data points of 
each MSPs were normalized with the corresponding 0.1--100\,GeV energy flux 
given in 4FGL-DR2, and the normalized 104 spectra are shown in 
Figure~\ref{fig:mspec}. We fit these data points with a  
power law with a sub-exponential cutoff (PLSEC), 
$dN/dE = N_{0} E^{-\Gamma} \exp[-(E/E_{c})^{b}]$,
where $E_{c}$ is the cutoff energy and $b$ is fixed to 2/3, a characteristic 
value for \gr\ pulsars listed in 4FGL-DR2. $N_{0}$ was obtained 
for given $\Gamma$ and $E_{c}$ values by requiring the integrated energy flux
to be 1. 

In our fitting to the spectral data points, a systematic uncertainty parameter 
of 0.1 was added to the flux uncertainties in quadrature, possibly representing
the intrinsic spectral differences of the targets. With this value added,
the minimum reduced $\chi^{2}$ was approximately equal to 1. At a 3-$\sigma$
confidence level, $\Gamma= 1.35_{-0.16}^{+0.15}$ and
$E_{c}= 1.45_{-0.35}^{+0.6}$\,GeV were obtained. 
The 3$\sigma$ region is shown in Figure~\ref{fig:mspec}.

\subsection{\gr\ efficiency and characteristic age distribution
of \gr\ MSPs}
\label{sec:sdis}

For the 104 \gr\ MSPs, we searched their spin-down rate $\dot{P}$ and
distance information in literature, and 85 of them were found to have
the information (listed in Table~\ref{tab:msp}). From spin period $P$
and $\dot{P}$, the spin-down luminosities $\dot{E}$ and 
$\tau_c$ were calculated. 
The \gr\ luminosities $L_{\gamma}$ were also obtained, where the 
energy fluxes in 0.1--100\,GeV were from 4FGL-DR2 and most of
the 85 MSPs only had distances estimated from the dispersion measures (DMs). 
For the DM distances, a 30\% uncertainty was assumed.
Then \gr\ efficiency $\eta$ was calculated from $L_{\gamma}/\dot{E}$.

Values of $\eta$ and $\tau_c$ of the 85 MSPs are shown in the left panel of
Figure~\ref{fig:et}. 
We note that four MSPs have face values of $\eta$ greater than 1. The 
unphysical values could be caused by over-estimation of distance values or
the uncertain beam correction factor $f_{\Omega}$ (we followed \citealt{2fpsr13}
and assumed $f_{\Omega}=1$). There appears to have a trend between
$\eta$ and $\tau_c$: older pulsars tend to have larger $\eta$. Using
the linear least squares method, we found a relationship of 
$\log\eta = 0.45\log\tau_c - 5.1$ (see Figure~\ref{fig:et}).

Nearly an order of magnitude scatters are present around the relationship. 
In order to have them
accounted for, we calculated the perpendicular distance (denoted by $x$) of 
each MSPs in
the left panel of Figure~\ref{fig:et} from the relationship,
and their distribution along the direction perpendicular to the relationship 
was obtained (right panel of Figure~\ref{fig:et}). While the distribution
is close to being Gaussian, we found that it was slightly better fit
with a log-normal 
function $\sim \exp[-(\ln x - \mu)^2/2\sigma_l^2]$, where $\mu$ is the $\ln x$
value at the distribution peak. From the fitting,
we obtained $\sigma_l\simeq 0.093$ and $\mu=-0.021$, 
where the minimum $\chi^2\simeq 4.2$ for 5 degrees of freedom (DoF).

\subsection{Characteristic ages of the GC MSPs}
\label{sec:tauc}

We collected the characteristic ages of the MSPs listed in the GC radio pulsar 
catalog\footnote{\footnotesize http://www.naic.edu/$\sim$pfreire/GCpsr.html}, 
and compared them with the ages $\tau_{\rm GC}$ of their corresponding GCs. 
The ratio distribution between the former and the latter is shown in 
Figure~\ref{fig:tauc}. As can be seen, the ratios are in a range of 
$\sim$0.01--0.4. There is a possible trend: the younger the MSPs, the more
there are. We tested to fit the distribution with one simple function, which
would return the number of the MSPs to be 
$-2.56\log(\tau_c/\tau_{\rm GC})-1.99$. This simple function obviously
would over- or under-predicts the numbers of MSPs at different
ratio values.  Alternatively we noted that the ranges
of $\log(\tau_c/\tau_{\rm GC})$ from $-1.765$ to $-$1.18 can be described with
a function of $-5.11\log(\tau_c/\tau_{\rm GC})-0.01$, which constitute
66.7\% of the MSPs, and the remaining 33.3\% MSPs be described with a 
constant distribution (cf., Figure~\ref{fig:tauc}).

We tested to generate pulsars based on the two ways of describing the 
ratio distribution, and the resulting numbers of \gr\ MSPs were similar
(cf., Section~\ref{sec:gs} \& \ref{sec:sfp}). We chose to use the second one 
in this work, as it more accurately describes the current GC MSP sample.

\subsection{Period distribution of \gr\ MSPs}

The final piece of information needed is $P$ of a pulsar. We constructed
the $P$ distribution for the \gr\ MSPs, which is shown in Figure~\ref{fig:pdis}.
Comparing to the whole MSP sample listed in the Australia Telescope 
National Facility (ATNF) pulsar catalogue (more than 300 MSPs; 
\citealt{man+05}), the \gr\ MSPs
tend to have shorter $P$. We fit the distribution with a log-normal function,
and obtained $P=2.5$\,ms at the peak of the function and $\sigma_l = 0.38$
(the minimum $\chi^2\simeq 6.9$ for DoF$=7$).
\begin{figure*}
\begin{center}
\includegraphics[width=0.73\linewidth]{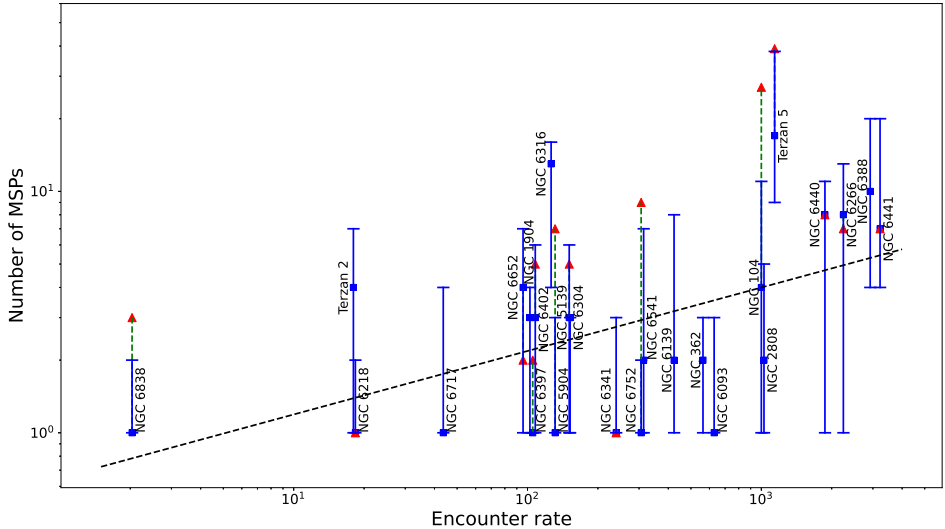}
\end{center}
	\caption{Numbers of \gr\ MSPs (blue squares) resulting from spectral 
fitting, with the error bars indicating the ranges estimated from
the 5\% best fitting results. The numbers of radio MSPs in the GCs
	are marked with red triangles (connected with green dashed lines to 
	corresponding blue squares) . The black dashed line indicates a possible
	relationship $\log N_{\rm MSP} \sim 0.26\log\Gamma_e $ 
(cf., Section~\ref{sec:comp}).
	}
\label{fig:ne}
\end{figure*}

\subsection{\gr\ spectrum generation for an MSP in a GC}
\label{sec:gs}

Using the models set up above, we generated \gr\ spectra of MSPs in each GC 
targets.  The procedure is listed below:
\begin{itemize}
	\item Generate $P$ for a pulsar based on the $P$ distribution of the \gr\ MSPs;
	\item $\tau_c$ of this pulsar is generated based on the 
$\log(\tau_c/\tau_{\rm GC})$ distribution that depends on the age of a GC target;
	\item $\dot{P}$ is estimated from $P$ and $\tau_c$ using $\tau_c=P/2\dot{P}$, and $\dot{E}$ is also calculated for this pulsar;
	\item Given $\tau_c$, the $\log \eta$--$\log \tau_c$ relationship is used to obtain an initial $\eta_i$ value, and for this pulsar $\eta = \eta_i + \Delta\eta$, where the latter is generated based on the distribution for the scatters around the relationship (cf., Section~\ref{sec:sdis}); 
	\item $L_{\gamma}$ of this pulsar is obtained from $\dot{E}$ and $\eta$;
	\item Based on the spectral parameter ranges (we used 3$\sigma$ ranges) determined for the \gr\ MSPs in Section~\ref{sec:ssd}, $\Gamma$ and $E_c$ are randomly generated for this pulsar, and its \gr\ spectrum is obtained given $L_{\gamma}$ and the distance to the GC target.
\end{itemize}

\subsection{Spectral fitting procedure for the GCs}
\label{sec:sfp}

The \gr\ spectrum (or spectral upper limits) obtained for each \gr\ GCs 
(or nearby GCs) was fitted with spectra of MSPs generated according to 
the steps given above in Section~\ref{sec:gs}. For a GC target, we started 
from one pulsar, obtained
its spectrum, fit the spectrum (or spectral upper limits) of the GC with it, 
obtained $\chi^2$, and then repeated the process of fitting with the spectrum 
added from that of two, three, and more pulsars. In this run of fitting, 
the minimum $\chi^2$ was found and the corresponding number of MSPs was 
obtained. For each GC targets, 1000 such runs were conducted.
Since there are flux upper limits in the GC spectra, one condition was 
set: when a model spectrum at the energy of an upper limit has a flux
greater than the upper limit, the fitting was considered to have reached
a limit and the run was stopped.
When such a case was met, if the last number of MSPs (before reaching over 
an upper limit) provided the smallest $\chi^2$, the $\chi^2$ value was taken 
as the minimum one and the number as that of MSPs.

\section{Results}
\label{sec:res}
We carried out fitting to the spectrum of each GC targets, using the procedure 
given above. A number for MSPs in a given GC that results in a minimum 
$\chi^2$ value was obtained for most cases.  In order to provide a range for 
the MSP number, 
we defined 5\% lowest $\chi^2$ values in 1000 runs and used their corresponding
MSP numbers as the representative range values. Figure~\ref{fig:tuc} shows 
an example of our fitting results, in which the \gr\ spectrum of the GC 
NGC~104 (47~Tuc) was best fitted with 4 MSPs and a range of 1--11 was 
estimated from the lowest 5\% $\chi^2$ values. 
The results for 30 \gr\ GCs are summarized in Table~\ref{tab:gc}. The
estimated numbers of \gr\ MSPs can also be seen in Figure~\ref{fig:ne}, in
which GLIMPSE-C01, GLIMPSE-C02, 2MS-GC01, NGC~7078, and Terzan~1 are not
included because they do not either have $\Gamma_e$ or have a MSP-type spectrum
(for the latter case, see below).
Most of the figures showing the spectra and fitting results for the GCs are 
presented in Appendix Section~\ref{sec:sf}. 

\begin{figure*}
\begin{center}
        \includegraphics[width=0.41\linewidth]{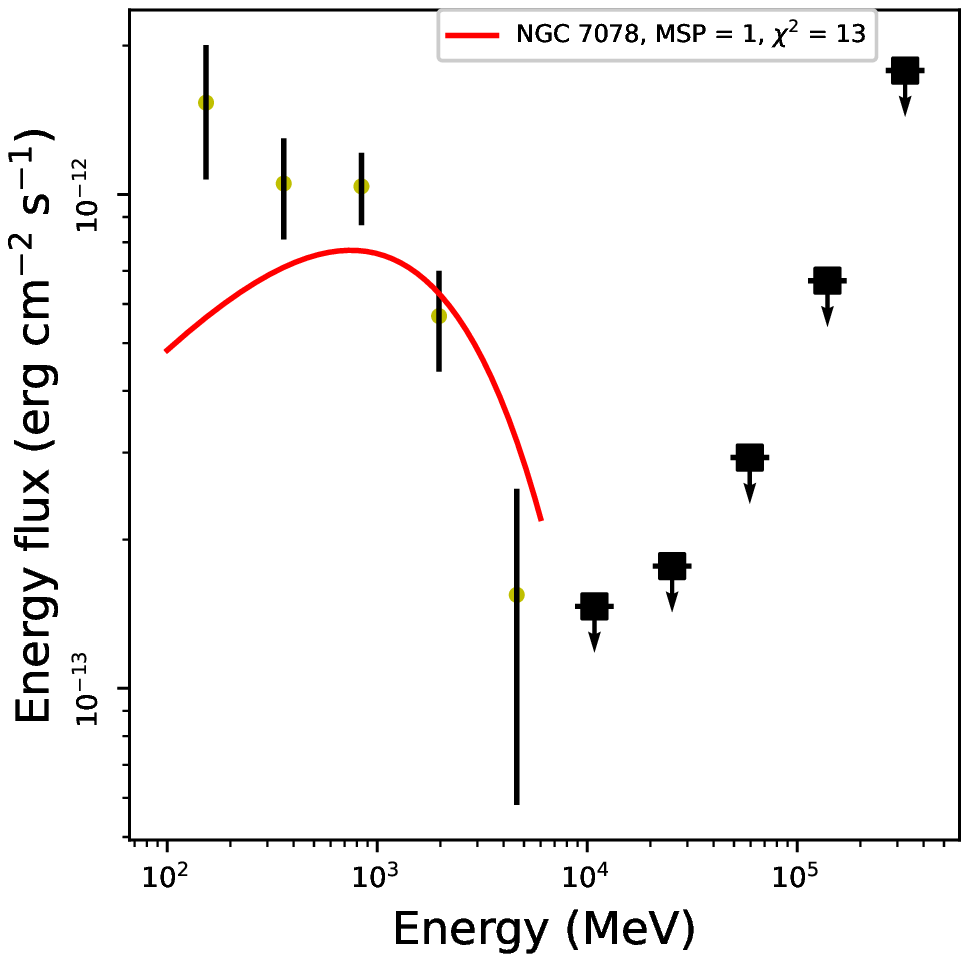}
\includegraphics[width=0.41\linewidth]{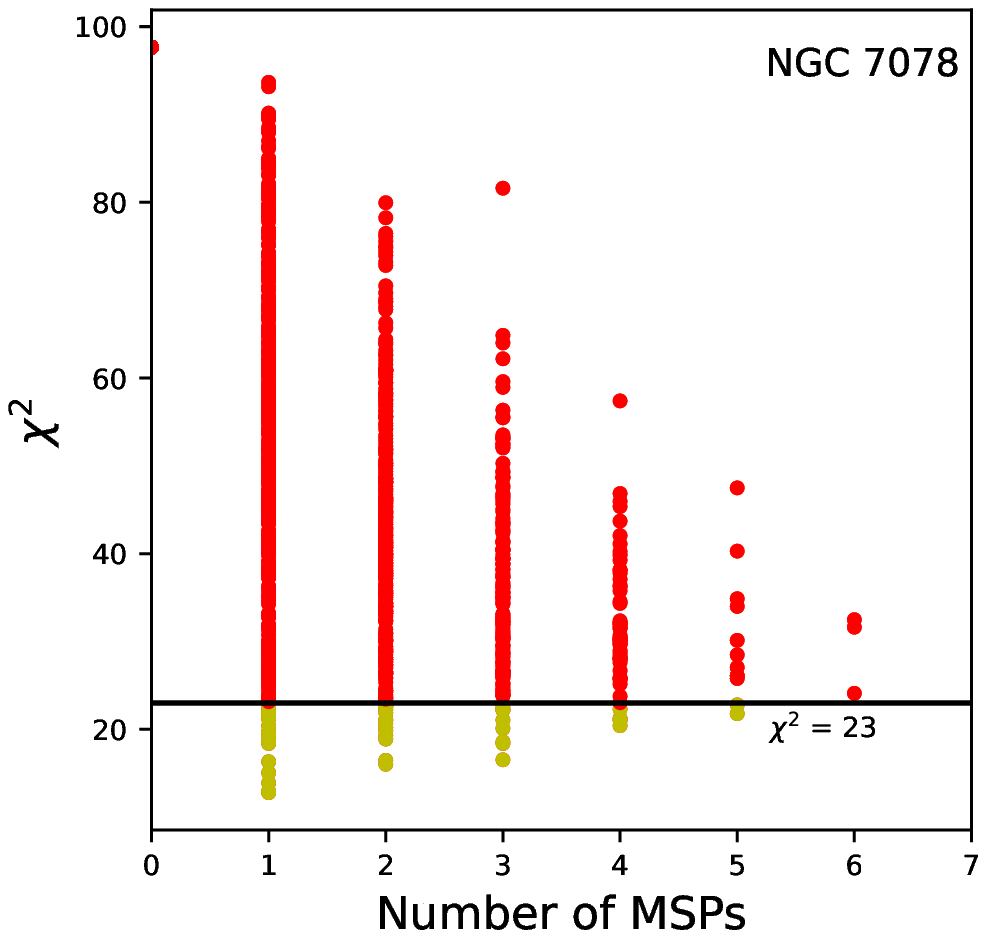}

\includegraphics[width=0.41\linewidth]{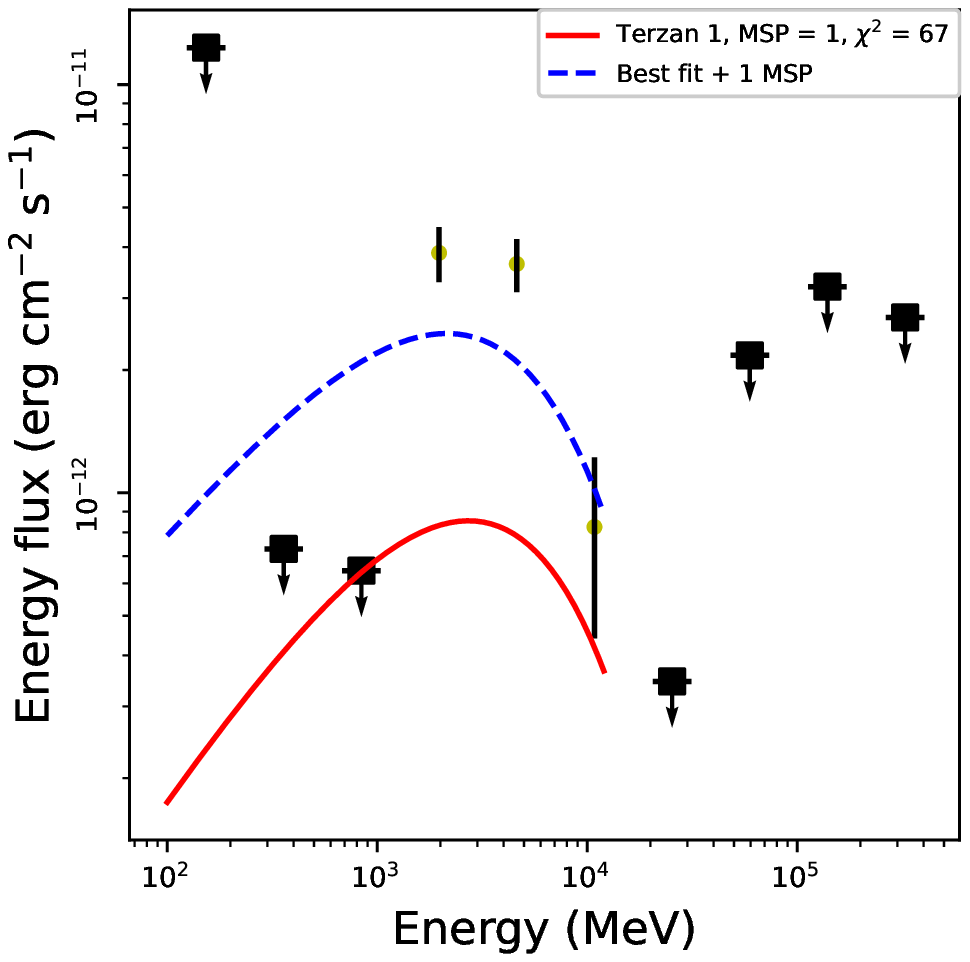}
\includegraphics[width=0.41\linewidth]{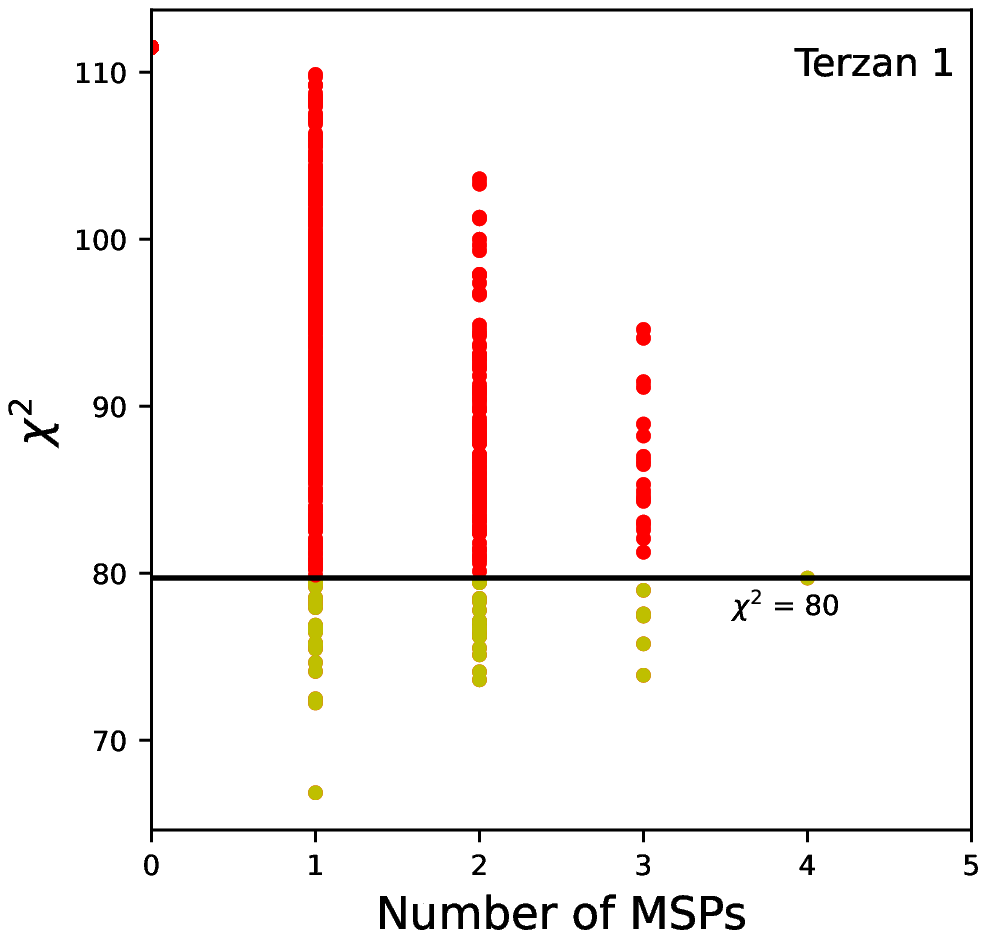}

\includegraphics[width=0.41\linewidth]{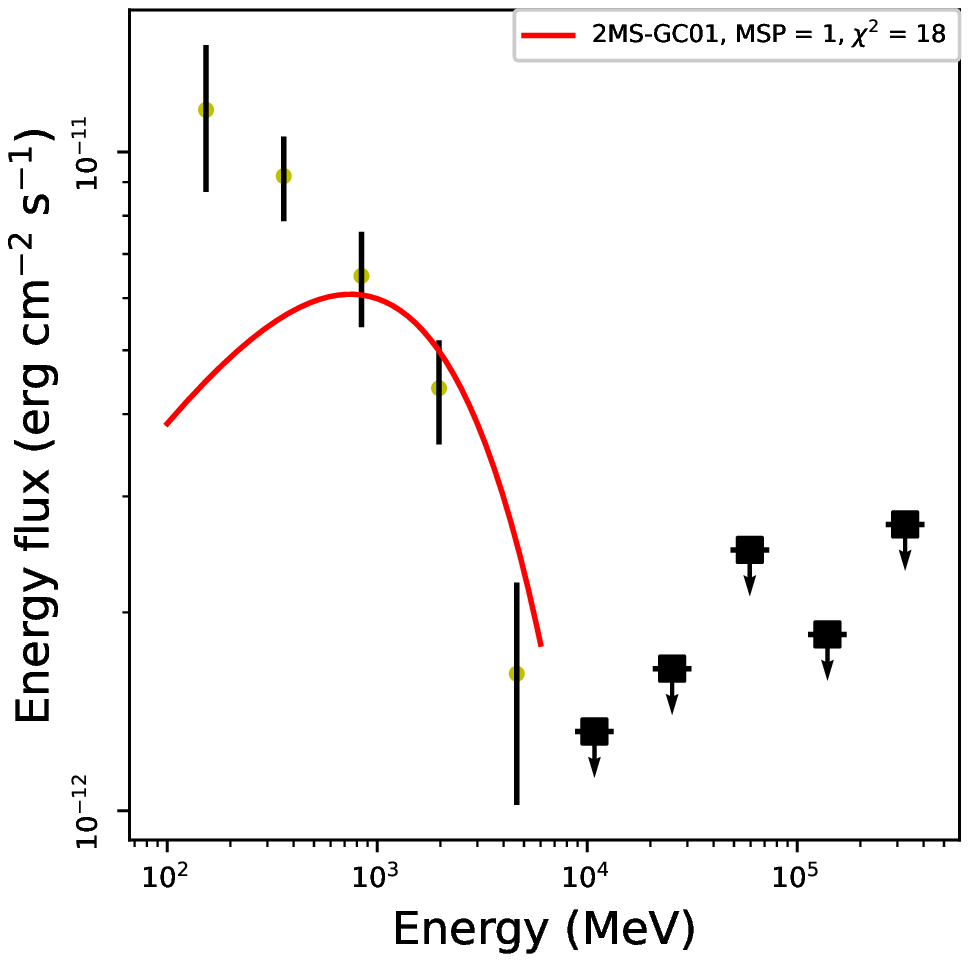}
\includegraphics[width=0.41\linewidth]{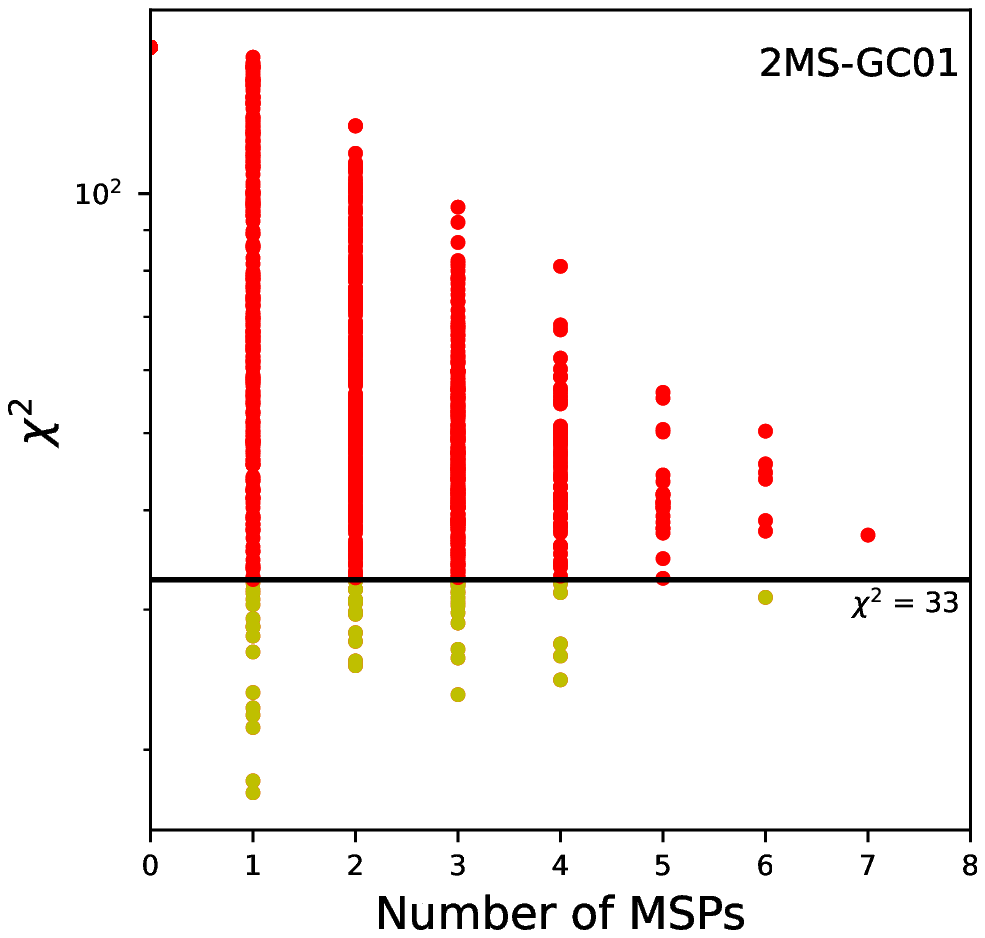}
\end{center}
        \caption{Same as Figure~\ref{fig:tuc}, but the spectra 
of these three GCs (NGC~7078, Terzan~1, and 2MS-GC01) can not be well fitted 
	with our MSP model.}
\label{fig:bad}
\end{figure*}

The spectra we obtained for the \gr\ GCs mostly contain 3--6 flux data 
points,
with Terzan~5 as the exception having 8 data points. The minimum $\chi^2$ 
values from
spectral fitting can generally indicate the goodness of how well a spectrum is 
fitted. Based on the $\chi^2$ values (see Table~\ref{fig:gc} and figures in 
Appendix Section~\ref{sec:sf}), we conclude that the \gr\ emission from most
of the GCs can be explained with the containment of \gr\ MSPs 
($\chi^2\lesssim 6$), although it should be noted that some of the spectra 
have large flux uncertainties. Four GCs, NGC~104, NGC~1904, Terzan~5, and 
GLIMPSE-C02, have the minimum $\chi^2$ values of 7--10. The relatively 
large values
suggest that MSPs may not be the only \gr\ sources in these GCs; 
additional components may
be needed. Finally there are three GCs, NGC~7078, Terzan~1, and 2MS-GC01, 
that have the minimum $\chi^2\geq 13$.  Their spectra and fitting results are 
shown in Figure~\ref{fig:bad}.  Examining their spectral fitting,  
whether their \gr\ emissions are due to MSPs is not certain. 
 
\section{Discussion and Summary}
\label{sec:dis}

From the whole-sky observations for more than 10 years with \fermi\ LAT,
highly representative samples for different types of high-energy objects have
been established. We took advantage of the progress, and built a model to 
study the GCs' \gr\ emission based
on the presumption that the \gr\ MSPs in the GCs are the main source
of the \gr\ emission.
The model produces \gr\ spectra of MSPs randomly but
within the properties of the known \gr\ MSP sample. Using spectra generated
from the model, we have
fitted the \gr\ spectra of the known \gr\ GCs and determined the numbers of
\gr\ MSPs in the GCs. Among 30 known \gr\ GCs, we have found that
23 have \gr\ spectra consistent with arising from \gr\ MSPs, 4 have
the relatively large minimum $\chi^2$ values, and 3 are not certain to
have MSPs as the
main \gr\ sources in them (non-MSP type). 
Below we first discuss the latter two types of the cases.
As we also investigated \gr\ properties of six nearby GCs that have 
relatively high $\Gamma_e$, we discuss our test results with our model 
applying to fitting of their \gr\ spectrum or spectral upper limits.
We then compare our results with those of previous studies of 
the \gr\ GCs and related radio MSPs. In the end, we provide a summary of
this work.

\subsection{GCs with the relatively large minimum $\chi^2$}

In our fitting, NGC 104, NGC 1904, Terzan 5, and GLIMPSE-C02 have the minimum
$\chi^2$ values of 7--10, which possibly indicate that the fitting for 
the four sources may not be satisfactory. In 
particular, Terzan~5 is required to contain the most, 9--38 \gr\ MSPs 
based on our fitting results, and NGC~104 (47~Tuc) has been reported to show 
an 18.4\,hr
periodic signal in its \gr\ emission \citep{zzy+20}. Both facts suggest that
these two GCs could be more complicated than the MSP scenario considered
in this work. 

\begin{figure}
\begin{center}
        \includegraphics[width=0.81\linewidth]{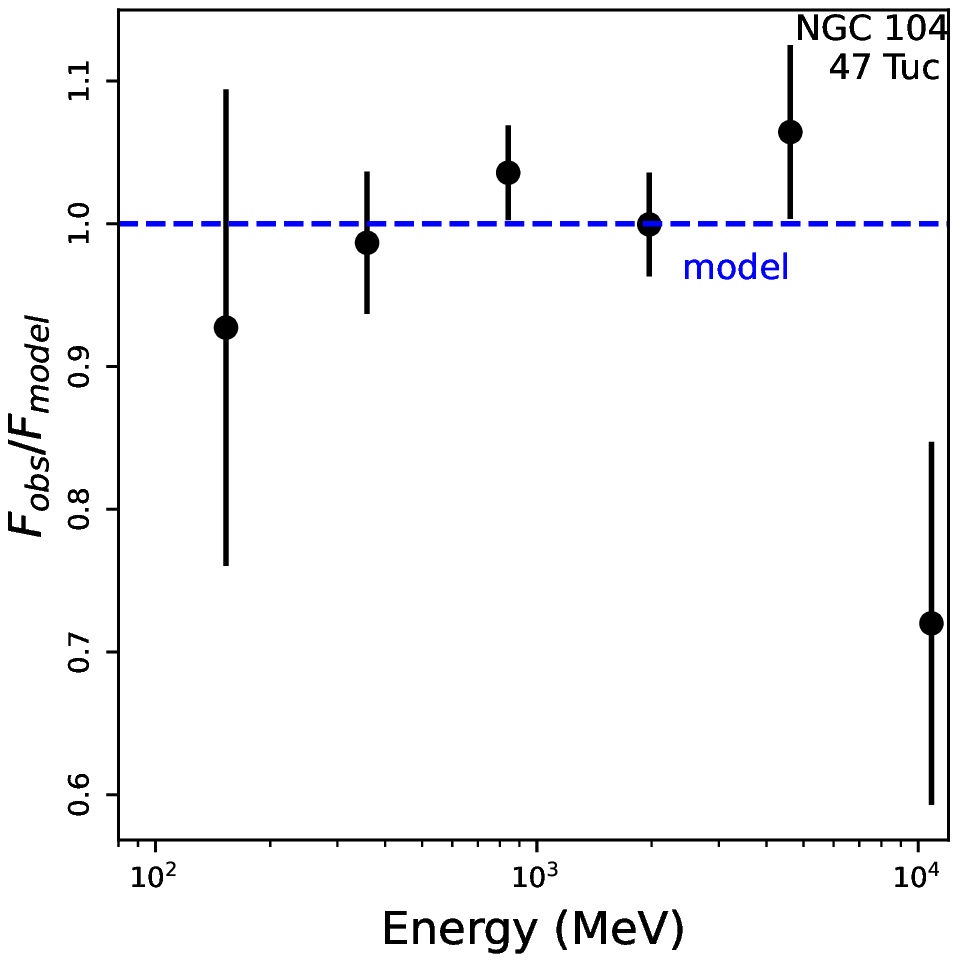}
\includegraphics[width=0.81\linewidth]{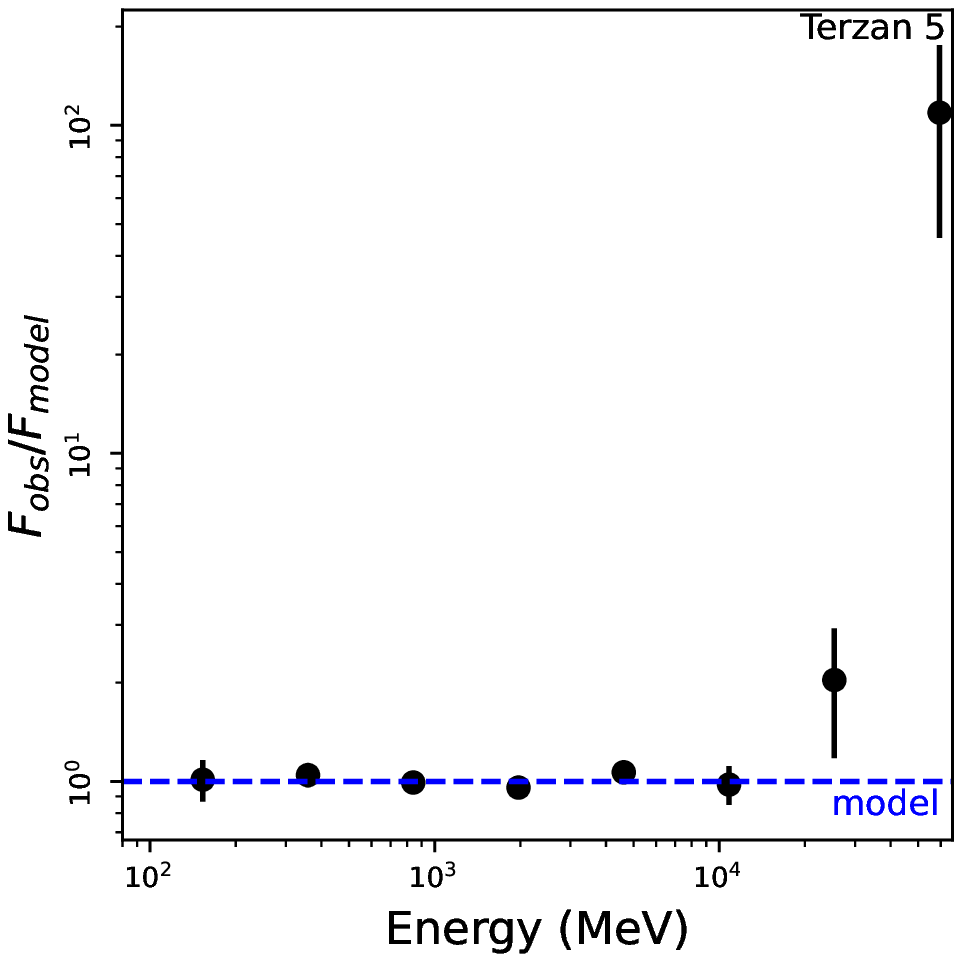}
\end{center}
	\caption{{\it Upper} panel: spectral data points normalized by the 
	best-fitting model of NGC~104. {\it Bottom} panel: the same as the upper
	panel for Terzan~5.}
\label{fig:res}
\end{figure}

We checked the residuals of the spectral fitting for the four GCs, but no 
significant deviations from the respective best-fitting model spectrum were 
found, partly due to the large flux uncertainties of the observed spectra. 
In Figure~\ref{fig:res}, the spectral data points of NGC~104 and Terzan~5,
normalized by the respective best-fitting model spectra, are
shown, as these two have the last data point appearing away from the model.
In particular for Terzan~5, the data point is nearly 100 times the model flux. 
However, because the flux uncertainties are large, the significances
of the deviations of the last data points are 2.3$\sigma$ in NGC~104
and 1.7$\sigma$ in Terzan~5 
respectively. Thus both deviations are not sufficiently 
significant such that additional source of \gr\ emission would be definitely
needed. Recently \citet{son+21} have applied a
double-component (PLEC plus power law) model to spectral fitting for 
the \gr\ GCs and stated that a power law component is significantly present. 
Although the significances in our studies
are not high, the deviations at the high-energy tails of the
spectra of NGC~104 and Terzan~5 do hint the existence of an extra component 
in \gr\ emission from these two GCs.

\begin{figure*}
\begin{center}
        \includegraphics[width=0.30\linewidth]{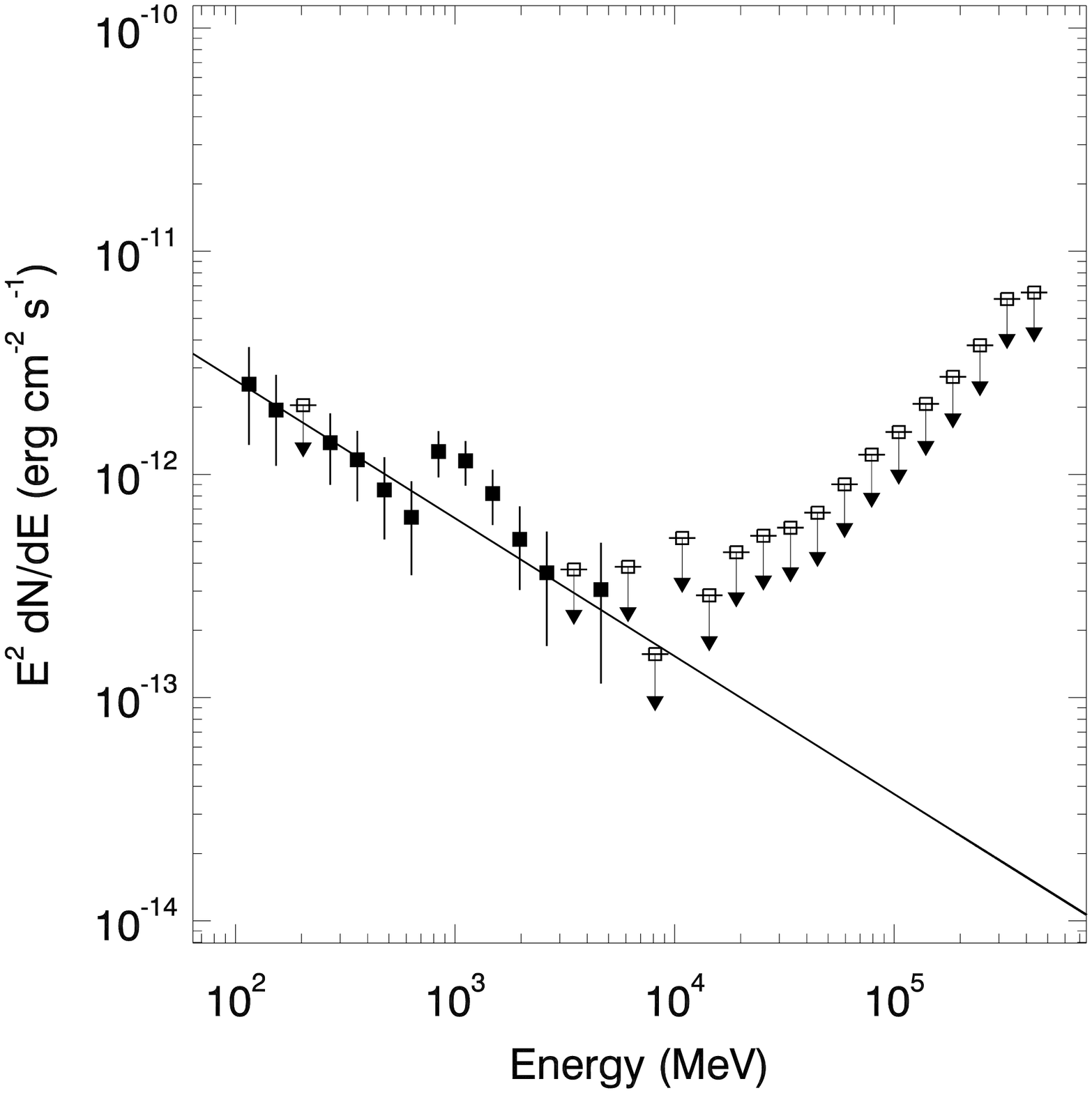}
        \includegraphics[width=0.30\linewidth]{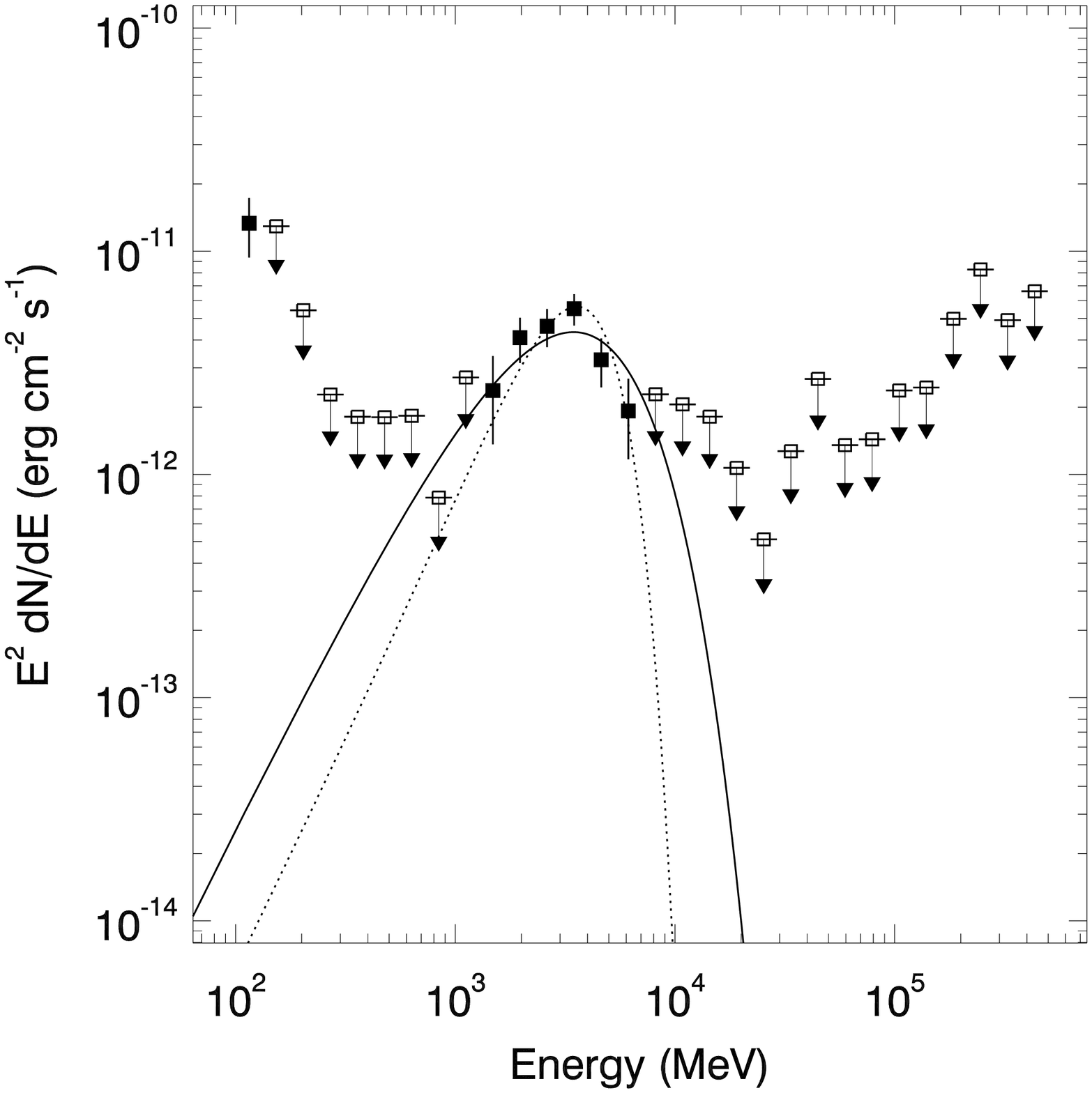}
        \includegraphics[width=0.31\linewidth]{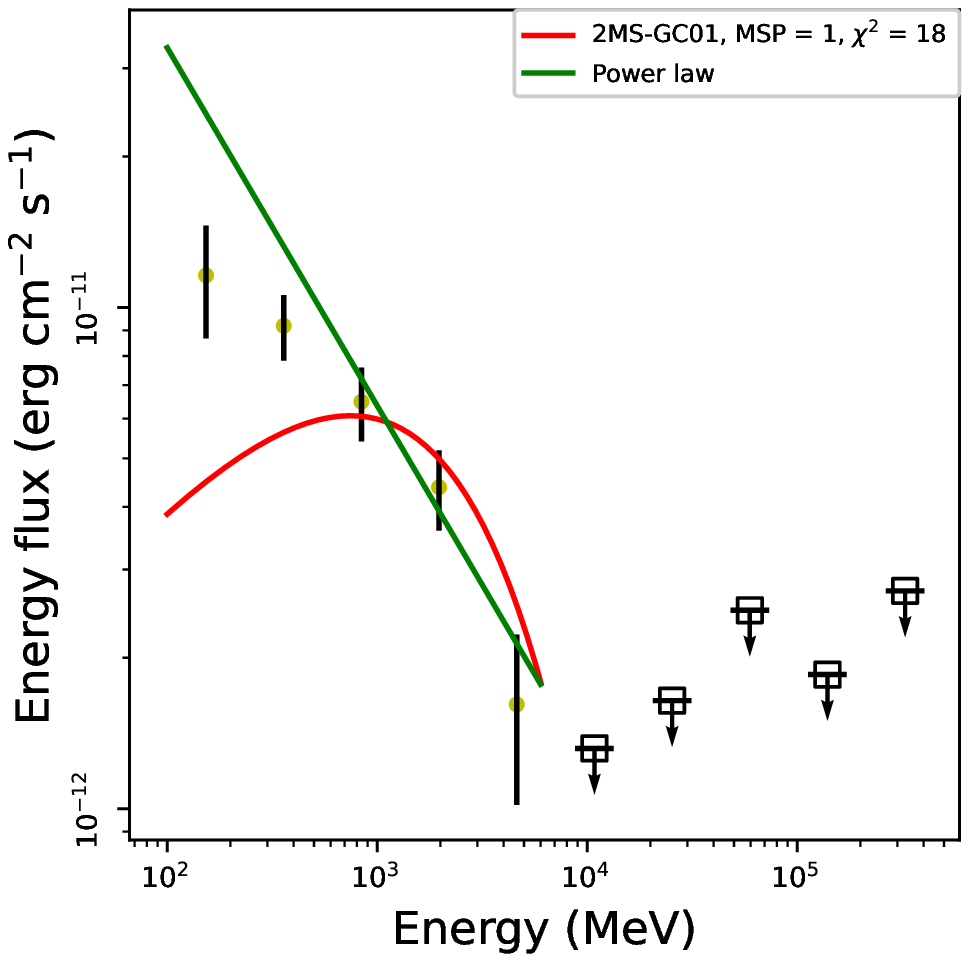}
\end{center}
	\caption{{\it Left} panel: spectrum of NGC~7078 with fine energy bins. 
	A power law with 2.6$\pm0.1$ photon index, given in 
	4FGL-DR2 \citep{bal+20}, can fit the spectrum approximately.
	{\it Middle} panel: spectrum of Terzan 1, for which either a PLEC 
	(solid curve)
	or a Gaussian (dashed curve) model can provide a fit. 
	{\it Right} panel: a power law fit (green line), with 2.714$\pm0.007$ 
	photon index, to the spectrum of 2MS-GC01.}
\label{fig:pls}
\end{figure*}
\subsection{Non-MSP type}

As can be seen in Figure~\ref{fig:bad}, the spectra of 
NGC~7078, Terzan~1, and 2MS-GC01 do not have pulsars' PLEC spectral
shape and can not be well fit with our model spectra of MSPs. The
first seemingly has a sharp drop (cutoff) after $\sim$1~GeV, the second has 
a relatively narrow spectral range of 1--10\,GeV, and the third is more 
consistent with a power law.
For these three sources, we constructed TS maps and confirmed the detection
of them and in particular the narrow spectral range of Terzan~1. For example,
in the energy ranges of $\leq 1$\,GeV or $\geq 10$\,GeV, no sources were seen
at the position of Terzan~1 in the TS maps. In order to check if the \gr\ 
sources at the positions of the three GCs could be background sources such as
blazars, we also constructed their light curves (60-d a bin). No obvious 
variability showing flares or significant flux changes was seen from each 
of them. 

We extracted fine energy-bin spectra of NGC~7078 and Terzan~1, for the purpose 
of better understanding their spectral shapes. The spectrum of NGC~7078 is shown
in the left panel of Figure~\ref{fig:pls}, which can approximately be described 
with a power law. In fact, power-law emission (with $\Gamma=2.6\pm0.1$) 
is given in 4FGL-DR2 for the source 
and it is consistent with the spectrum we obtained (Figure~\ref{fig:pls}). 

The fine energy-bin spectrum of Terzan~1 is shown in the middle panel of
Figure~\ref{fig:pls}.
This spectrum is actually not typical among those of the known Galactic \gr\ 
sources such as MSPs or other classes; MSPs generally have a PLEC shape with
the power-law components detectable from 0.1\,GeV to several GeV (cf., Figure~\ref{fig:mspec}), and
others are often described with a power law (e.g., \citealt{4fgl20}). 
We conducted global fits to the emission of Terzan~1 with a PLEC model and 
a Gaussian one. For the first model, 
the TS value is 148 with $\Gamma\simeq 0$ and cutoff energy 
$E_c=1.726\pm0.005$\,GeV. The difference of the emission from that of a typical
MSP is its hardness,
as it is detectable only above $\sim$1\,GeV. For the second model, the fitting
resulted in a slightly higher TS value, 151, with the peak value at 
1.492$\pm0.012$\,GeV and the standard deviation of the Gaussian function 
being 1.933$\pm0.012$\,GeV.
The narrow energy emission, likely described with a Gaussian function, may 
suggest an abnormal origin for the emission, such as the dark matter particle
annihilations (see, e.g., \citealt{bhl15}). We note that 
Terzan~1 is located along the direction towards the Galactic center 
and it is the only special case we found
among the GCs. Further discussion for the origin of its emission is beyond 
the scope of this paper and will be presented elsewhere (Huang, Wang, et al.
in preparation).

For 2MS-GC01, we conducted a global fit to its emission. To avoid possible 
contamination from the background since the GC is located close to the
Galactic center, the energy range we used was 0.3--500\,GeV. We obtained
a power law with $\Gamma = 2.714\pm0.007$. This power law,
shown in the right panel of Figure~\ref{fig:pls}, can approximately describe 
the source's spectrum, although at the low energies, the spectral data points 
are slightly
lower than the model fit. We conclude that a power law at least better
fits the spectrum of the GC.

\subsection{Nearby GCs with high encounter rates}
\label{sec:ngcs}

\begin{figure*}
\begin{center}
        \includegraphics[width=0.41\linewidth]{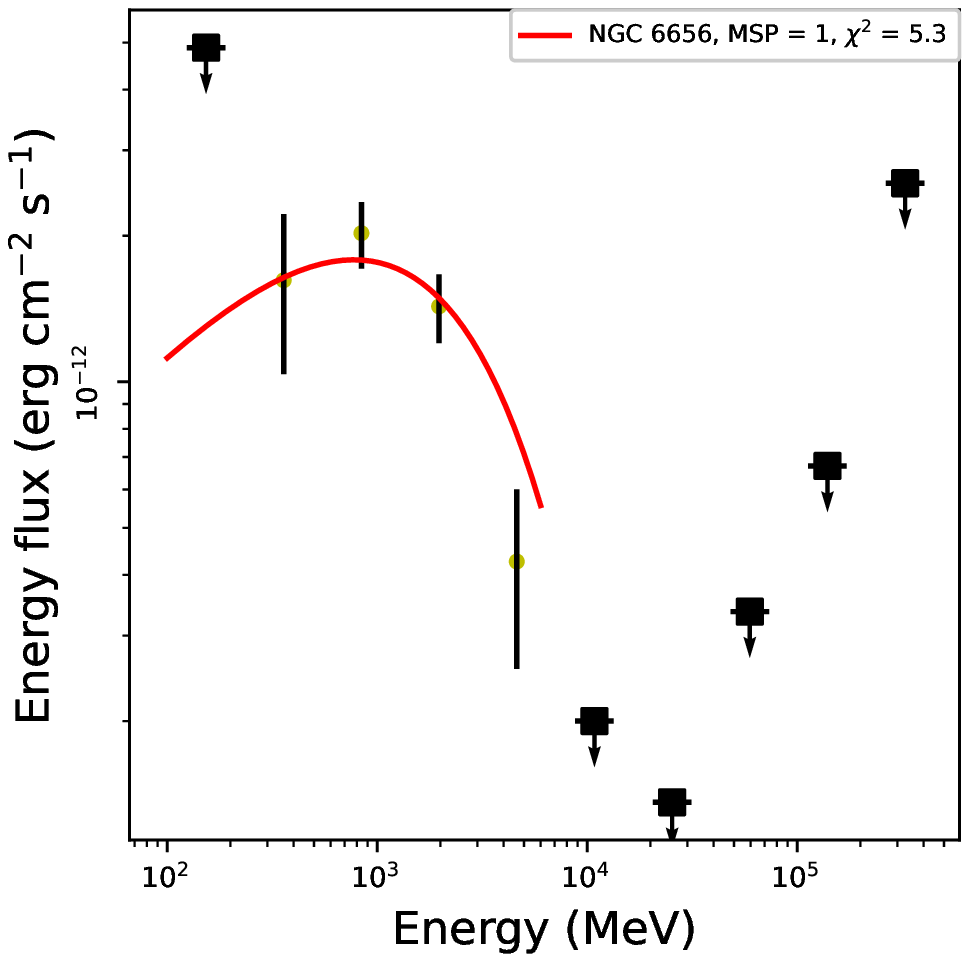}
\includegraphics[width=0.41\linewidth]{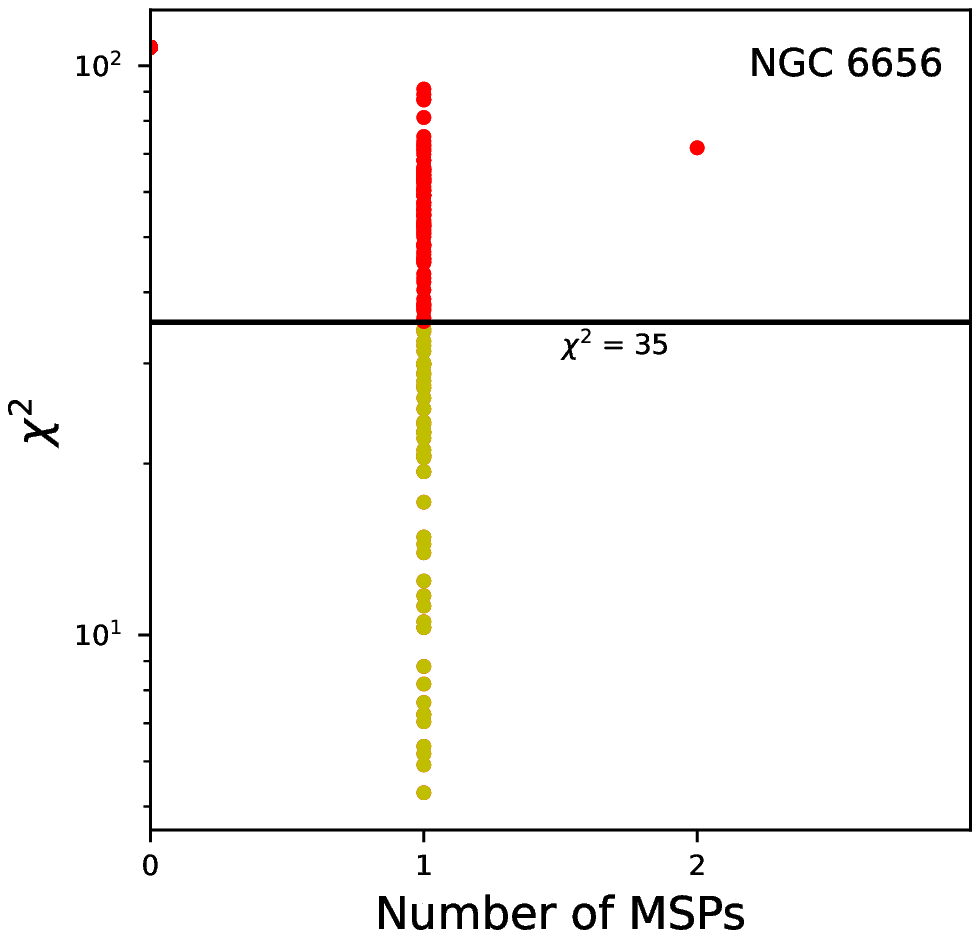}
\end{center}
	\caption{Same as Figure~\ref{fig:tuc}, for the candidate 
counterpart of NGC~6656 (whose TS=157 in 0.1--500\,GeV).
}
\label{fig:n6656}
\end{figure*}

For six nearby GCs with high $\Gamma_e$, only NGC~6656 has a possible \gr\
counterpart, but the counterpart does not have a sufficiently high association
probability according to 4FGL-DR2. In any case, we fit its \gr\ spectrum with
our model and the fitting results are shown in Figure~\ref{fig:n6656}. 
Interestingly, we found that the spectrum can be fit with a single MSP
and the number of MSPs in this GC would be limited to be 1. 
The fitting results thus
support the association, if we consider MSPs are the main sources of 
\gr\ emission in GCs.

For other five GCs, we obtained their spectral upper limits and tested our
model fitting on the upper limits (see Figure~\ref{fig:ul} in Appendix 
Section~\ref{sec:ul}). 
All of them are limited to have 
$\leq 1$ MSP but with the dominant numbers being zero.
Therefore,
the likely reason why these five GCs have not been detected
at $\gamma$-rays is because they do not contain detectable \gr\ MSPs.

Given the close distances of these five GCs, it is hard to miss the detections
with \fermi\ LAT if they have detectable \gr\ emission. The limits
on mostly zero \gr\ MSPs in them we set should be reliable. One possibility
is that they contain a few MSPs but the MSPs do not have pulsed \gr\ emission 
towards the Earth by chance. 
We checked the GC radio pulsar catalog,
the reported numbers for them are only 1 or 2 (NGC~6121, NGC~6254, NGC~6544, 
and NGC~6656). The radio results seem to be consistent with our upper-limit 
results, suggesting that the GCs probably do not have many MSPs. 

We tested the $\log N_{\rm MSP}=0.64\log\Gamma_e+0.80$ relationship
derived in \citet{dcn19}, which was based on \gr\ luminosities of 20 GCs.
The relationship would predict $\sim$46 \gr\ MSPs in NGC~6544 and $\sim$2--9
in other four GCs. 
Considering their relatively high $\Gamma_e$ values, the non-detection of \gr\ 
emission
in the five GCs (in particular NGC~6544) suggests that
the number of \gr\ MSPs in a GC may not be directly indicated by $\Gamma_e$;
other factors could play important roles in determining the number of \gr\
MSPs a GC contains (e.g., \citealt{dcn19,son+21}). 

\subsection{Comparison with previous studies and radio results}
\label{sec:comp}

In our studies, we took into account the detailed properties of the \gr\ MSPs 
that included the spectral information, and based on the properties we 
constructed a model that produces spectra of GCs by adding the
generated MSPs' spectra in a GC. We found that the model spectra can generally 
well fit most of the observed spectra of the \gr\ GCs. The fitting thus 
indicates that \gr\ emission in 27 of the GCs can be explained with MSPs 
contained 
in them. This result is different from that found by \citet{son+21}, since
our fitting results do not show the need of an additional power-law component 
(due to inverse-Compton scattering of infrared and optical photons). We note
that several power-law like spectra (e.g., NGC~6093, NGC~6139, NGC~6304; see
Appendix Figures~\ref{fig:6093}, \ref{fig:6139} \& \ref{fig:6304}) can 
actually be fit with our PLEC-like spectral
form, partly due to the large flux uncertainties, and in some cases
the flux upper limits do require a curved spectrum (e.g., NGC~6304). In the
future when more data are collected to improve the spectrum measurements,
the picture whether these spectra are better described with a PLEC type 
spectral form would become clear.

We did identify three GCs whose spectra do not appear to have
 a PLEC shape that can be described with our model spectra of MSPs,
among which NGC~7078 and 2MS-GC01 have power-law like spectra. One possibility
could be that the \gr\ sources are background blazars due to positional
coincidence, which can be verified by monitoring their variability.
While if they are indeed the true counterparts to the two GCs, our fitting 
indicates that MSPs' emission is not the 
dominant one at least. These results are consistent with that reported in
\citet{lcb18} and \citet{son+21}. Thus the power-law \gr\ emission from
the two GCs is intriguing and may be the inverse-Compton component proposed
by \citet{son+21}.

Different from previous estimations for numbers of \gr\ MSPs in the GCs,
which used typical luminosity parameters for \gr\ MSPs and then compared with
\gr\ luminosities of the GCs, we worked through spectral fitting. 
The numbers of \gr\ MSPs for 25 GCs
we estimated are shown in Figure~\ref{fig:ne}. We also found the
numbers of radio pulsars in these GCs (listed in the GC radio pulsar catalog)
and marked them in the figure. As can be seen, the numbers from our
spectral fitting generally match those reported at radio, while several
GCs (e.g., NGC~104, NGC~5904, NGC~6752) notably have significantly more
radio MSPs. We note that recently observations with
Five-hundred-meter Aperture Spherical radio Telescope (FAST) have found many
new radio pulsars in GCs (e.g., \citealt{pan+21,pmq+12}). Hopefully 
high-sensitivity observations with FAST will reveal more pulsars in the
near future and the comparison of our results with the updated 
numbers of radio pulsars would provide a clear understanding for 
the relation of \gr\ MSPs to radio ones in the GCs.

As the correlation between the numbers of MSPs and $\Gamma_e$ have been 
studied, we checked our results. If we consider the ranges of numbers
as the uncertainties, we would find a relationship 
of $\log N_{\rm MSP} \simeq 0.26\log\Gamma_e - 0.19$.
This relationship, shown in
Figure~\ref{fig:ne}, depends more weakly on $\Gamma_e$ than that 
derived by \citet{dcn19} from \gr\ luminosities of 20 GCs.
It would predict $\sim$1--3 \gr\ MSPs in the five nearby GCs with relatively high
$\Gamma_e$ values (cf., Section~\ref{sec:ngcs}). Specifically for NGC~6544,
$\sim$3 \gr\ MSPs would be expected. Considering uncertainties on
the estimation of $\Gamma_e$ and the large ranges for our estimated 
numbers of \gr\ MSPs, the relationship likely represents more closely 
the observational facts (i.e., the non-detection of \gr\ emission in the 
five GCs).  Then again as discussed in 
Section~\ref{sec:ngcs},
it is likely that high $\Gamma_e$ GCs tend to have more \gr\ MSPs, but whether 
there
is a relationship that can be well defined for the numbers of \gr\ MSPs 
in GCs is uncertain.

\subsection{Summary}

We have studied high-energy properties of 30 \gr\ GCs and 6 nearby, high
$\Gamma_e$ GCs by considering MSPs as the main high-energy sources
in them. From fitting the spectra or spectral upper limits of the GCs with
that of the MSPs presumed in them, we have found that \gr\ emission of 27 GCs
can be explained as due to MSPs. If we consider that the \gr\ sources
at the positions of NGC~7078, Terzan~1, and 2MS-GC01 are their true 
counterparts, \gr\ emission of them needs explanations, as their spectra
can not be well fit with the MSP-type spectral model. For the 6 nearby
GCs, we have found evidence supporting the possible detection of \gr\ emission
from NGC~6656 and set limits of $\leq 1$ \gr\ MSPs in the other 5 GCs
from our model fitting. 

Numbers of \gr\ MSPs in the 27 \gr\ GCs have been estimated, and they 
generally match well those of radio MSPs found in the GCs. 
Deep radio searches for finding new MSPs in the GCs are 
warranted. The samples of radio MSPs, if built as fully as possible from 
the searches, could be compared with those estimated from the \gr\ properties 
of the GCs and the comparisons would help improve our understanding of 
the properties of \gr\ MSPs among the total MSPs in each of the GCs.

\acknowledgements
We thank anonymous referee for critical comments that helped greatly
improving this work's results.
This research made use of the High Performance Computing Resource in the Core
Facility for Advanced Research Computing at Shanghai Astronomical Observatory. 
This work was supported by the National Natural
Science Foundation of China (11633007).
Z.W. acknowledges the support by the Original
Innovation Program of the Chinese Academy of Sciences (E085021002).

\bibliographystyle{aasjournal}

\clearpage
\begin{table}
\begin{center}
\caption{Properties of the known \gr\ globular clusters and estimated numbers of \gr\ MSPs}
	\label{tab:gc}
\begin{tabular}{lcccccccc}
\hline
	Name      & Age     & Distance$^a$ & $\rho_{0}^a$ & $r_{c}^a$ & $N_{\rm MSP}$ & $\chi^{2}_{\rm min}/N^b$ & $N^{5\%}_{\rm MSP}$ \\
	    & (Gyr) & (kpc) & ($L_{\sun}$ pc$^{-3}$) & (arcmin) &  &  &  \\\hline 
	NGC 104$^1$   &  11.8 & 4.5  & 4.9   & 0.36 &  4 & 7.7/6 & 1--11\\
	NGC 362$^1$   &  10.8  & 8.6  & 4.7   & 0.18 &  2 & 0.9/3 & 1--3 \\
	NGC 1904$^2$ &  11.0 & 12.9 & 4.1   & 0.16  & 3 & 7.9/5 & 1--4\\
	NGC 2808$^1$ &  11.0 & 9.6  & 4.7  & 0.25 & 2 & 1.6/4 & 1--5  \\
	NGC 5139$^3$ &  11.5 & 5.2& 3.2 & 2.4  & 3 & 3.6/5 & 1--6 \\
	NGC 5904$^1$ &  11.5  & 7.5 & 3.9 & 0.44 & 1 & 0.012/3 & 1-3 \\
	NGC 6093$^3$ &  12.5  & 10 & 4.8 & 0.15 & 1 & 2.0/6 & 1--3 \\
	NGC 6139$^4$ &  12.6    & 10.1 & 4.7  & 0.15 & 2 & 1.4/4 & 1--8 \\
	NGC 6218$^1$ &  13.0    & 4.8 & 3.2 & 0.79 & 1 & 5.0/4 & 1--2 \\
	NGC 6266$^3$ &  11.8 & 6.8  & 5.2 & 0.22 & 8 & 5.5/7 & 1--13 \\
	NGC 6304$^1$  &  11.3  & 5.9 & 4.5 & 0.21 & 3 & 2.3/4 & 1--3 \\
	NGC 6316$^5$ &  14.8  & 10.4   & 4.2 & 0.17 & 13 & 5.2/6 & 4--16 \\
	NGC 6341$^1$    &  12.8    & 8.3  & 4.3 & 0.26  & 1 & 5.3/5 & 1--3 \\
	NGC 6388$^3$    &  12.0 & 9.9  & 5.4  & 0.12 & 10 & 2.9/6 & 4--20 \\
	NGC 6397$^1$     &  13.0  & 2.3 & 5.8 & 0.05 & 1 & 2.3/5 & 1--1 \\
	NGC 6402$^6$ &  11.0       & 9.3 & 3.4 & 0.79 & 3 & 0.8/5 & 1--6 \\
	NGC 6440$^7$ &  11.0  & 8.5   & 5.2 & 0.14  & 8 & 4.8/6 & 1--11\\
	NGC 6441$^3$ &  11.3  & 11.6 & 5.3 & 0.13 & 7 & 6.3/5 & 4--20 \\
	NGC 6541$^1$    &  12.5    & 7.5 & 4.7 & 0.18 & 2 & 1.2/5 & 1--7 \\
	NGC 6652$^1$    &  11.3  & 10 & 4.5 & 0.10 & 4 & 0.34/5 & 1--7\\
	NGC 6717$^1$    &  12.5  & 7.1 & 4.6  & 0.08 & 1 & 0.71/4 & 1--4 \\
	NGC 6752$^1$    & 12.5     & 4.0 & 5.0 & 0.17 & 1 & 2.5/4 & 1--2 \\
	NGC 6838$^1$    &  11.0     & 4.0 & 2.8  & 0.63 & 1 & 2.5/4 & 1--2 \\
	NGC 7078$^1$    &  12.8 & 10.4  & 5.1   & 0.14 & 1 & 13/5 & 1--5 \\
	Terzan 1$^4$    &  12.6    & 6.7  & 3.9 & 0.04 & 1 & 67/3 & 1--3 \\
	Terzan 2$^4$    &  8.9     & 7.5  & 4.9 & 0.03 & 4 & 2.6/5 & 1--7\\
	Terzan 5$^8$    &  12.0 & 6.9 & 5.1   & 0.16 &  17 & 9.5/8 & 9--38\\ 
	2MS-GC01$^4$    & 15.9    & 3.6 & \nodata  & 0.85 & 1 & 18/5 & 1--6 \\
	GLIMPSE-C01$^{9,10}$   &  2.0        & 4.2  &   \nodata & 0.59 & 3 & 5.3/5 & 1--6\\
	GLIMPSE-C02$^9$   &  2.0        & 5.5 & \nodata   & 0.70 & 2 & 8.1/5 & 1--4 \\ \hline
	NGC 6656$^{1,c}$    &  12.5    & 3.2 & 3.6 & 1.3 & 1 & 5.3/3 & 1--1 \\\hline
\end{tabular}
\end{center}
	\footnotesize{References for the ages of the GCs: $^1$\citet{van+13}; $^2$\citet{car+18}; $^3$\citet{fb10}; $^4$\citet{kha+13}; $^5$\citet{cez+13}; $^6$\citet{con+13}; $^7$\citet{ori+08}; $^8$\citet{fer+16}; $^9$\citet{dav+16}; $^{10}$\citet{hkr18}. $^a$Distance, central luminosity density, and core radius values are from the GC catalog of \citet{har96}. $^b$$N$ is the number of spectral data points in each spectrum. $^c$A candidate \gr\ counterpart is considered for this GC.}
\end{table}

\appendix
\label{sec:app}

\restartappendixnumbering

\section{Information for 85 \gr\ MSPs}
\label{sec:amsp}

Among 104 \gr\ MSPs used to determine the spectral shape 
(Section~\ref{sec:ssd}), we found $\dot{P}$ and distance information for
85 of them. They are listed in Table~\ref{tab:msp}. The fluxes $F_{0.1-100}$ 
of them are in 0.1--100\,GeV energy range, given in the 4FGL-DR2 \citep{bal+20}.
The luminosities were calculated from 
$L_{\gamma}=4\pi d^2f_{\Omega}F_{0.1-100}$, where $d$ is the distance and
the beam correction factor $f_{\Omega}$ was assumed to be 1 (following 
\citealt{2fpsr13}).

\begin{ThreePartTable}
\begin{TableNotes}
\item[$a$] Distances without errors are assumed to have 30\% uncertainties;
\item[$1$,$2$,$3$,$4$] Distances are from \citet{arz+18}, \citet{vl14}, \citet{rea+16}, and \citet{des+16}, respectively.
\end{TableNotes}
  \centering
    \begin{longtable}{lcccccc}
  \caption{Information for 85 \gr\ MSPs}
\label{tab:msp}\\
\hline
	    Name & $P$ & $\dot{E}/10^{34}$ & $\tau_{c}$ & $d^a$ & $F_{0.1-100}/10^{-12}$ & $L_{\gamma}/10^{33}$  \\
	    & (ms) & (erg\,s$^{-1}$) & (Gyr) & (kpc) & (erg\,s$^{-1}$\,cm$^{-2}$) & (erg\,s$^{-1}$) \\\hline
\endfirsthead
\hline
   Name & $P$ & $\dot{E}/10^{34}$ & $\tau_{c}$ & $d$ (kpc) & $F_{0.1-100}/10^{-12}$ & $L_{\gamma}/10^{33}$ \\
	    & (ms) & (erg\,s$^{-1}$) & (Gyr) & (kpc) & (erg\,s$^{-1}$\,cm$^{-2}$) & (erg\,s$^{-1}$) \\\hline
\endhead
\endfoot
\hline
\insertTableNotes
\endlastfoot
J0023+0923$^1$ & 3.05  & 1.6  & 4.23  & $1.1_{-0.2}^{+0.2}$ & 7.55  & 1.1  \\
    J0030+0451$^1$ & 4.87  & 0.35  & 7.58  & $0.325_{-0.009}^{+0.009}$ & 59.5  & 0.75  \\
    J0034$-$0534 & 1.88  & 3.0  & 5.99  & 1.348 & 20.7  & 4.5  \\
    J0218+4232$^2$ & 2.32  & 24  & 0.48  & $3.15_{-0.6}^{+0.85}$ & 48.1  & 57  \\
    J0248+4230 & 2.60  & 3.8  & 2.44  & 1.853 & 1.86  & 0.76  \\
    J0251+2606 & 2.54  & 1.8  & 5.32  & 1.170 & 4.48  & 0.73  \\
    J0340+4130 & 3.30  & 0.77  & 7.42  & 1.603 & 19.4  & 6.0  \\
    J0437$-$4715$^3$ & 5.76  & 1.2  & 1.59  & $0.1569_{-0.0022}^{+0.0022}$ & 17.3  & 0.051 \\
    J0533+6759 & 4.39  & 0.59  & 5.52  & 2.392 & 9.17  & 6.3  \\
    J0605+3757 & 2.73  & 0.93  & 9.01  & 0.215 & 6.46  & 0.036  \\
    J0610$-$2100 & 3.86  & 0.85  & 4.96  & 3.259 & 6.80  & 8.6  \\
    J0613$-$0200$^1$ & 3.06  & 1.3  & 5.06  & $1.1_{-0.2}^{+0.3}$ & 37.9  & 5.5  \\
    J0614$-$3329 & 3.15  & 2.2  & 2.85  & 2.690 & 114  & 98  \\
    J0621+2514 & 2.72  & 4.9  & 1.74  & 1.641 & 4.53  & 1.5  \\
    J0740+6620 & 2.89  & 2.0  & 3.75  & 0.929 & 3.16  & 0.33  \\
J0751+1807$^4$ & 3.48  & 0.73  & 7.08  & $1.07_{-0.17}^{+0.24}$ & 9.64  & 1.3 \\
    J0931$-$1902 & 4.64  & 0.14  & 20.26  & 3.722 & 1.87  & 3.1  \\
    J0952$-$0607 & 1.41  & 6.7  & 4.70  & 1.740 & 2.34  & 0.85  \\
    J0955$-$6150 & 2.00  & 7.0  & 2.22  & 2.170 & 7.13  & 4.0  \\
    J1012+5307 & 5.26  & 0.47  & 4.87  & 0.805 & 4.88  & 0.38  \\
    J1024$-$0719$^4$ & 5.16  & 0.53  & 4.41  & $1.08_{-0.16}^{+0.23}$ & 4.39  & 0.61 \\
    J1035$-$6720 & 2.87  & 7.7  & 0.98  & 1.460 & 18.6  & 4.7 \\
    J1048+2339 & 4.67  & 1.2  & 2.46  & 2.002 & 5.07  & 2.4  \\
    J1125$-$5825 & 3.10  & 8.1  & 0.81  & 1.744 & 5.99  & 2.2  \\
    J1125$-$6014 & 2.63  & 0.87  & 10.40  & 0.989 & 2.87  & 0.34 \\
    J1142+0119 & 5.08  & 0.45  & 5.37  & 2.169 & 6.20  & 3.5  \\
    J1207$-$5050 & 4.84  & 0.21  & 12.66  & 1.268 & 4.73  & 0.91 \\
    J1227$-$4853 & 1.69  & 9.1  & 2.41  & 1.244 & 22.6  & 4.2  \\
    J1231$-$1411 & 3.68  & 1.8  & 2.56  & 0.420 & 101  & 2.1 \\
    J1302$-$3258 & 3.77  & 0.48  & 9.12  & 1.432 & 10.9  & 2.7 \\
    J1311$-$3430 & 2.56  & 4.9  & 1.94  & 2.430 & 61.0  & 43 \\
    J1312+0051 & 4.23  & 0.92  & 3.82  & 1.471 & 14.4  & 3.8 \\
    J1431$-$4715 & 2.01  & 6.8  & 2.26  & 1.822 & 6.35  & 2.5 \\
    J1446$-$4701 & 2.19  & 3.7  & 3.55  & 1.569 & 6.80  & 2.0 \\
    J1513$-$2550 & 2.12  & 9.0  & 1.55  & 3.956 & 6.95  & 13  \\
    J1514$-$4946 & 3.59  & 1.6  & 3.05  & 0.909 & 40.8  & 4.0  \\
J1536$-$4948 & 3.08  & 2.9  & 2.30  & 0.978 & 78.5  & 9.0  \\
    J1543$-$5149 & 2.06  & 7.3  & 2.02  & 1.148 & 17.8  & 2.8  \\
    J1544+4937 & 2.16  & 1.2  & 11.67  & 2.991 & 3.38  & 3.6  \\
    J1552+5437 & 2.43  & 0.77  & 13.76  & 2.636 & 2.82  & 2.4  \\
    J1555$-$2908 & 1.79  & 31  & 0.64  & 7.555 & 6.97  & 48  \\
    J1600$-$3053$^1$ & 3.60  & 0.81  & 6.00  & $2_{-0.3}^{+0.3}$ & 7.38  & 3.5 \\
    J1614$-$2230$^1$ & 3.15  & 1.2  & 5.19  & $0.67_{-0.04}^{+0.05}$ & 25.3 & 1.4 \\
    J1622$-$0315 & 3.85  & 0.81  & 5.26  & 1.141 & 7.86  & 1.2 \\
    J1625$-$0021 & 2.83  & 3.7  & 2.11  & 0.951 & 20.3  & 2.2  \\
    J1630+3734 & 3.32  & 1.2  & 4.88  & 1.187 & 5.97  & 1.0  \\
    J1640+2224 & 3.16  & 0.35  & 17.80  & 1.507 & 2.33  & 0.63 \\
    J1658$-$5324 & 2.44  & 3.0  & 3.47  & 0.880 & 19.8  & 1.8  \\
    J1713+0747$^1$ & 4.57  & 0.35  & 8.50  & $1.22_{-0.04}^{+0.04}$ & 7.05  & 1.3 \\
    J1730$-$2304$^3$ & 8.12  & 0.15  & 6.38  & $0.62_{-0.1}^{+0.15}$ & 7.35  & 0.34 \\
    J1732$-$5049 & 5.31  & 0.37  & 5.93  & 1.875 & 6.00  & 2.5 \\
    J1741+1351$^1$ & 3.75  & 2.3  & 1.97  & $1.8_{-0.3}^{+0.5}$ & 4.14  & 1.6 \\
    J1744$-$1134$^1$ & 4.07  & 0.52  & 7.23  & $0.44_{-0.02}^{+0.02}$ & 37.4  & 0.87 \\
    J1745+1017 & 2.65  & 0.58  & 15.41  & 1.214 & 7.60  & 1.3 \\
    J1747$-$4036 & 1.65  & 12  & 1.99  & 7.151 & 12.2  & 74  \\
    J1805+0615 & 2.13  & 9.3  & 1.48  & 3.885 & 5.29  & 9.6  \\
    J1811$-$2405 & 2.66  & 2.9  & 3.06  & 1.830 & 18.6  & 7.5 \\
    J1816+4510 & 3.19  & 5.2  & 1.17  & 4.356 & 10.2  & 23  \\
    J1824+1014 & 4.07  & 0.32  & 12.00  & 2.904 & 6.68  & 6.7 \\
    J1832$-$0836 & 2.72  & 1.6  & 5.21  & 0.807 & 20.1  & 1.6  \\
    J1843$-$1113 & 1.85  & 6.0  & 3.06  & 1.706 & 21.8  & 7.6  \\
    J1855$-$1436 & 3.59  & 0.93  & 5.23  & 5.126 & 5.38  & 17  \\
    J1901$-$0125 & 2.79  & 6.5  & 1.24  & 2.360 & 19.5  & 13  \\
    J1902$-$5105 & 1.74  & 6.9  & 3.00  & 1.645 & 22.7  & 7.3  \\
    J1903$-$7051 & 3.60  & 0.88  & 5.47  & 0.930 & 5.26  & 0.54 \\
    J1908+2105 & 2.56  & 3.2  & 2.94  & 2.601 & 4.12  & 3.3  \\
    J1921+0137 & 2.50  & 4.9  & 2.05  & 5.085 & 11.1  & 34  \\
    J2006+0148 & 2.16  & 1.3  & 10.39  & 2.437 & 2.98  & 2.1  \\
    J2017$-$1614 & 2.31  & 0.78  & 14.98  & 1.444 & 6.20  & 1.6 \\
    J2017+0603 & 2.90  & 1.3  & 5.74  & 1.399 & 35.7  & 8.4  \\
    J2039$-$5617 & 2.65  & 3.0  & 2.97  & 1.707 & 15.1  & 5.3  \\
    J2043+1711 & 2.38  & 1.5  & 7.20  & 1.476 & 5.93  & 1.6 \\
    J2043+1711$^1$ & 2.38  & 1.5  & 7.20  & $1.6_{-0.2}^{+0.2}$ & 28.4  & 8.7 \\
    J2051$-$0827 & 4.51  & 0.55  & 5.61  & 1.469 & 2.79  & 0.72 \\
    J2052+1219 & 1.99  & 3.4  & 4.70  & 3.916 & 5.68  & 10 \\
    J2115+5448 & 2.60  & 17  & 0.55  & 3.106 & 7.10  & 8.2 \\
    J2124$-$3358$^4$ & 4.93  & 0.68  & 3.80  & $0.38_{-0.05}^{+0.06}$ & 38.6  & 0.67 \\
    J2214+3000 & 3.12  & 1.9  & 3.36  & 1.673 & 32.1  & 11 \\
    J2215+5135 & 2.61  & 7.4  & 1.24  & 2.773 & 17.0  & 16 \\
    J2234+0944 & 3.63  & 1.7  & 2.86  & 1.587 & 9.89  & 3.0  \\
    J2241$-$5236 & 2.19  & 2.5  & 5.22  & 0.963 & 25.4  & 2.8  \\
    J2256$-$1024 & 2.29  & 3.7  & 3.20  & 1.329 & 8.01  & 1.7 \\
    J2302+4442 & 5.19  & 0.39  & 5.97  & 0.859 & 38.2  & 3.4  \\
    J2310$-$0555 & 2.61  & 1.1  & 8.35  & 1.556 & 5.40  & 1.6  \\
    J2339$-$0533 & 2.88  & 2.3  & 3.24  & 0.751 & 29.2  & 2.0  \\\hline
    \end{longtable}%
\end{ThreePartTable}%

\section{Spectral fitting results for Gamma-Ray GCs}
\label{sec:sf}

Spectra of 26 \gr\ GCs and spectral fitting results are shown in the following
figures. We note that several spectra appear to have high fluxes at the low end 
(0.1--0.2\,GeV) of the spectral energy range, away from the other spectral data
points (and our model spectra). Checking them with TS maps, we 
found that they were due to the contamination by the strong background emission 
or nearby sources, since the point spread function of LAT at 100\,MeV is as
large as $\sim7\arcdeg$.

\begin{figure*}
\begin{center}
        \includegraphics[width=0.41\linewidth]{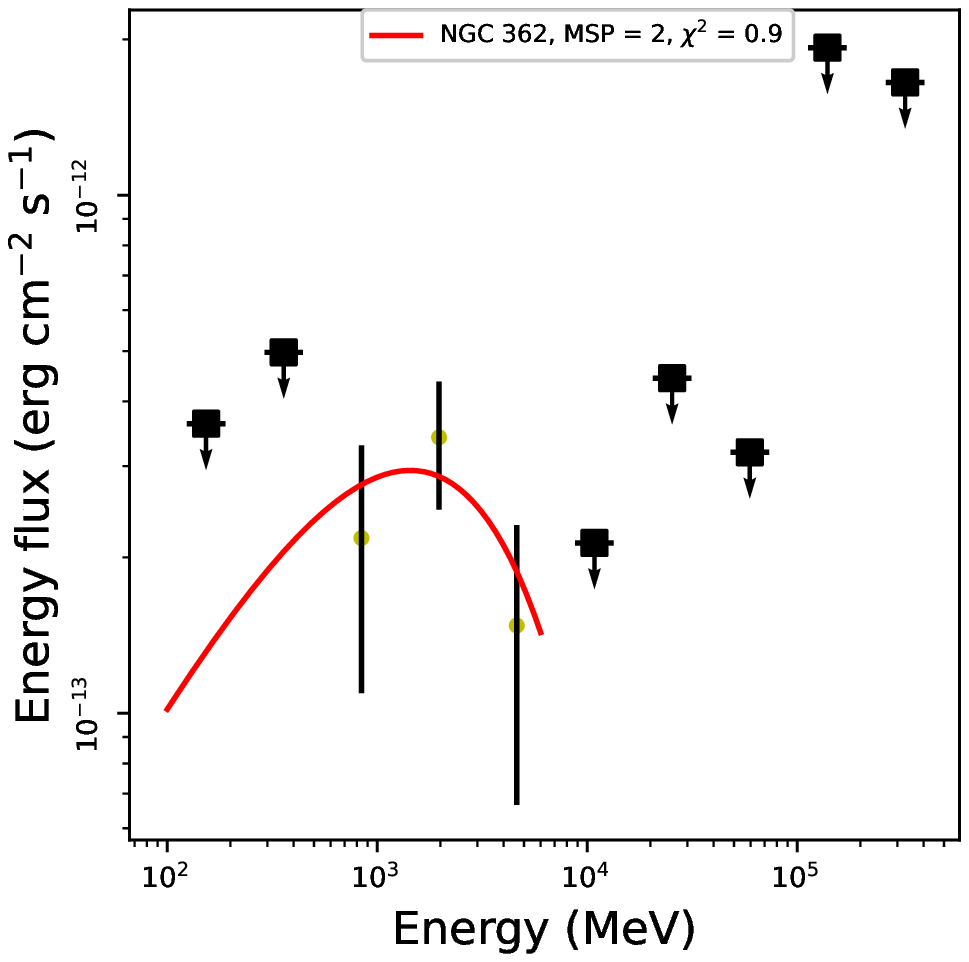}
\includegraphics[width=0.41\linewidth]{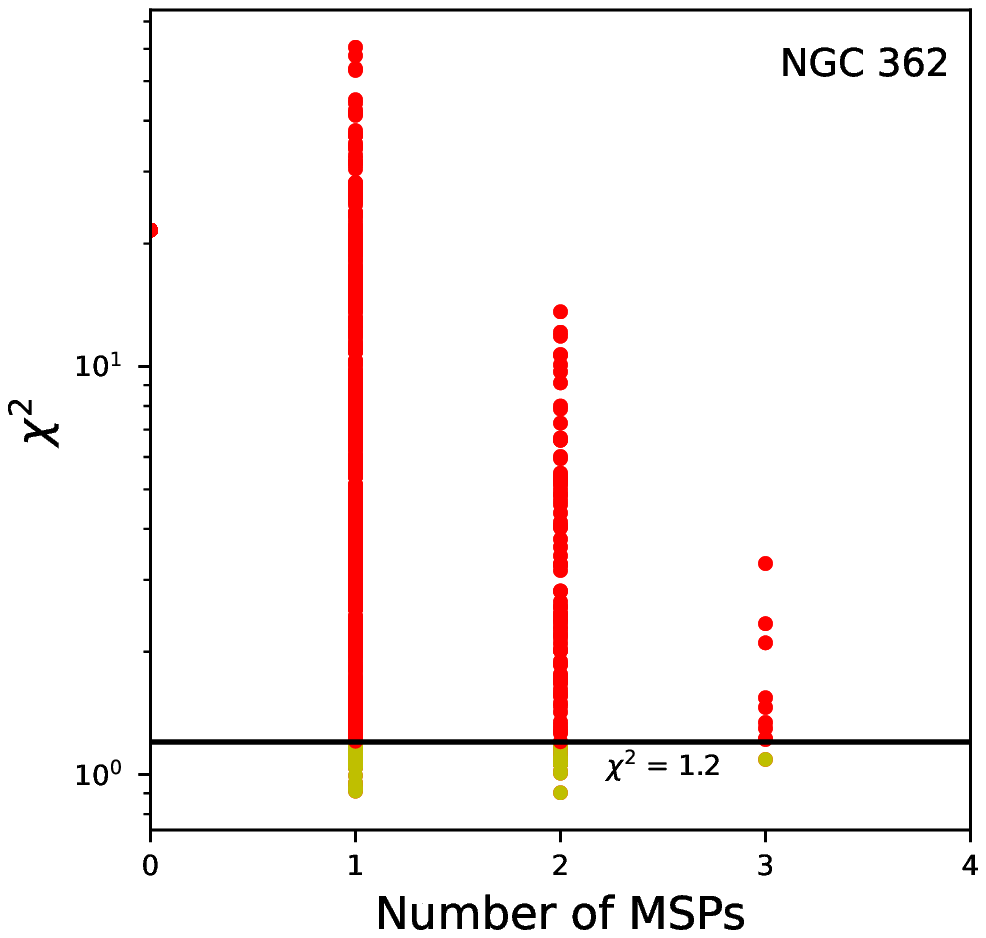}

        \includegraphics[width=0.41\linewidth]{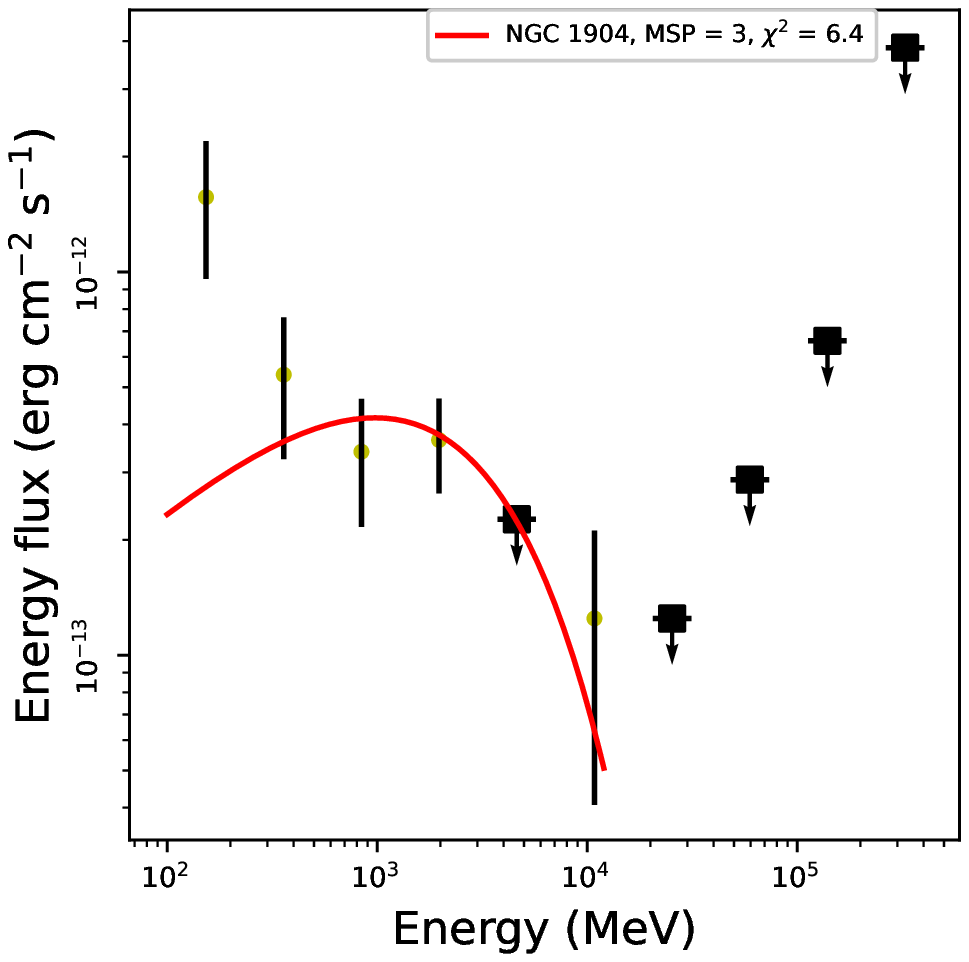}
\includegraphics[width=0.41\linewidth]{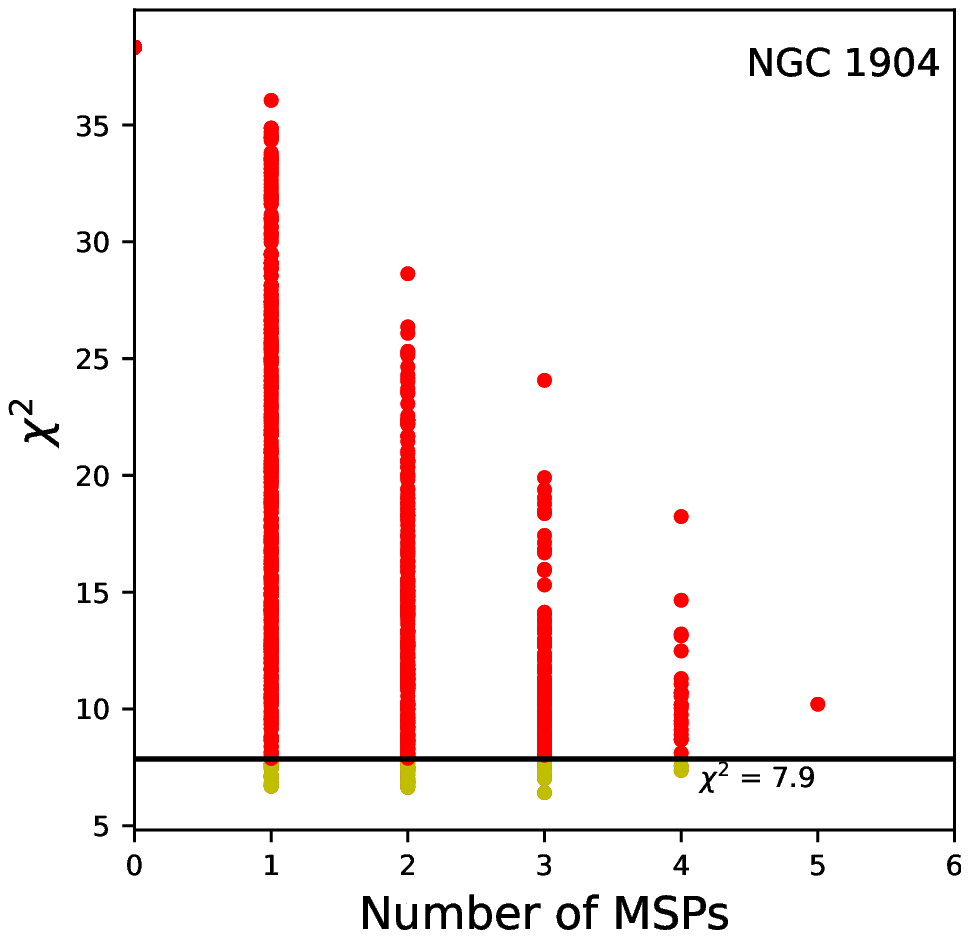}

\includegraphics[width=0.41\linewidth]{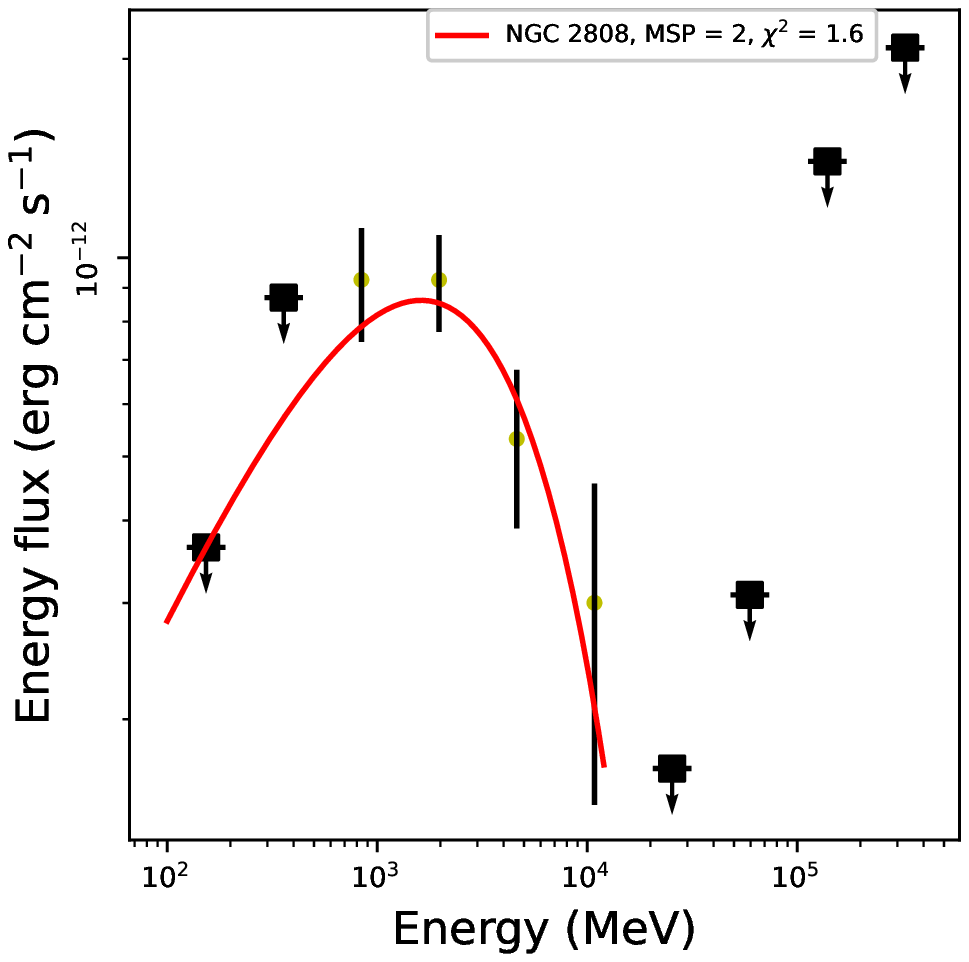}
\includegraphics[width=0.41\linewidth]{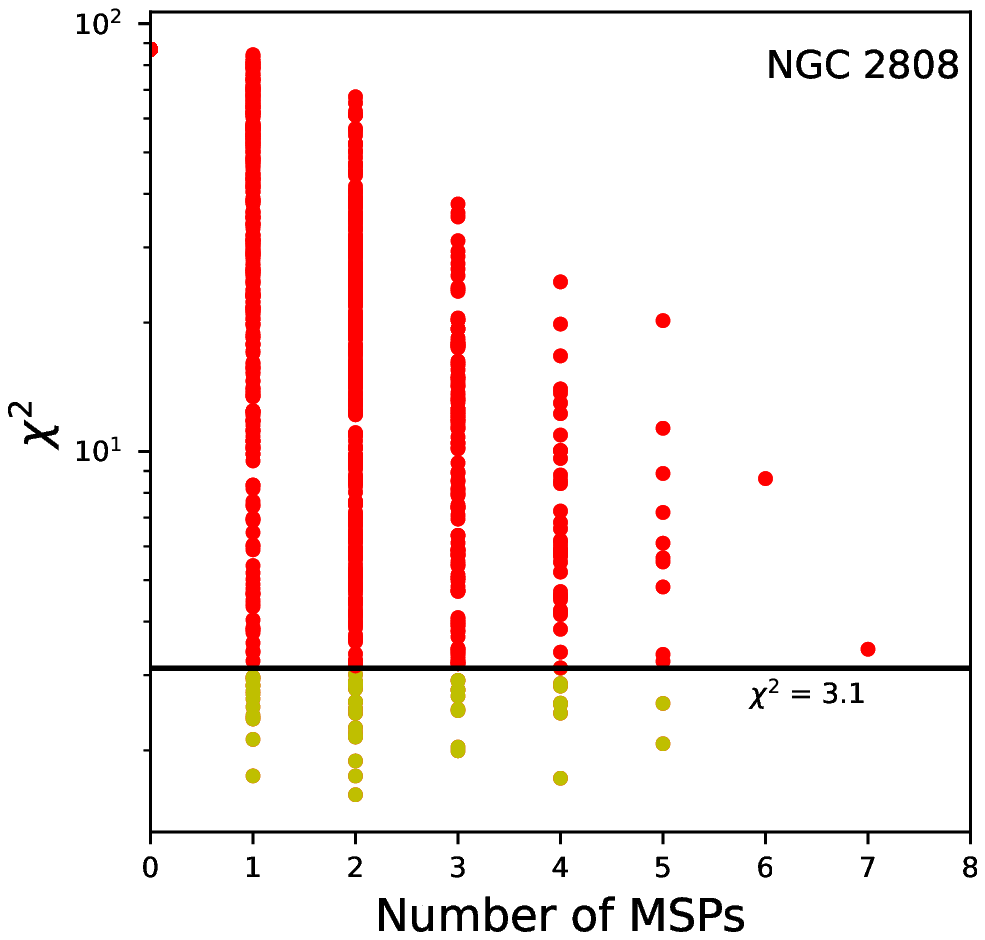}
\end{center}
        \caption{Same as Figure~\ref{fig:tuc}.}
\end{figure*}

\begin{figure*}
\begin{center}
        \includegraphics[width=0.41\linewidth]{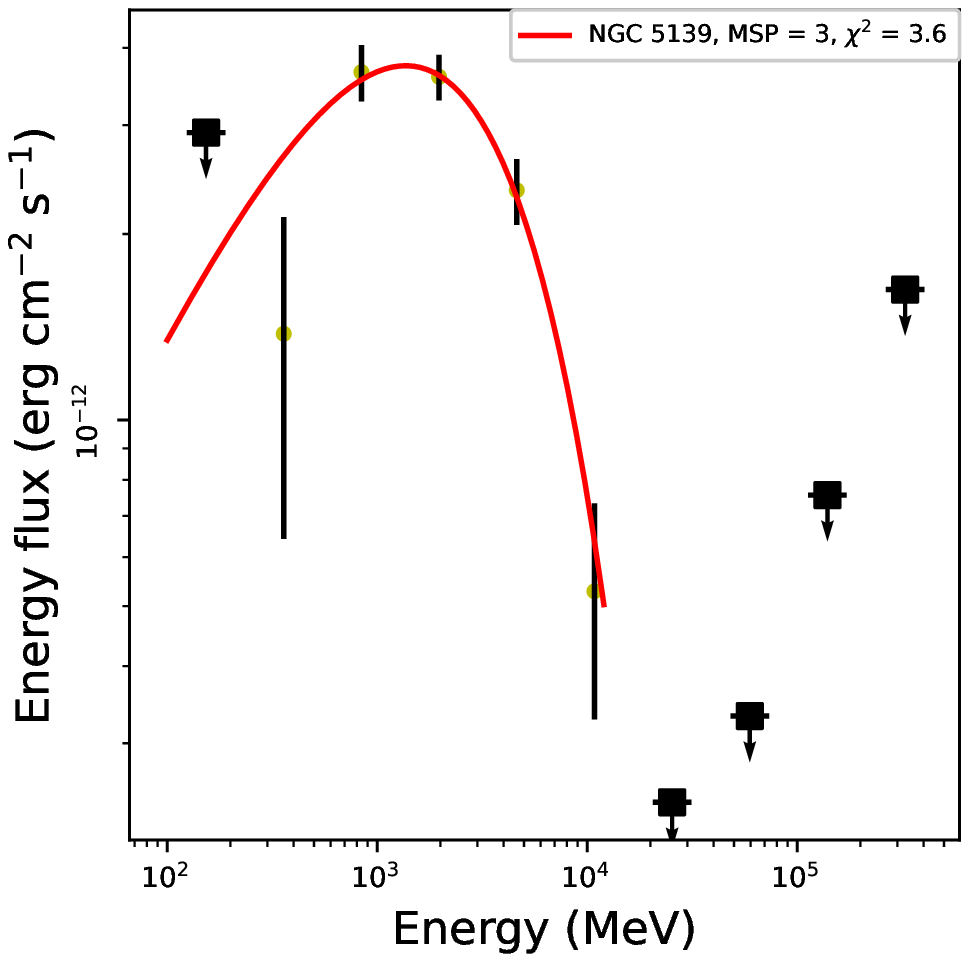}
\includegraphics[width=0.41\linewidth]{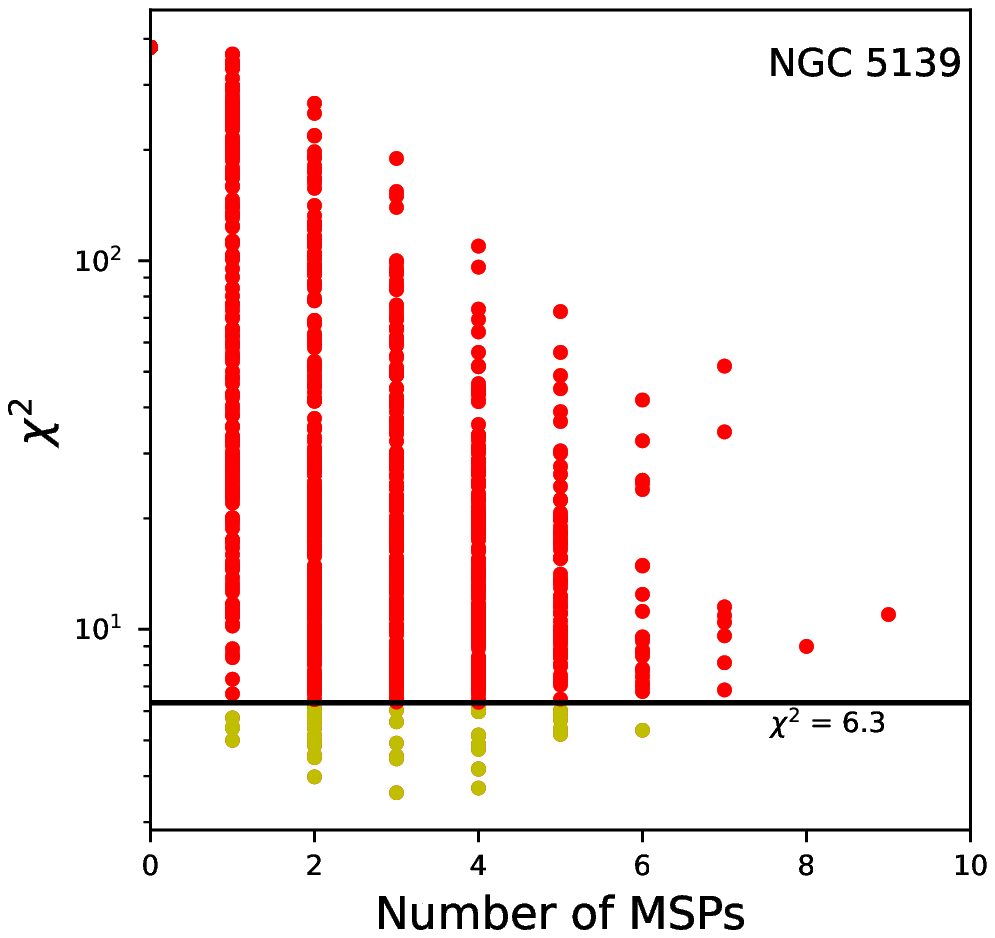}

        \includegraphics[width=0.41\linewidth]{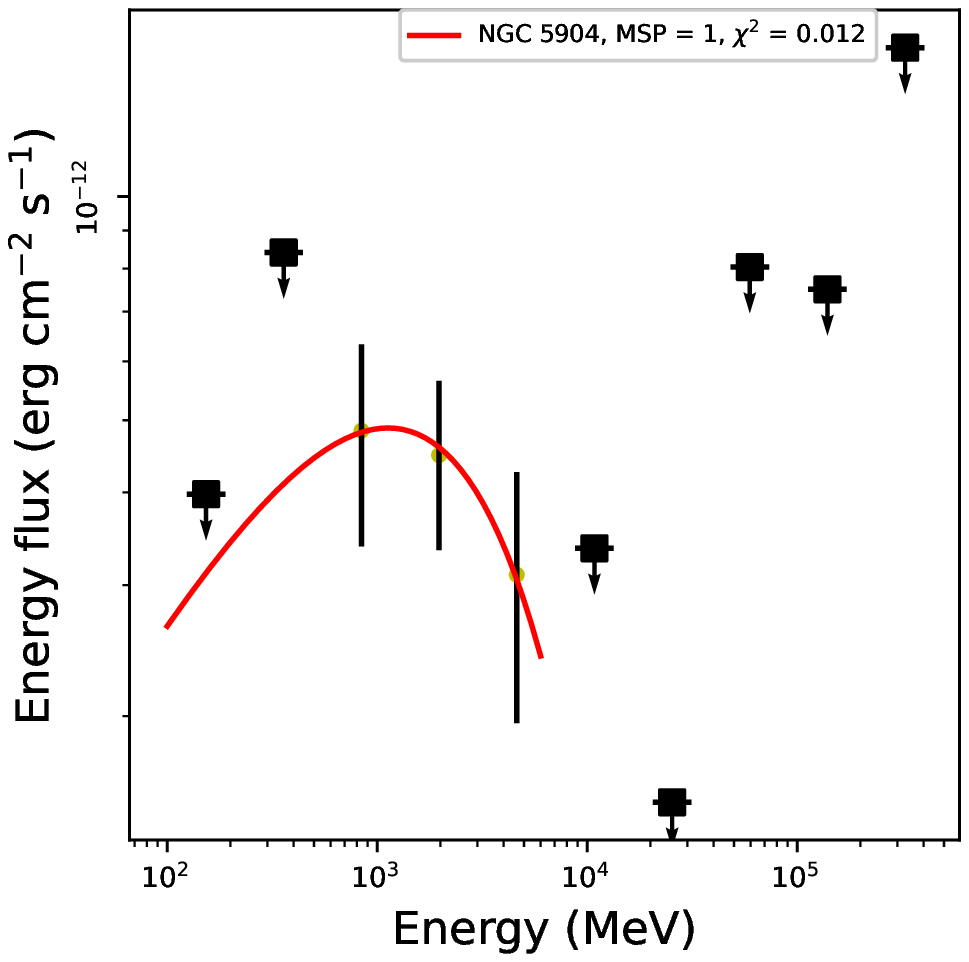}
\includegraphics[width=0.41\linewidth]{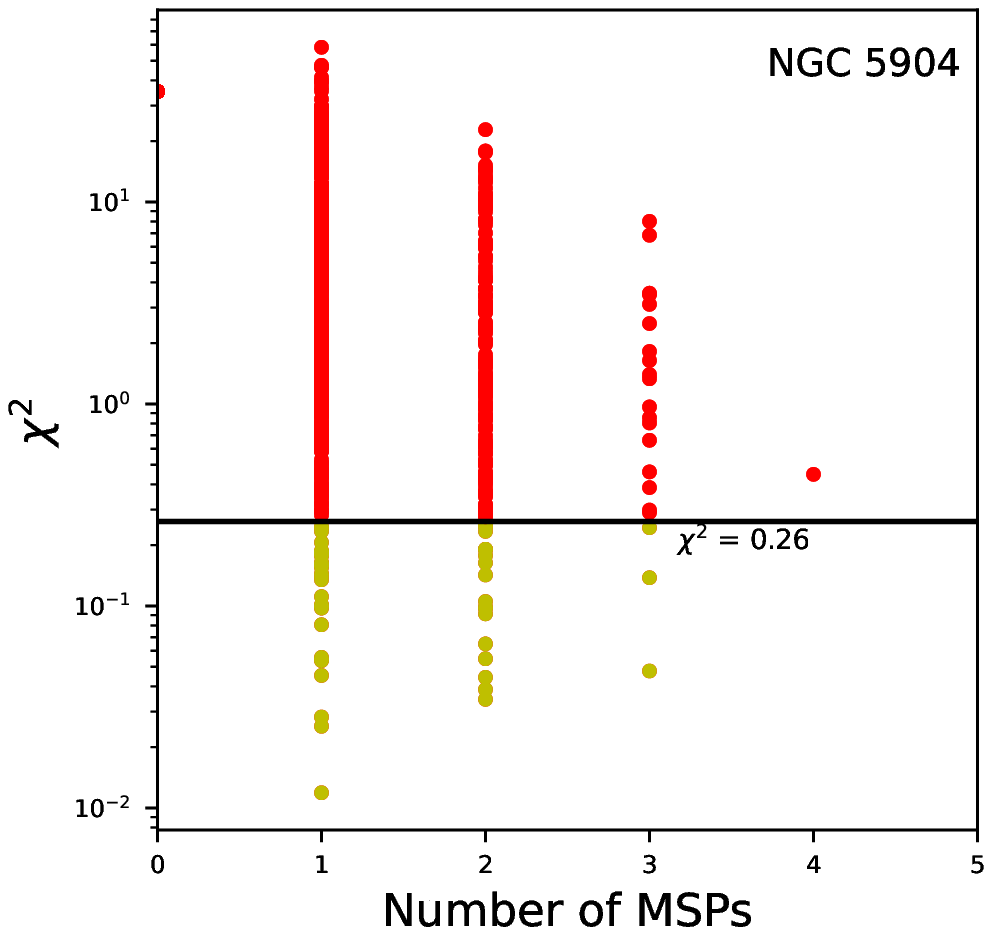}

        \includegraphics[width=0.41\linewidth]{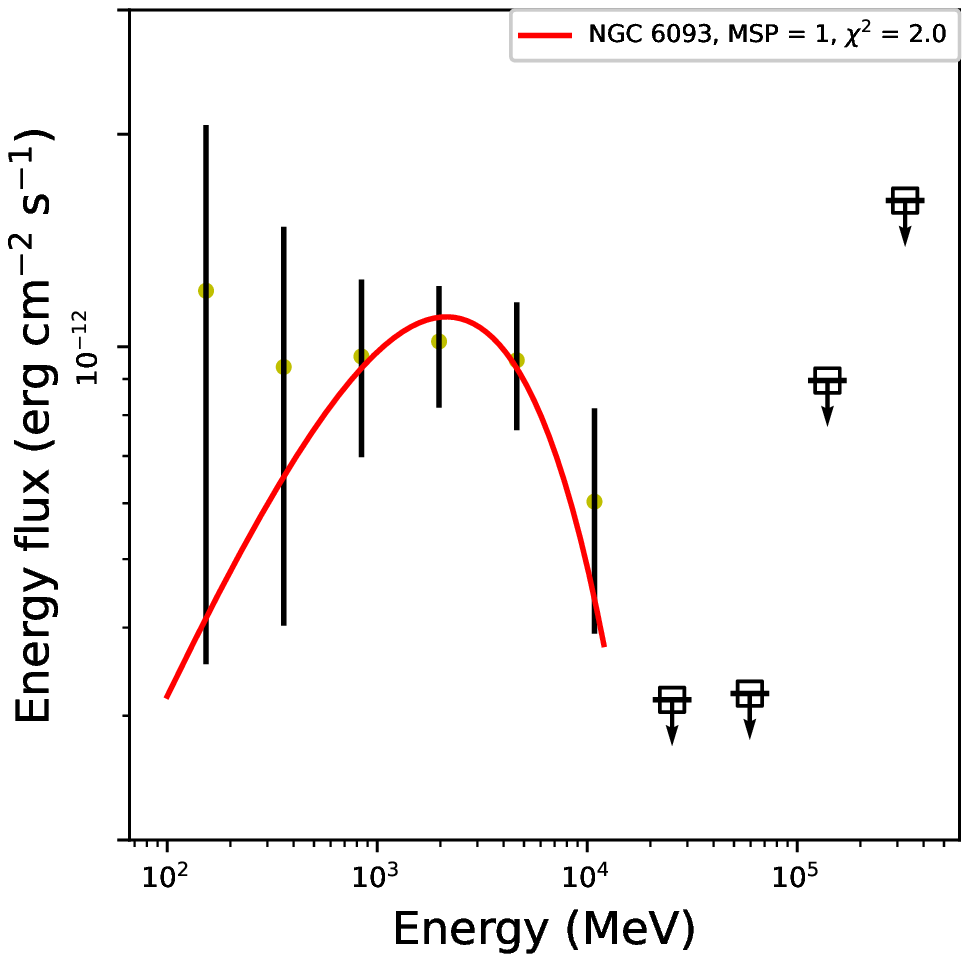}
\includegraphics[width=0.41\linewidth]{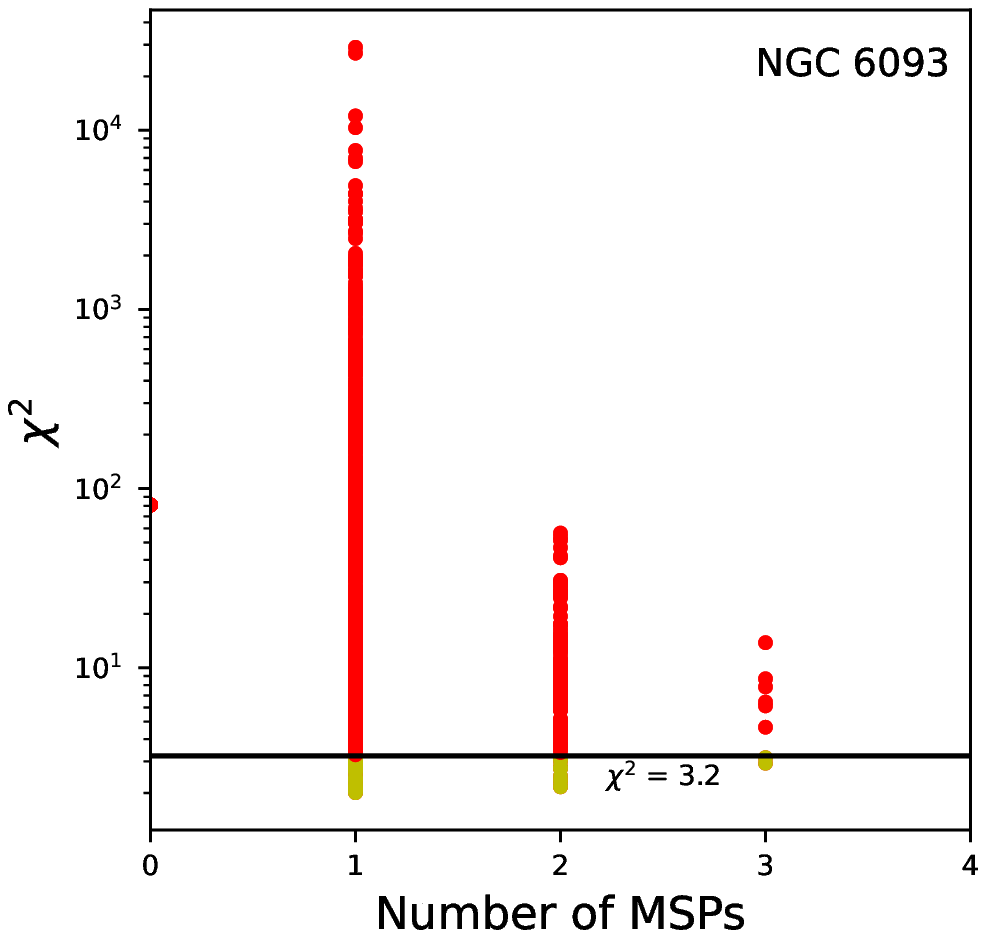}

\end{center}
        \caption{Same as Figure~\ref{fig:tuc}.}
\label{fig:6093}
\end{figure*}

\begin{figure*}
\begin{center}
\includegraphics[width=0.41\linewidth]{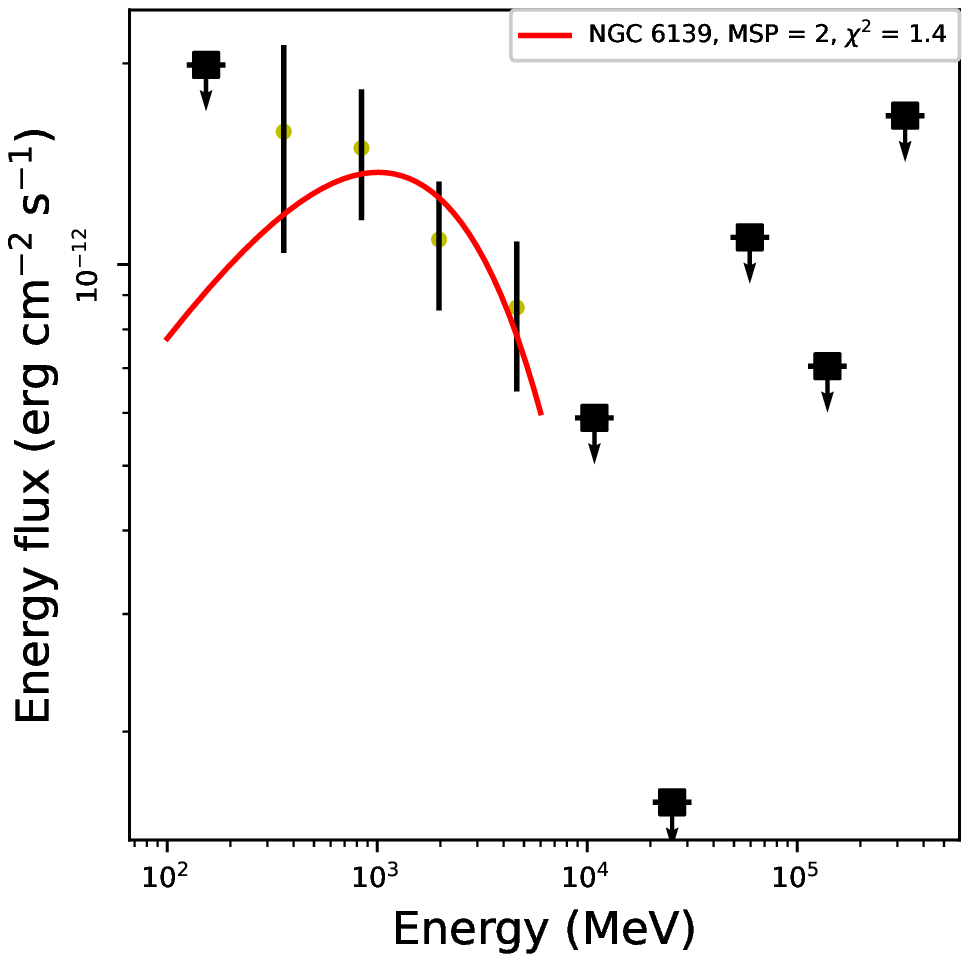}
\includegraphics[width=0.41\linewidth]{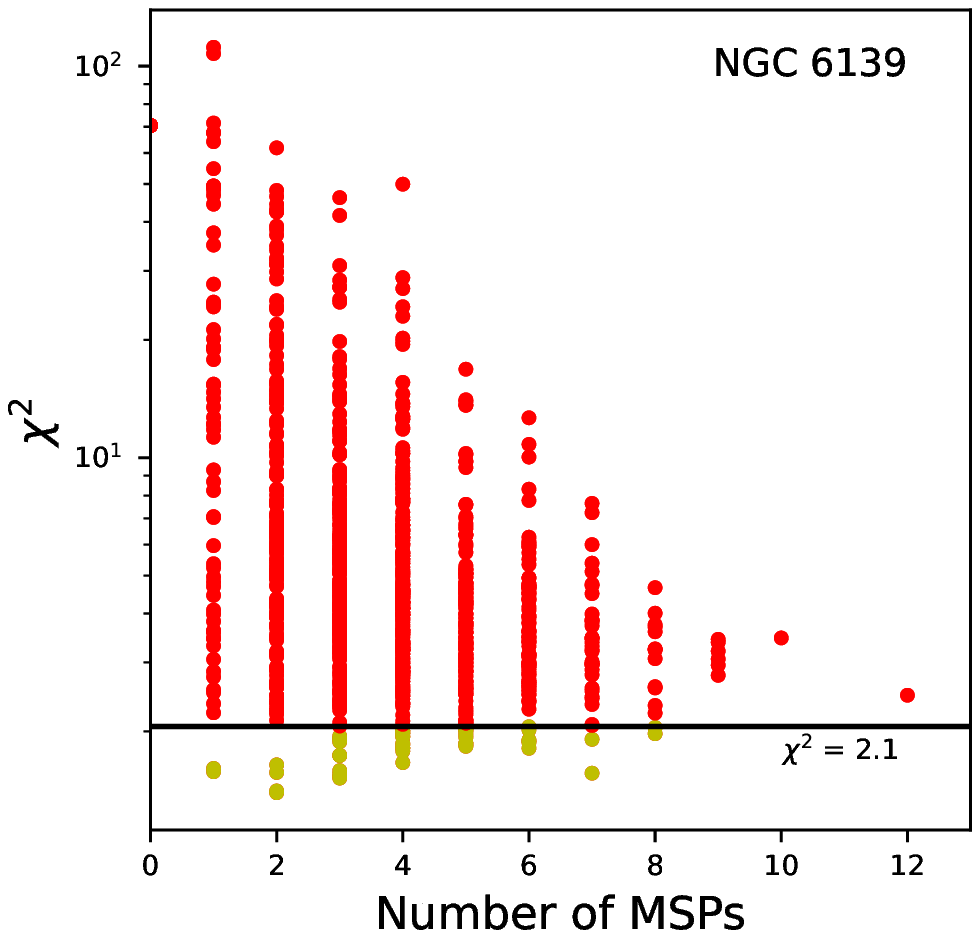}

        \includegraphics[width=0.41\linewidth]{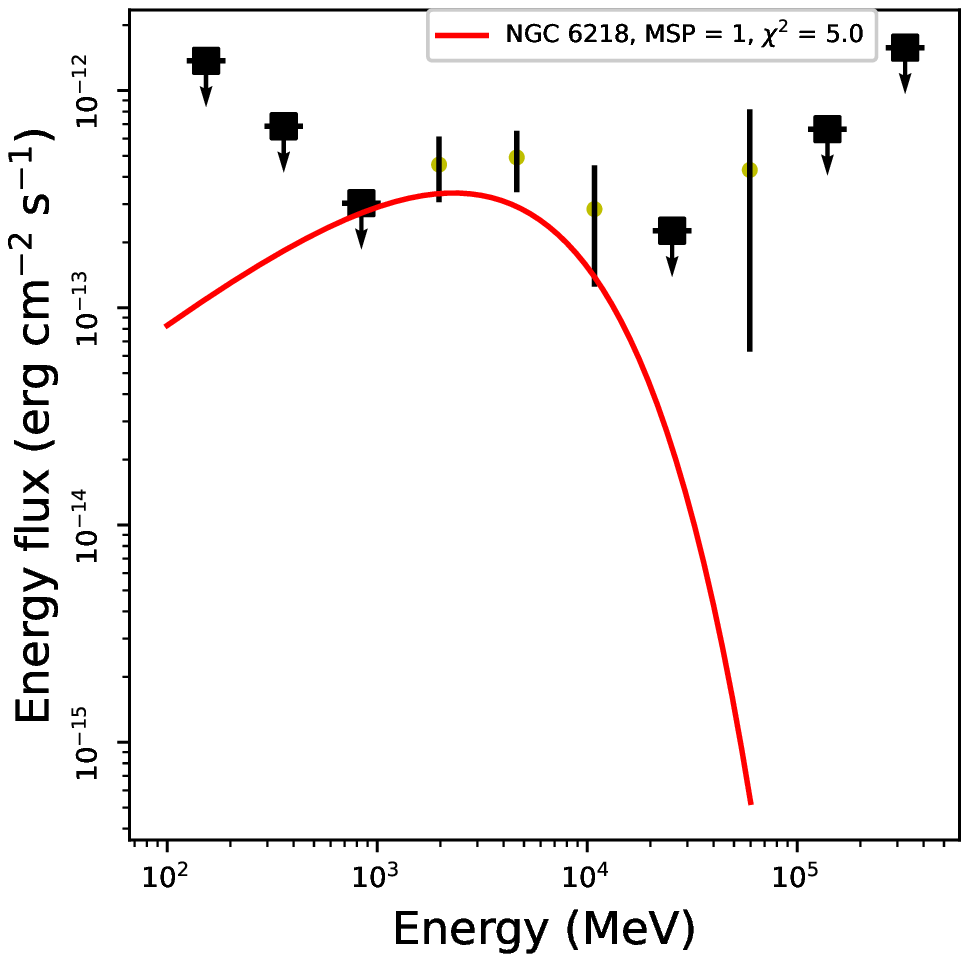}
\includegraphics[width=0.41\linewidth]{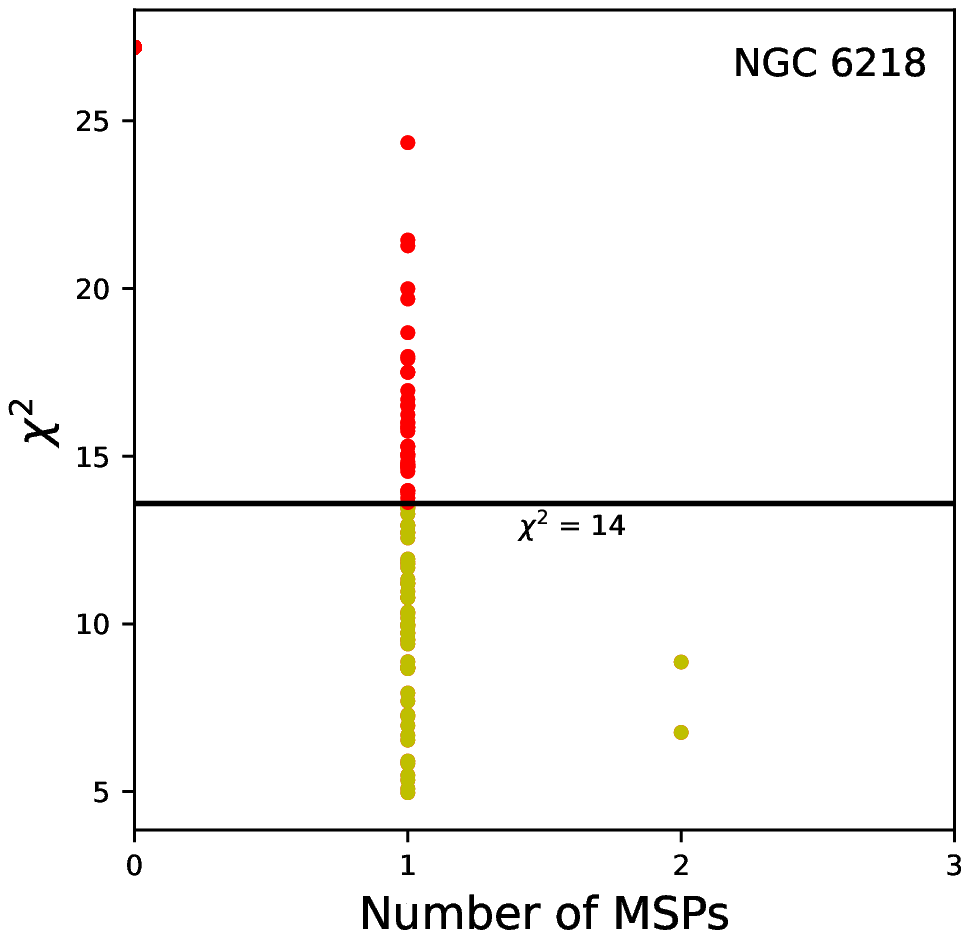}

\includegraphics[width=0.41\linewidth]{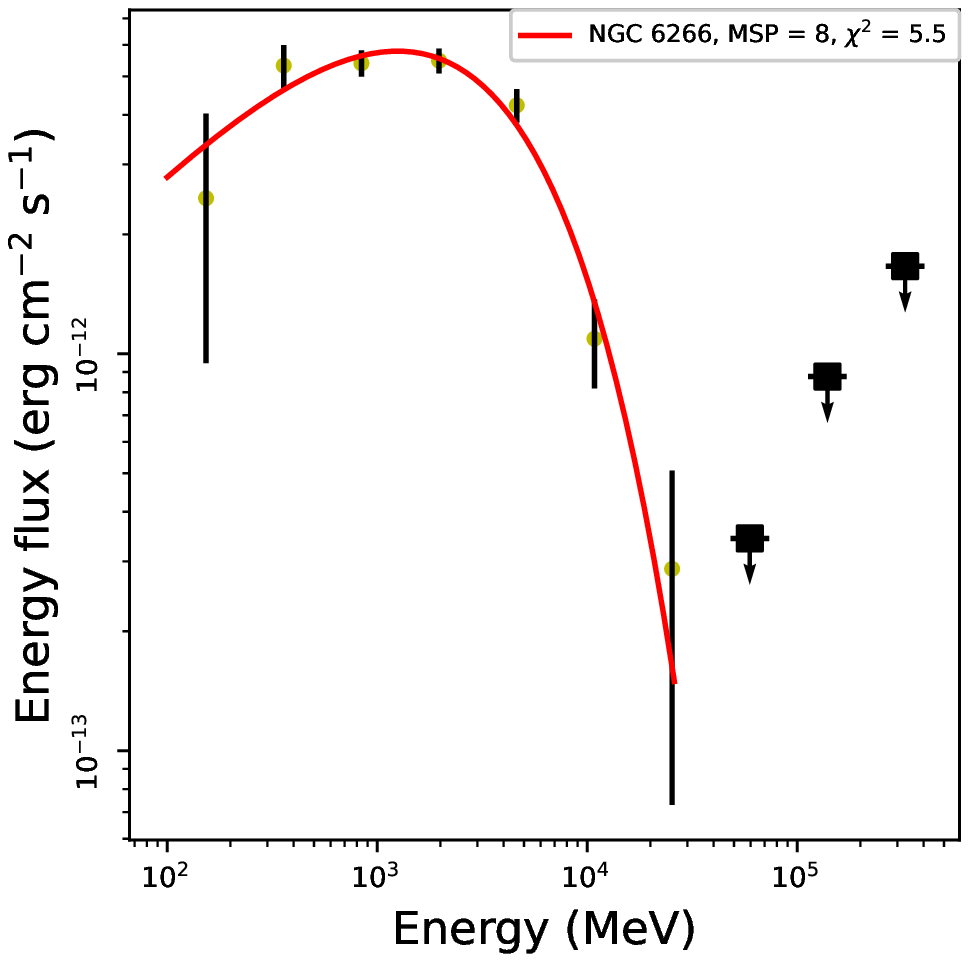}
\includegraphics[width=0.41\linewidth]{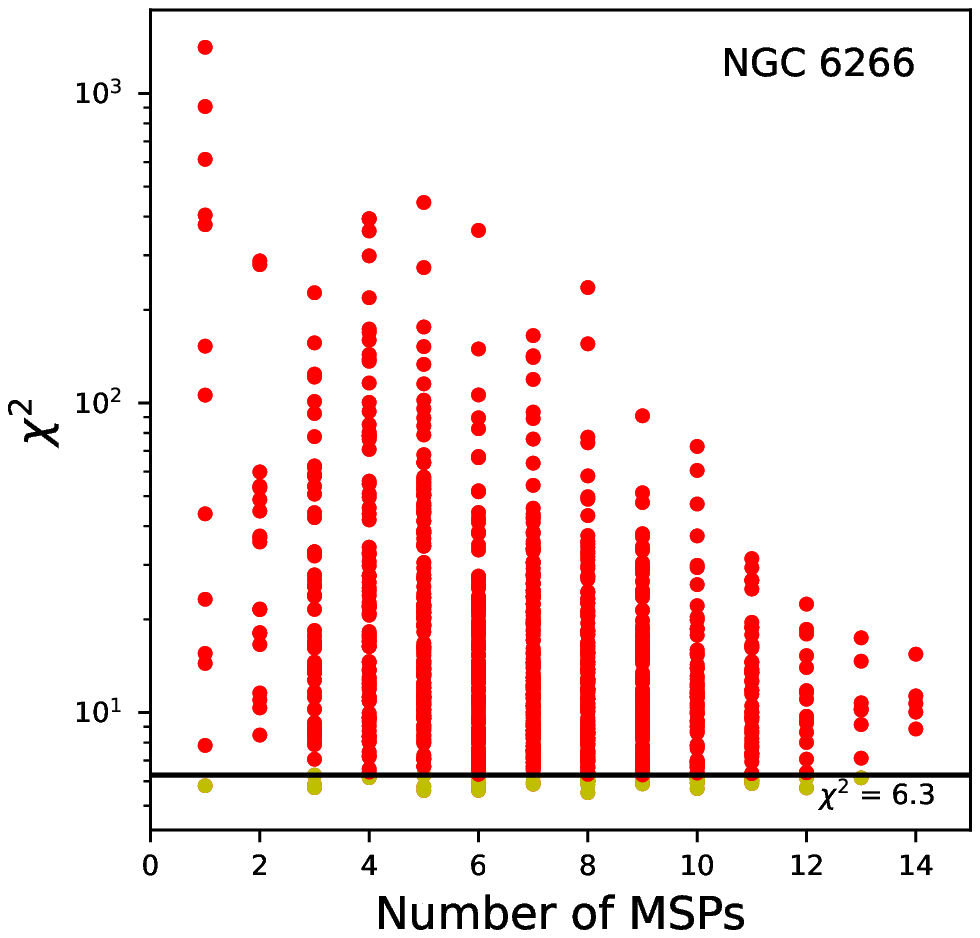}
\end{center}
        \caption{Same as Figure~\ref{fig:tuc}.}
\label{fig:6139}
\end{figure*}

\begin{figure*}
\begin{center}
        \includegraphics[width=0.41\linewidth]{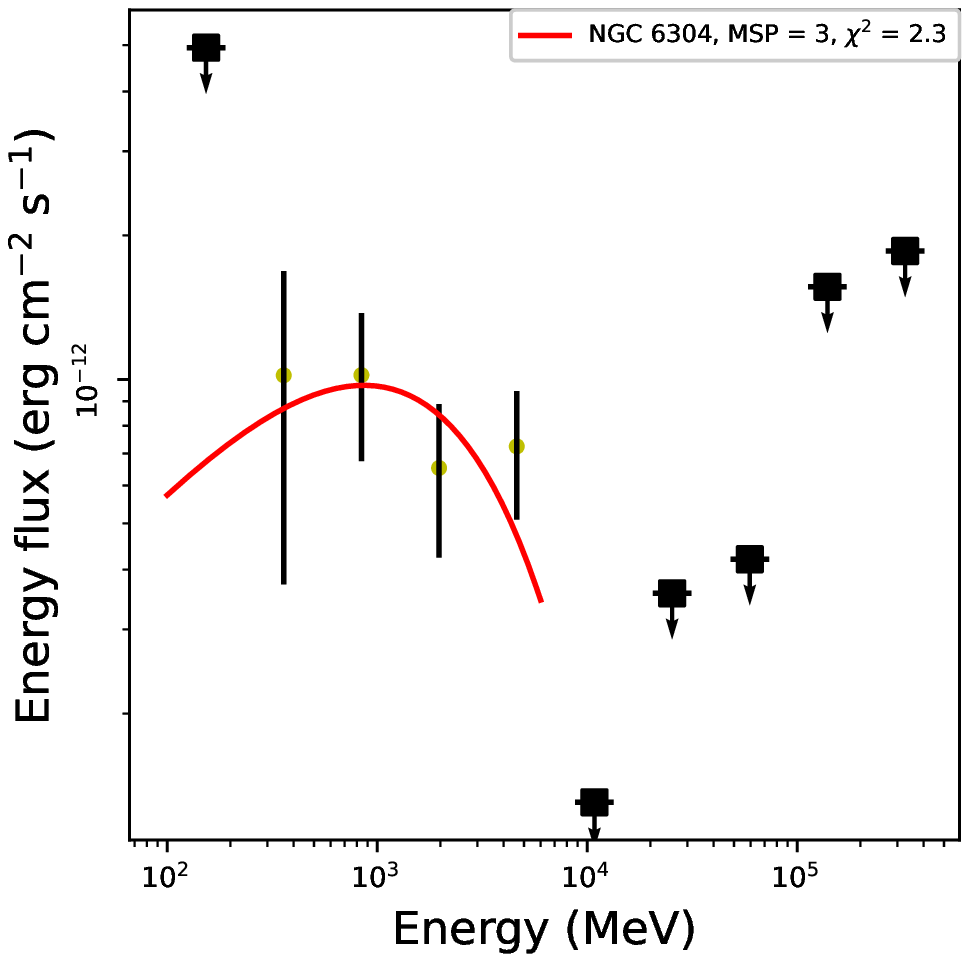}
\includegraphics[width=0.41\linewidth]{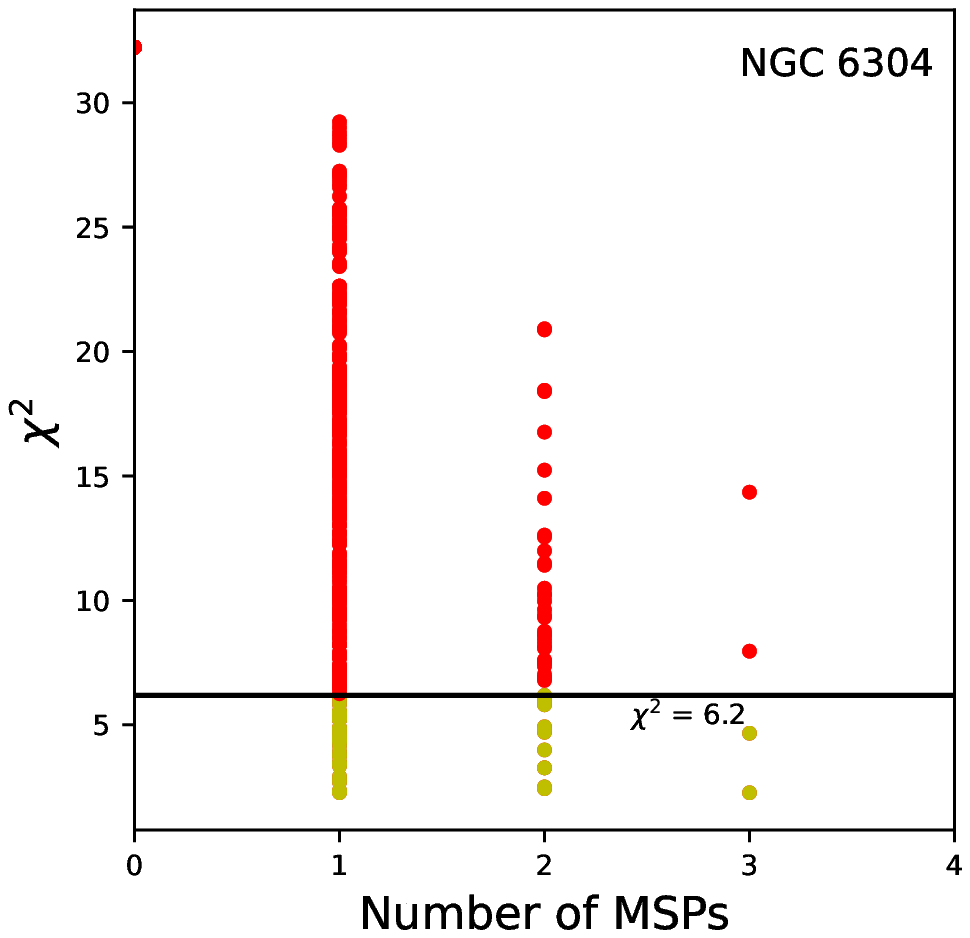}

\includegraphics[width=0.41\linewidth]{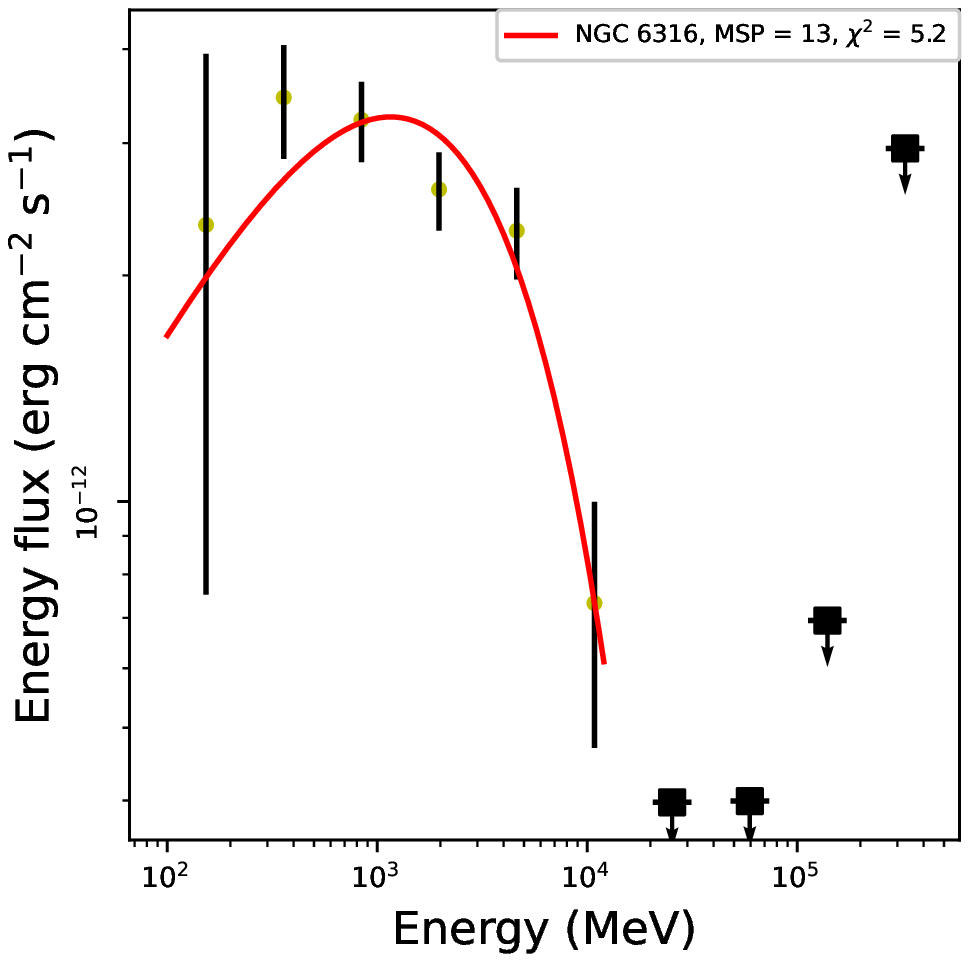}
\includegraphics[width=0.41\linewidth]{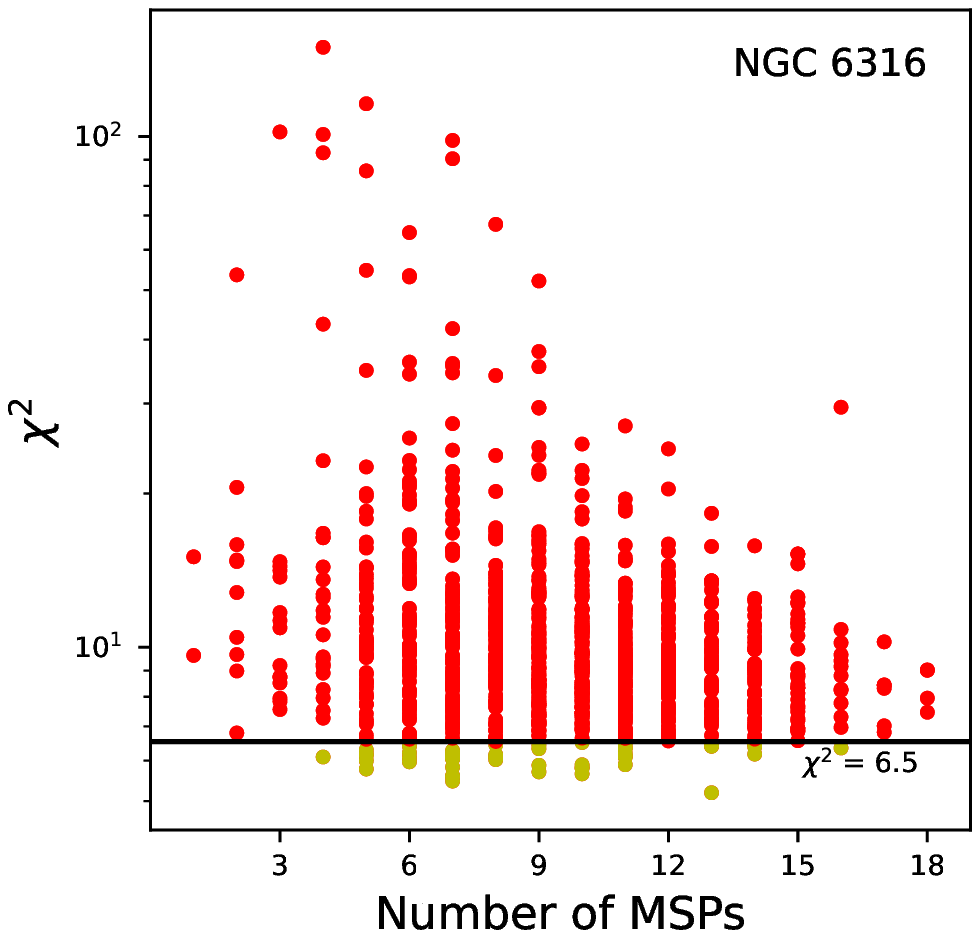}

        \includegraphics[width=0.41\linewidth]{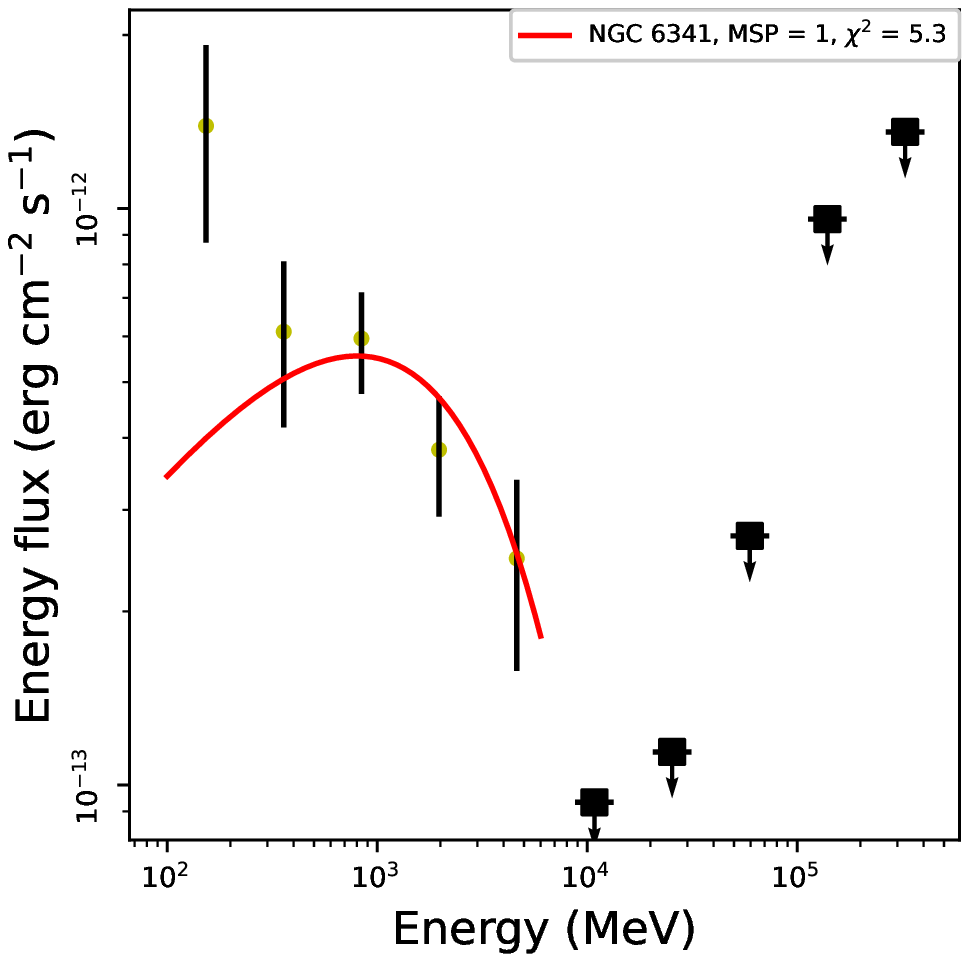}
\includegraphics[width=0.41\linewidth]{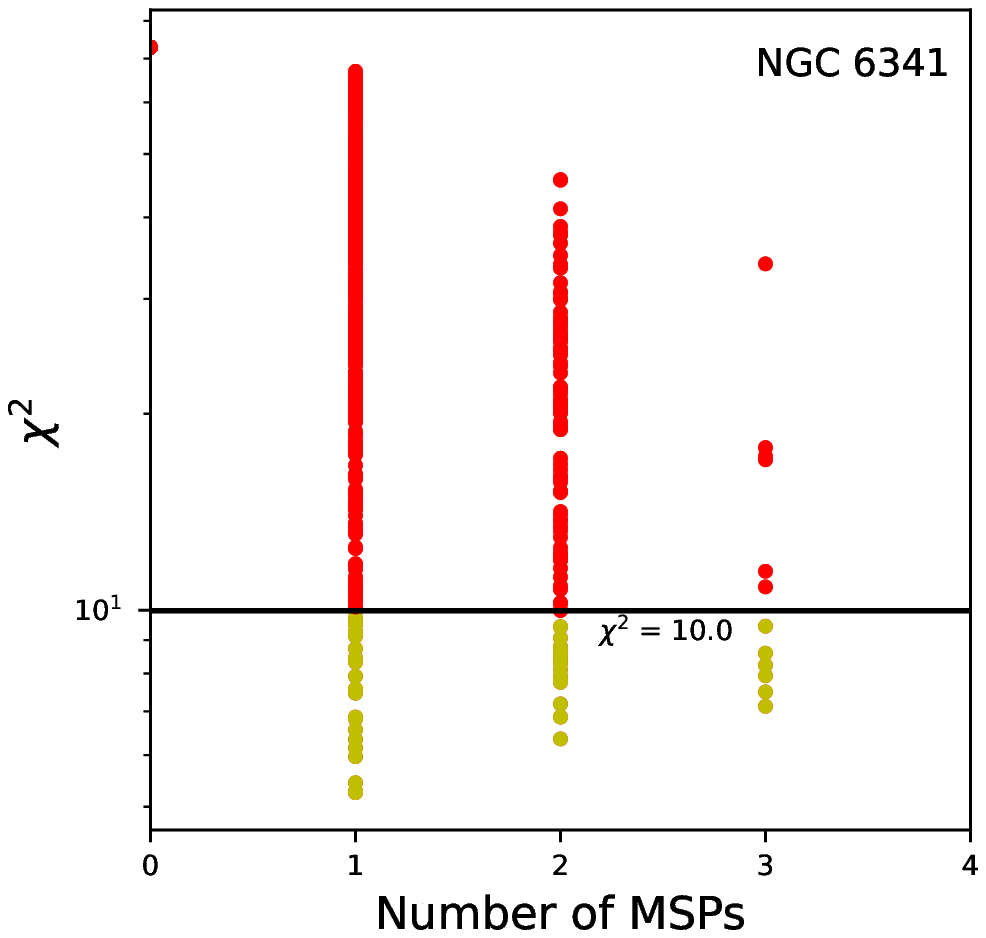}

\end{center}
        \caption{Same as Figure~\ref{fig:tuc}.}
\label{fig:6304}
\end{figure*}

\begin{figure*}
\begin{center}
        \includegraphics[width=0.41\linewidth]{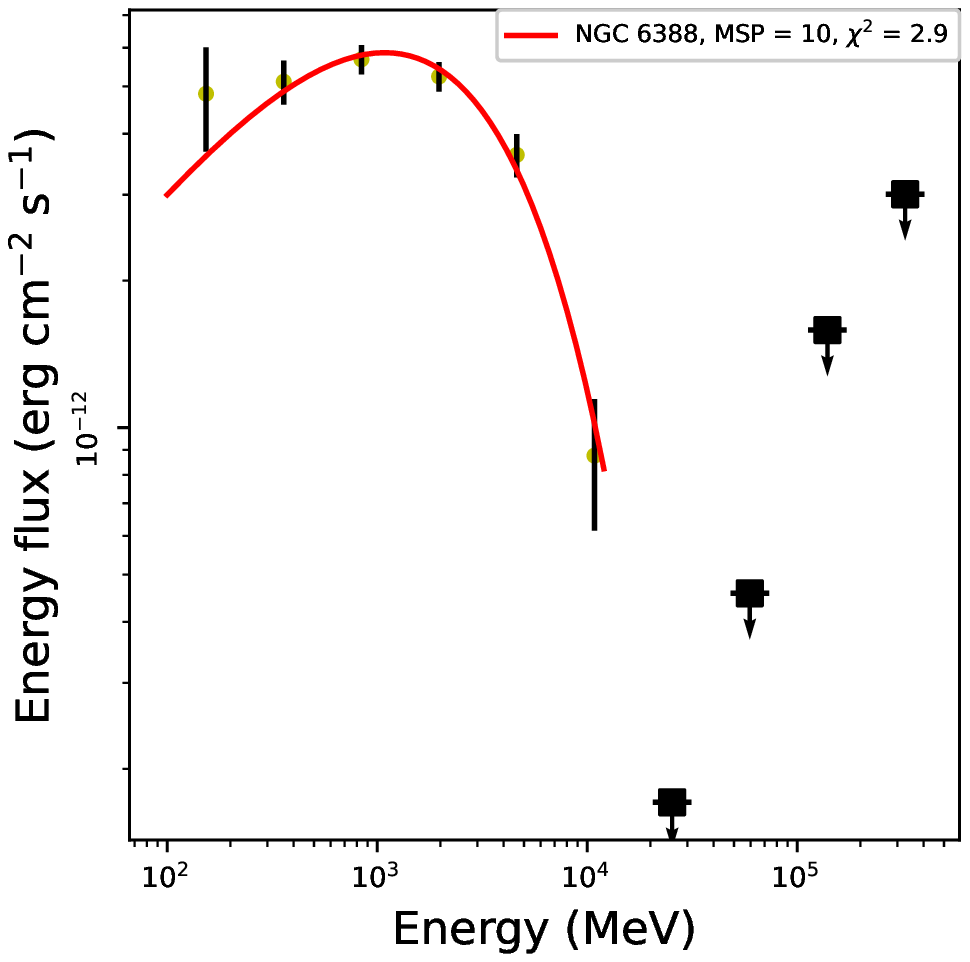}
\includegraphics[width=0.41\linewidth]{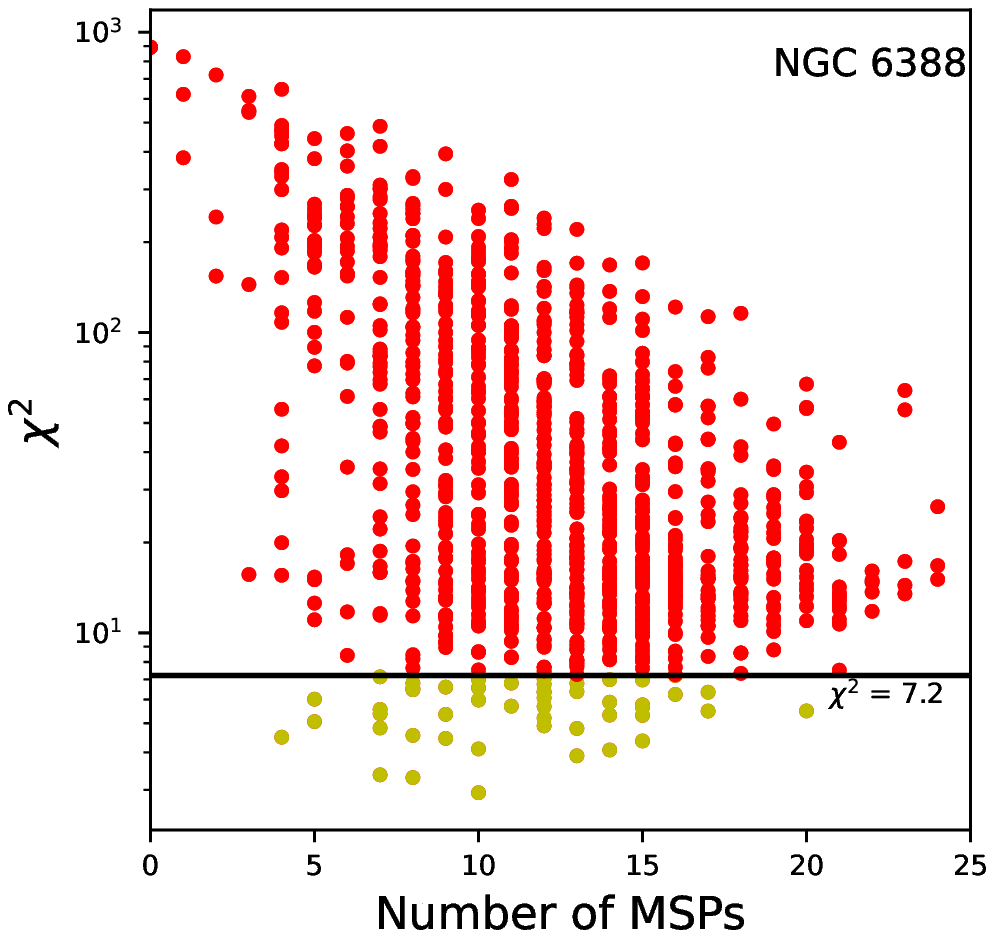}

\includegraphics[width=0.41\linewidth]{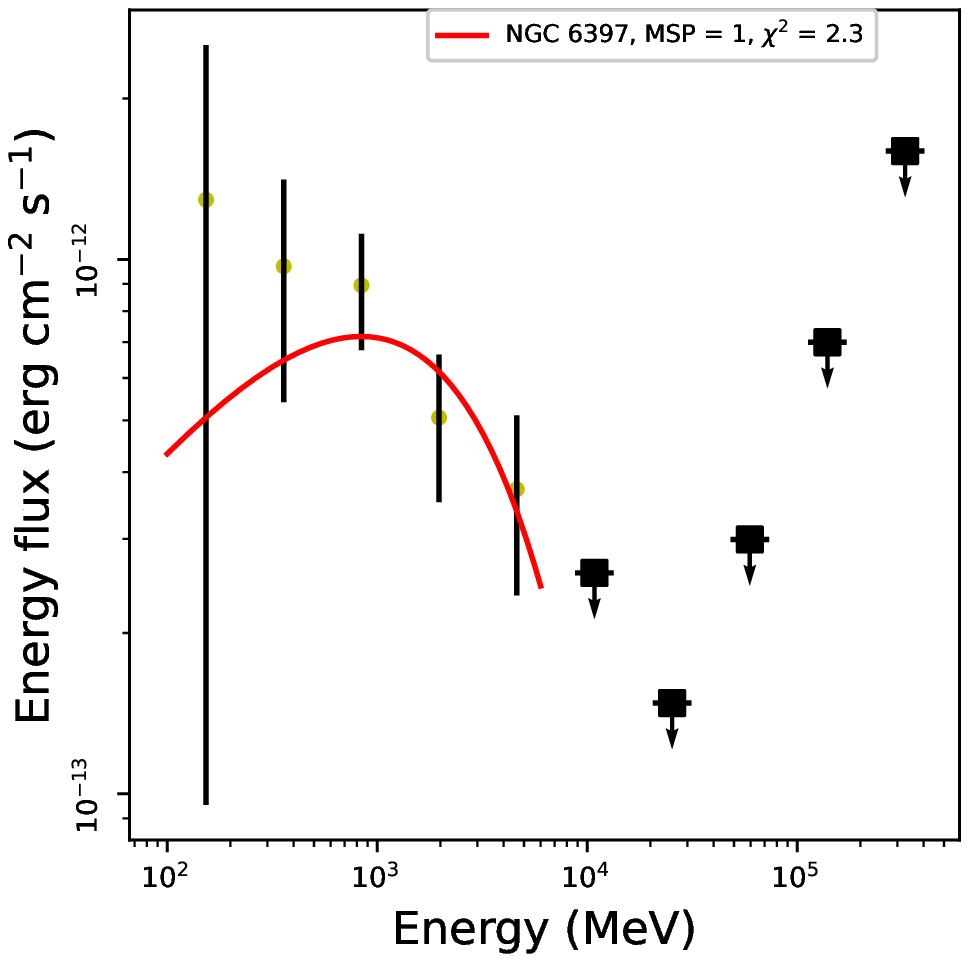}
\includegraphics[width=0.41\linewidth]{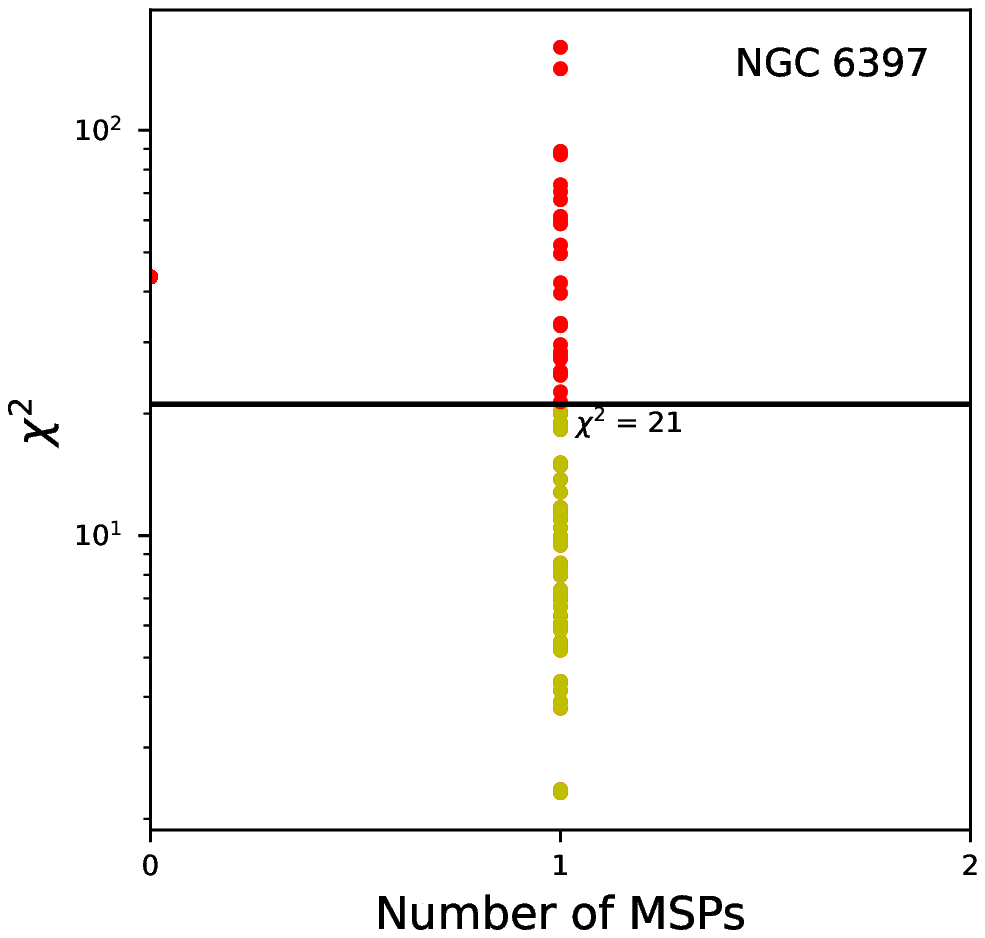}

        \includegraphics[width=0.41\linewidth]{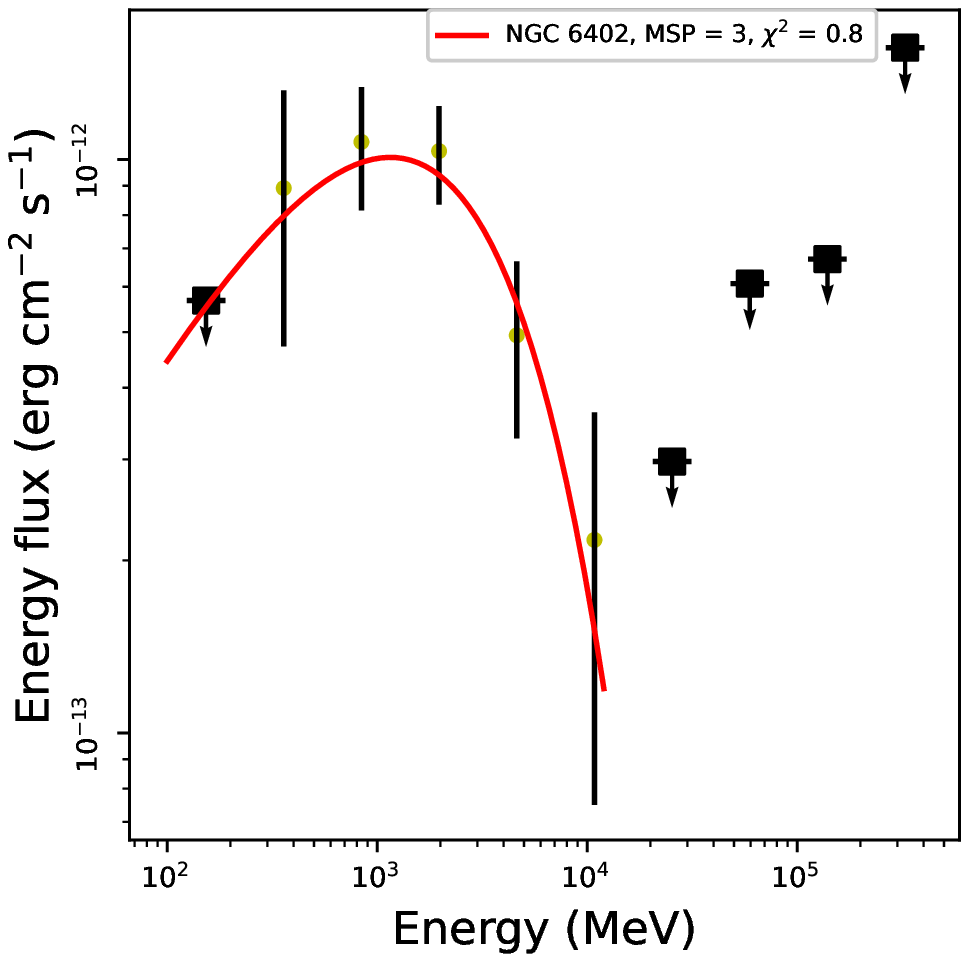}
\includegraphics[width=0.41\linewidth]{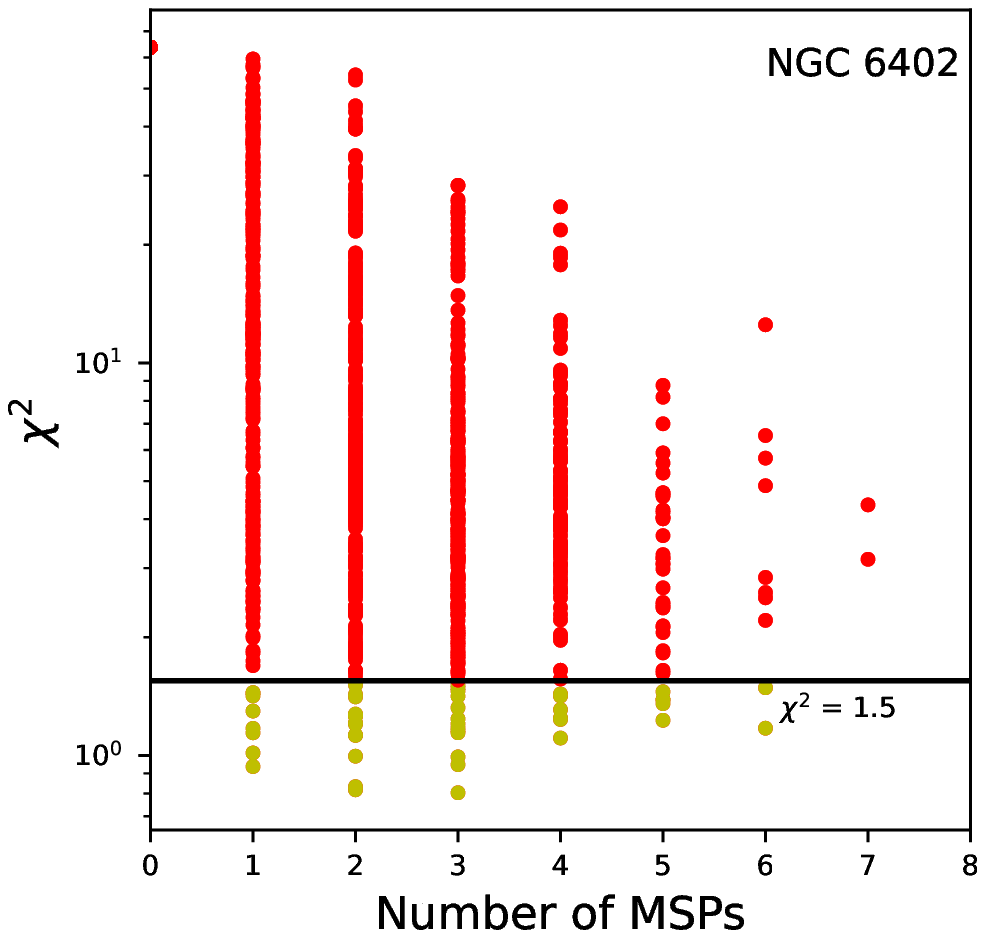}
\end{center}
        \caption{Same as Figure~\ref{fig:tuc}.}
\end{figure*}

\begin{figure*}
\begin{center}
        \includegraphics[width=0.41\linewidth]{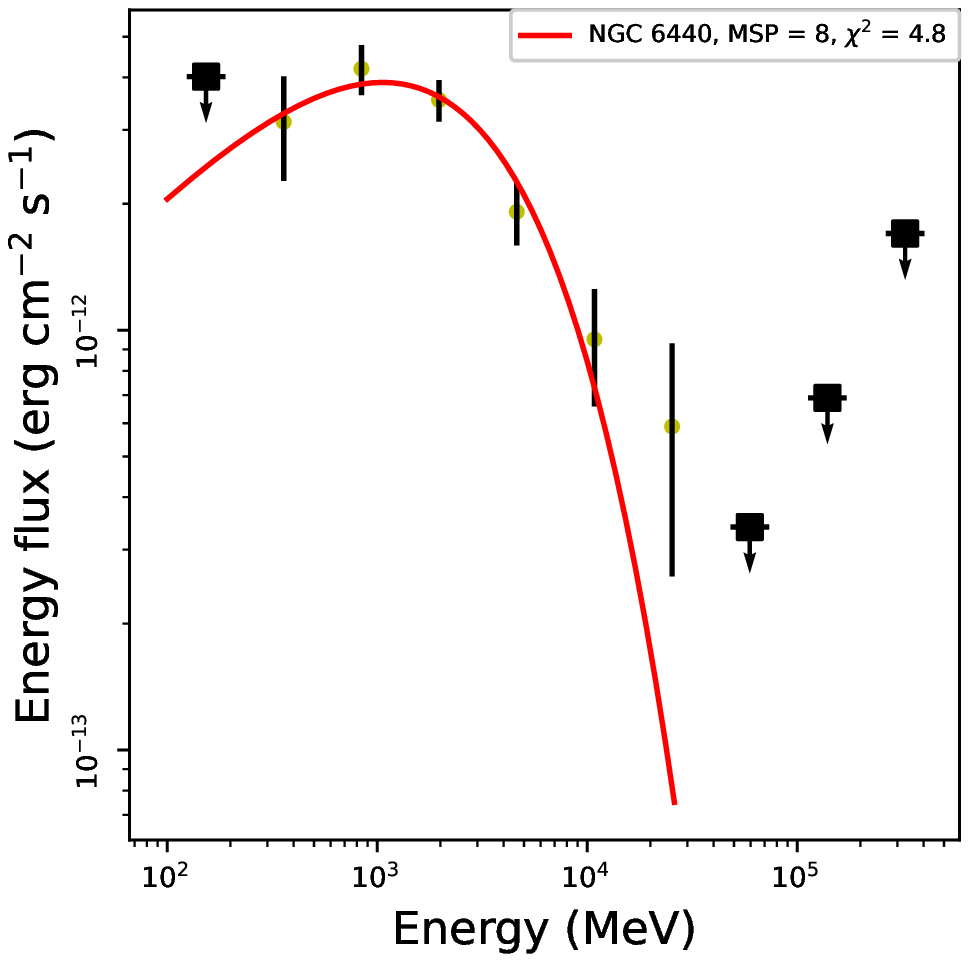}
\includegraphics[width=0.41\linewidth]{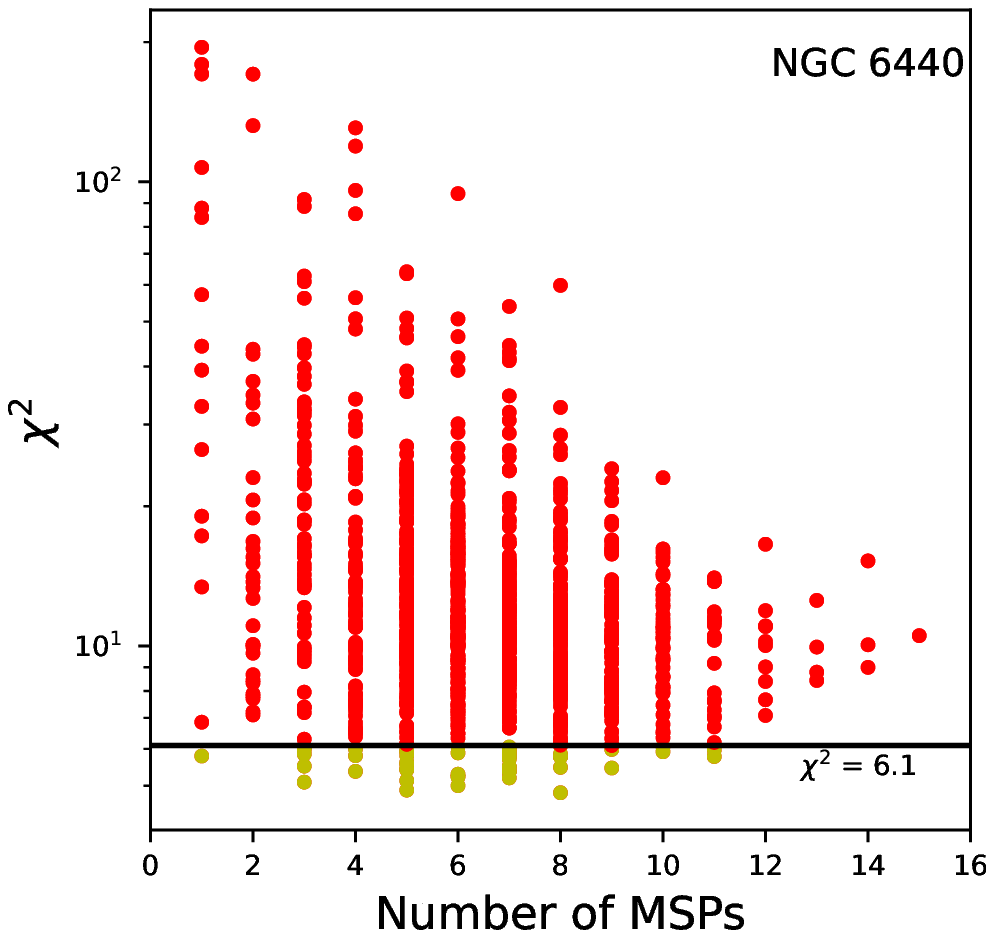}

\includegraphics[width=0.41\linewidth]{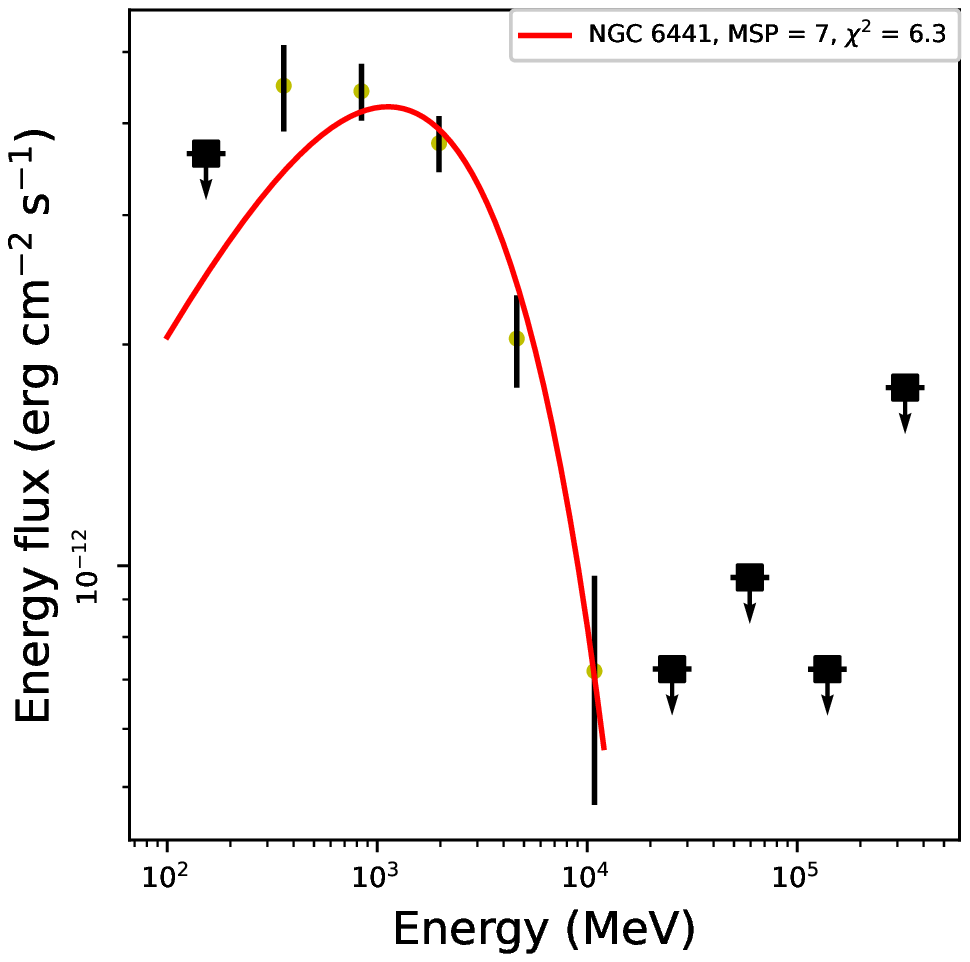}
\includegraphics[width=0.41\linewidth]{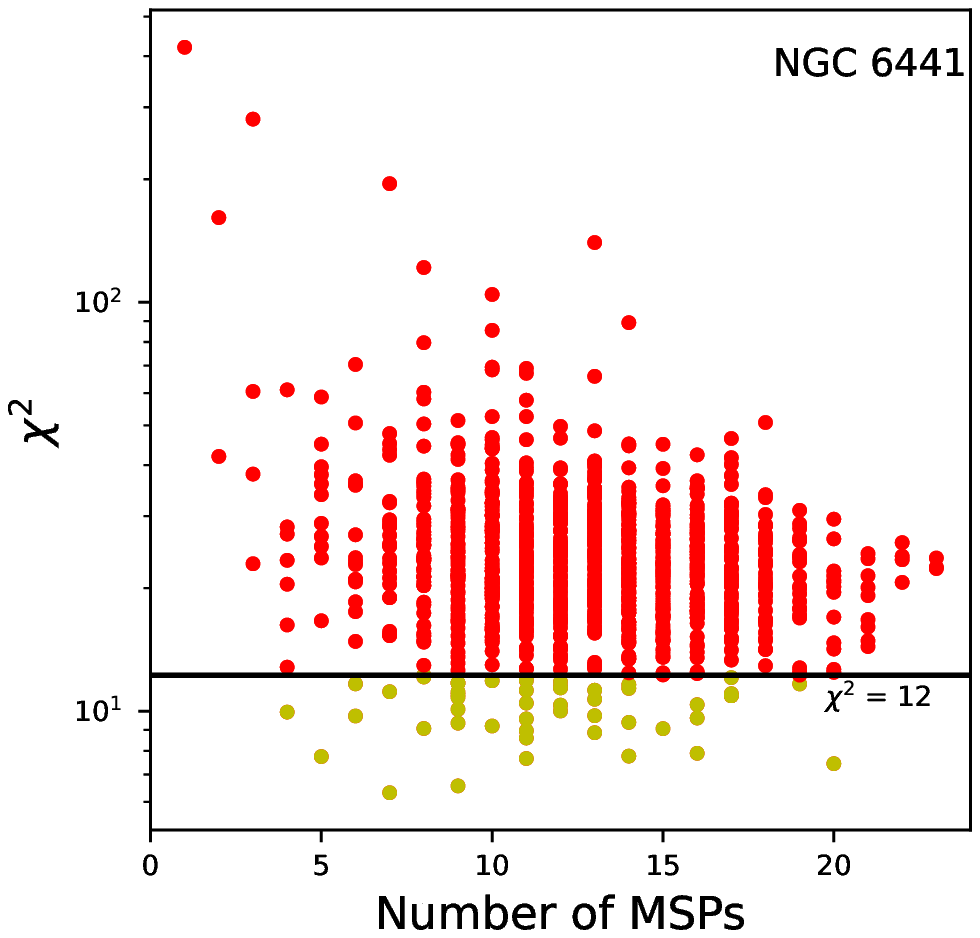}

        \includegraphics[width=0.41\linewidth]{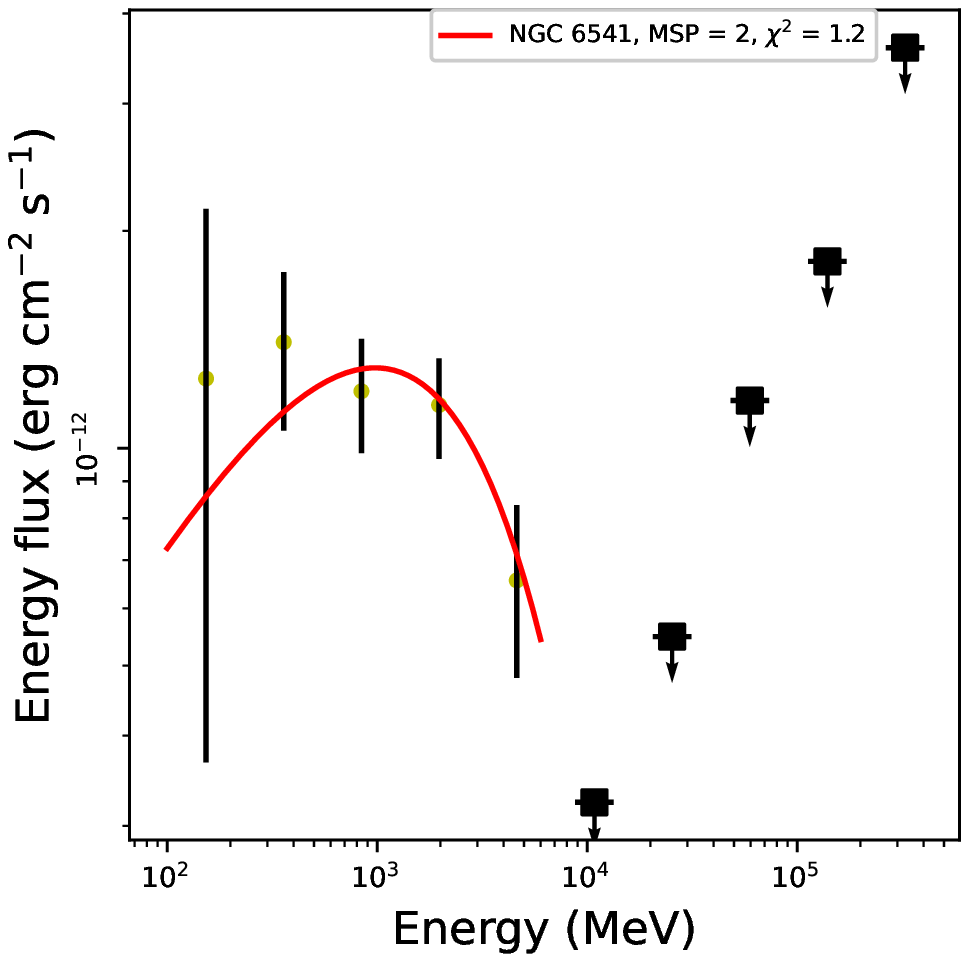}
\includegraphics[width=0.41\linewidth]{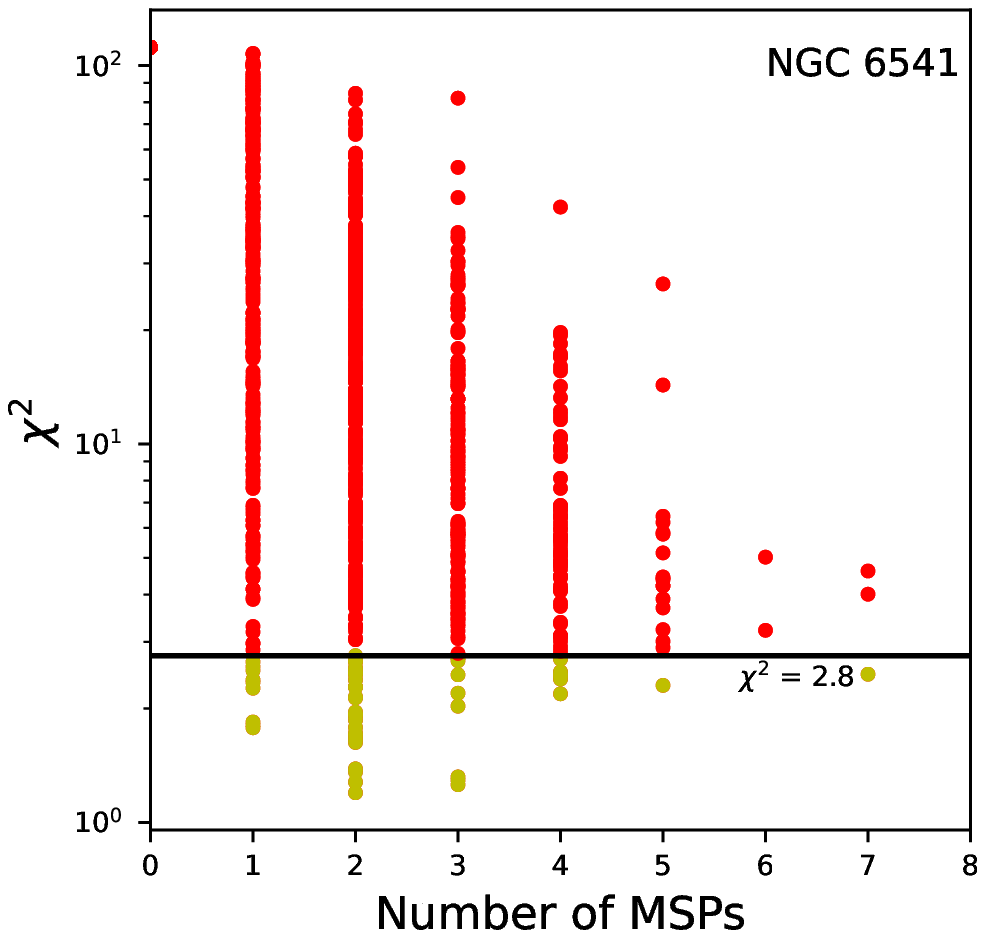}
\end{center}
        \caption{Same as Figure~\ref{fig:tuc}.}
\end{figure*}

\begin{figure*}
\begin{center}
        \includegraphics[width=0.41\linewidth]{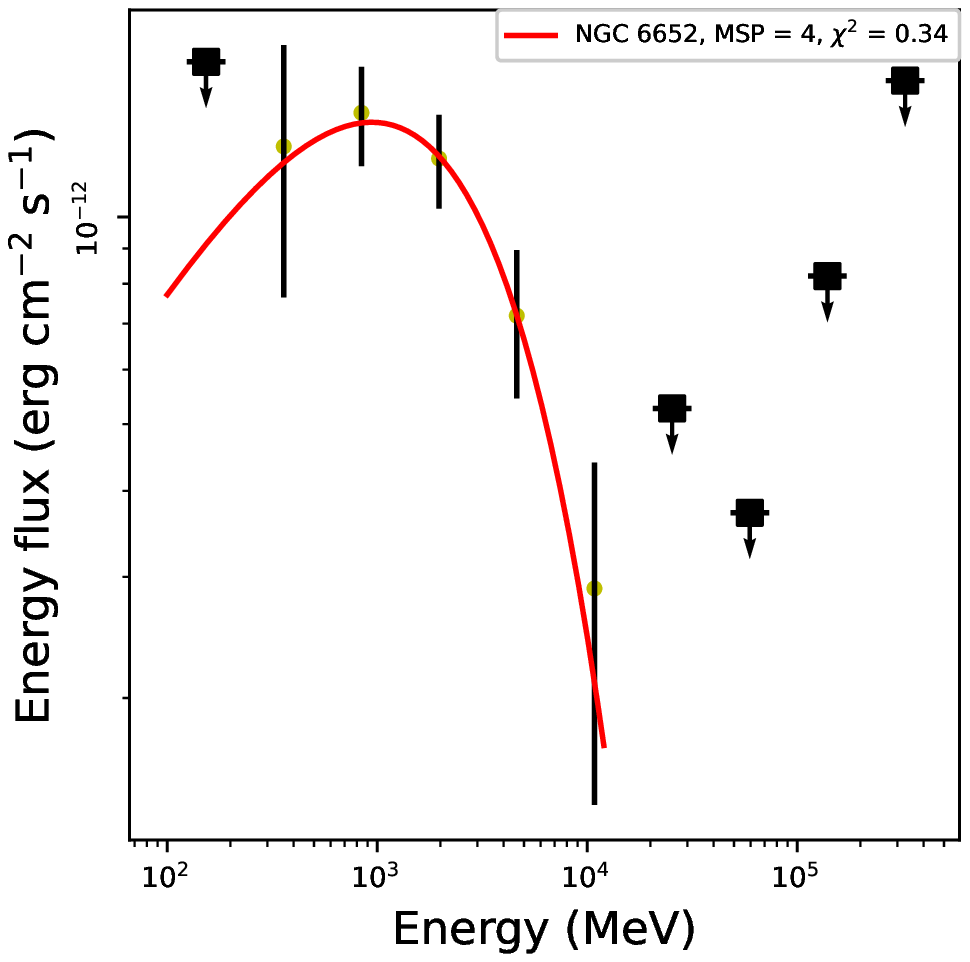}
\includegraphics[width=0.41\linewidth]{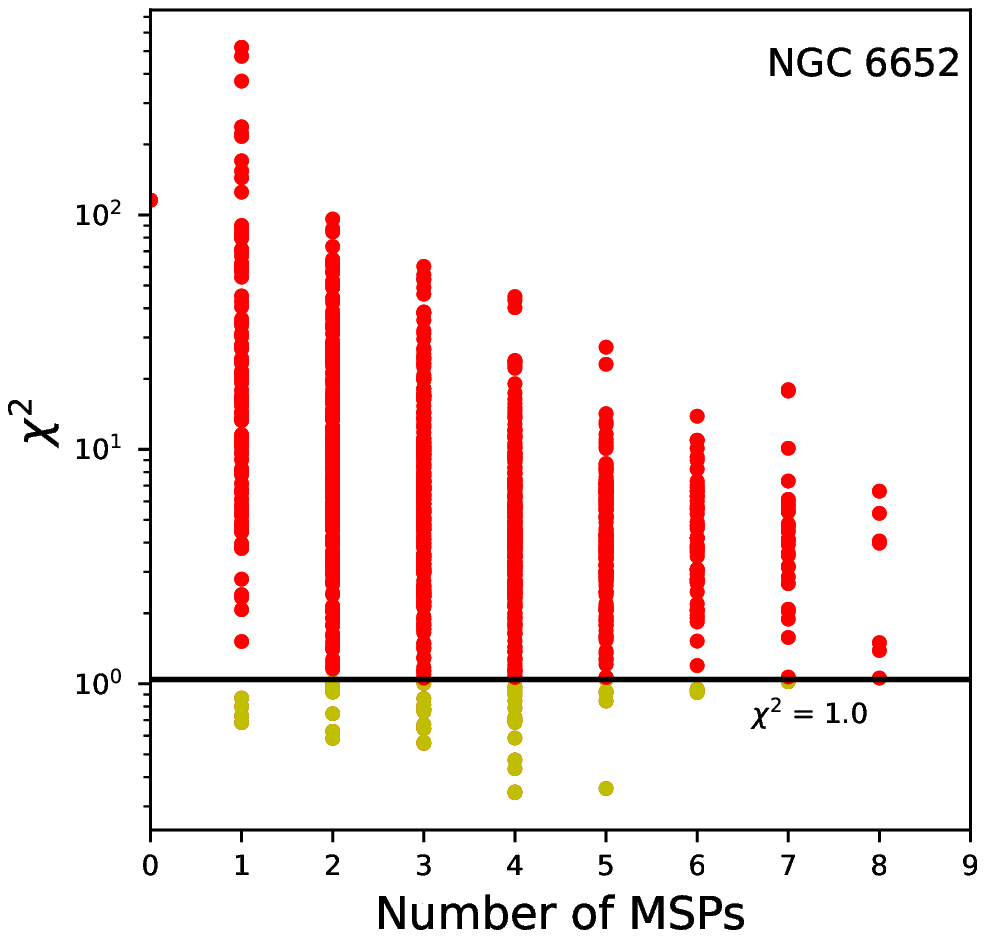}

        \includegraphics[width=0.41\linewidth]{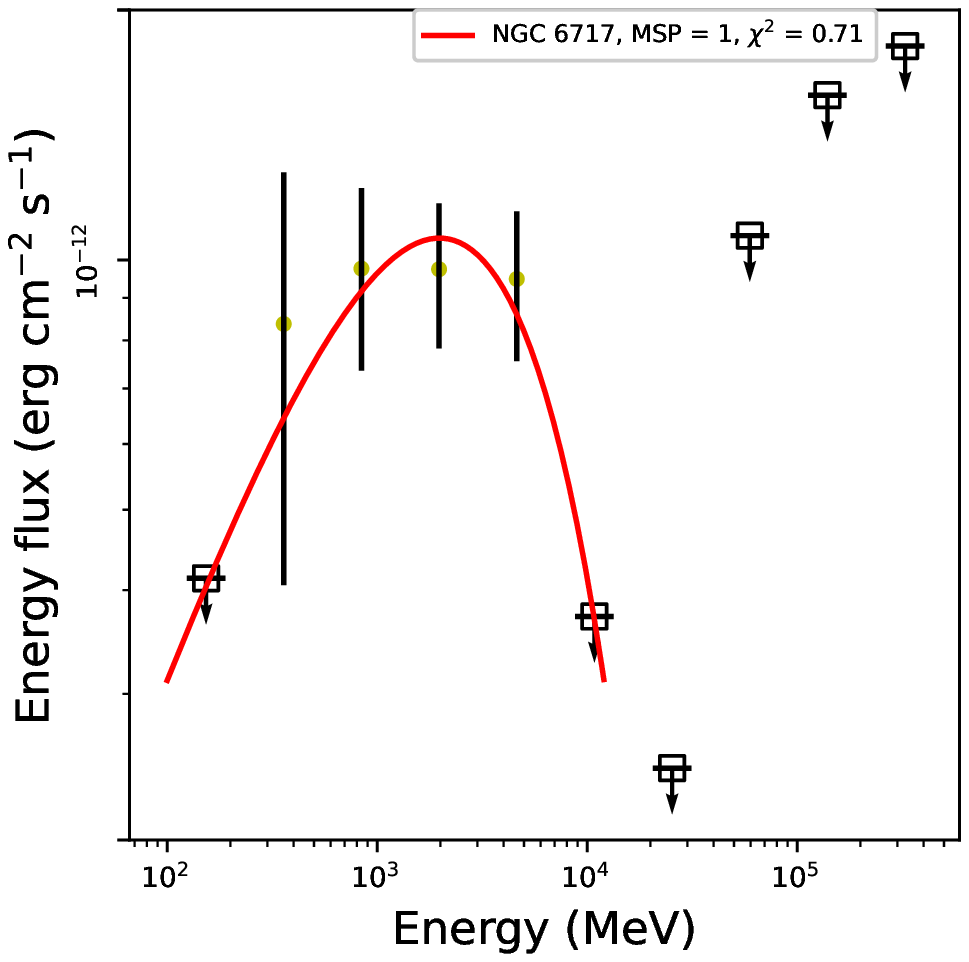}
\includegraphics[width=0.41\linewidth]{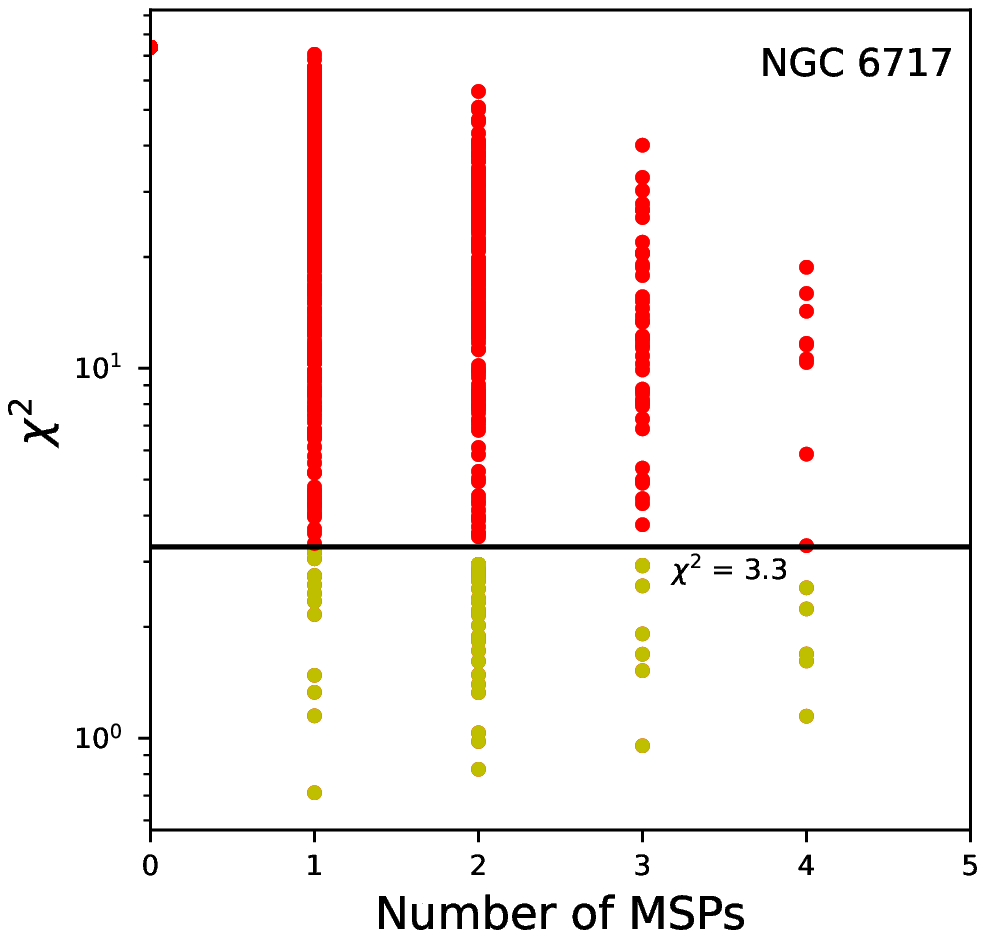}

 \includegraphics[width=0.41\linewidth]{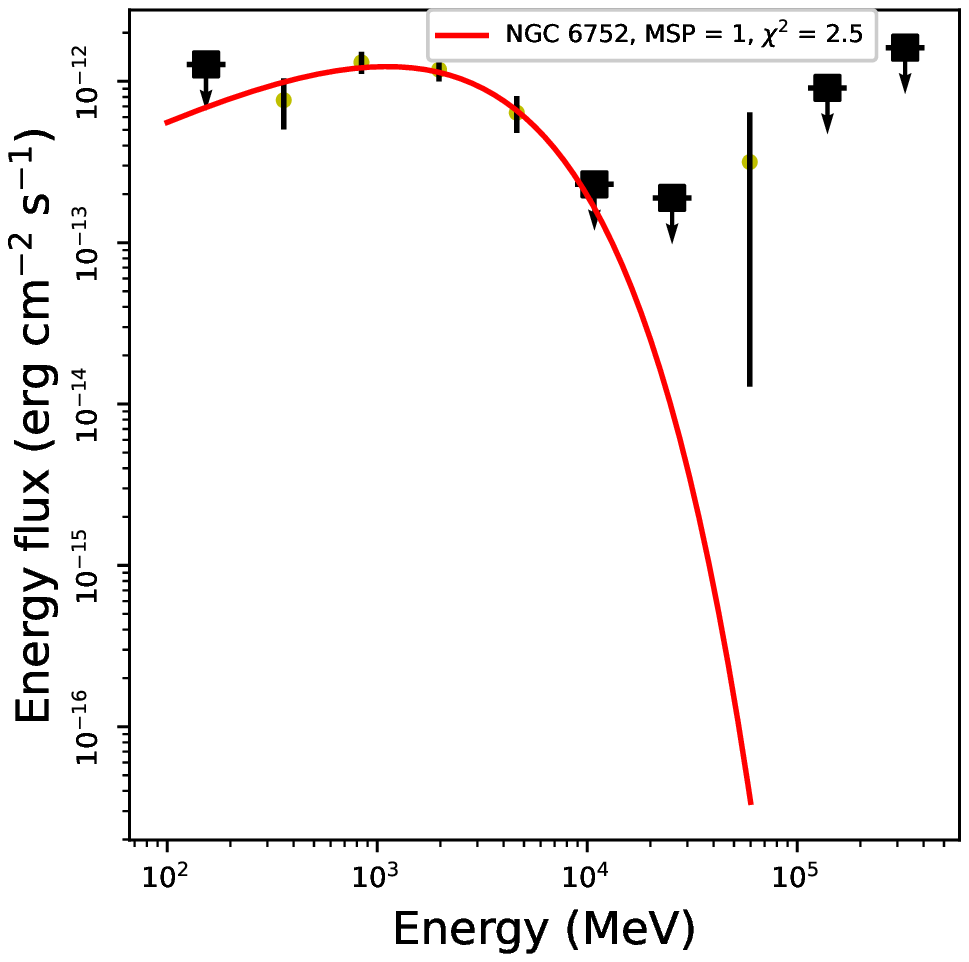}
\includegraphics[width=0.41\linewidth]{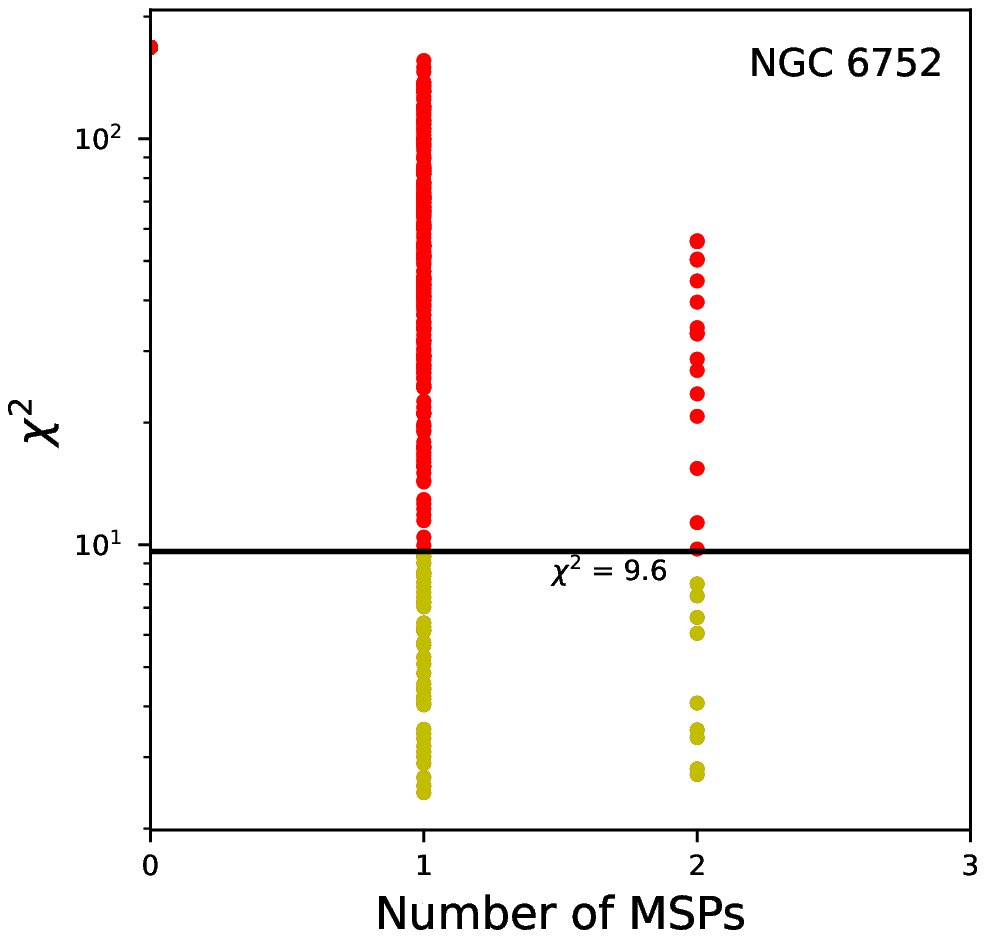}
\end{center}
        \caption{Same as Figure~\ref{fig:tuc}.}
\end{figure*}

\begin{figure*}
\begin{center}

\includegraphics[width=0.41\linewidth]{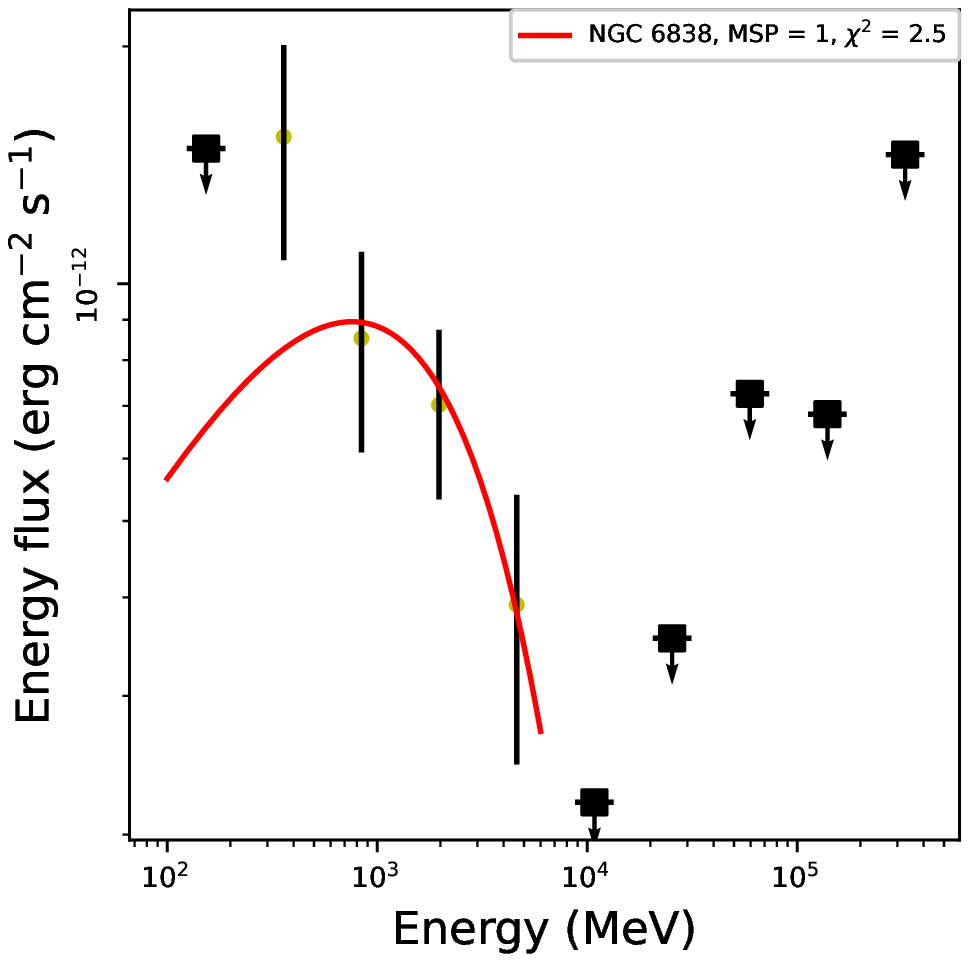}
\includegraphics[width=0.41\linewidth]{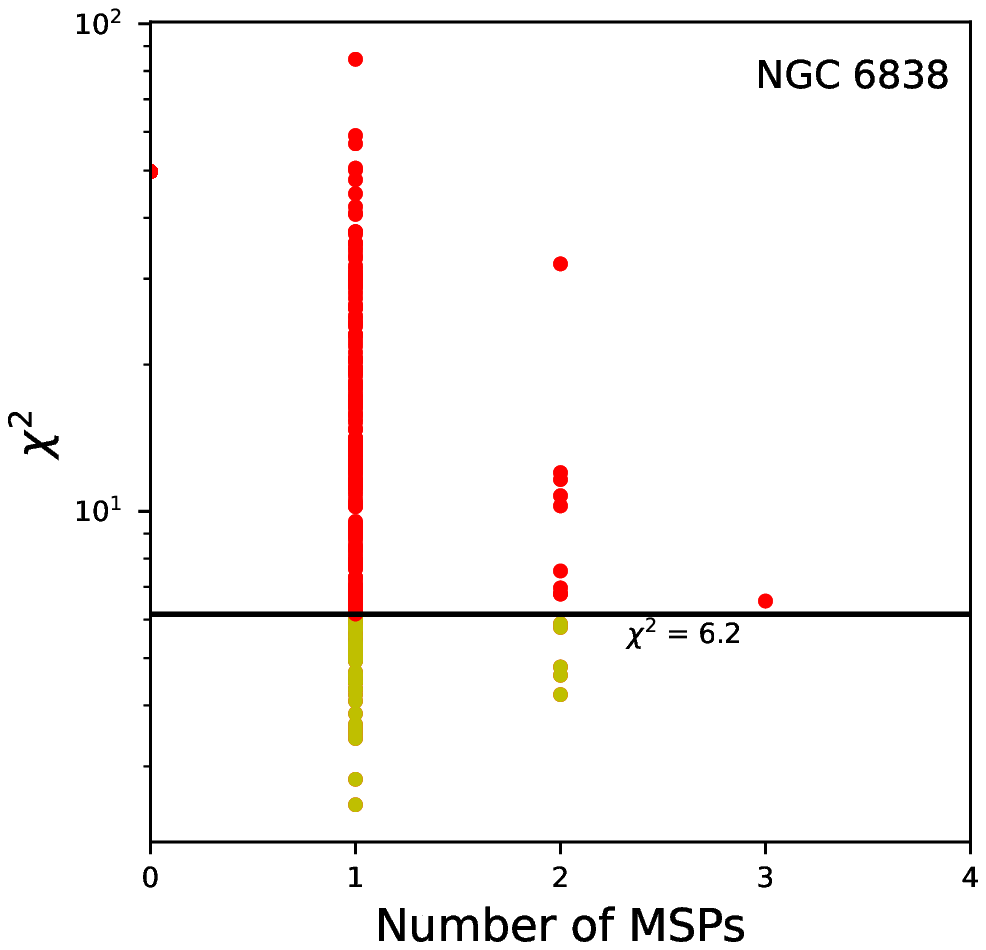}

        \includegraphics[width=0.41\linewidth]{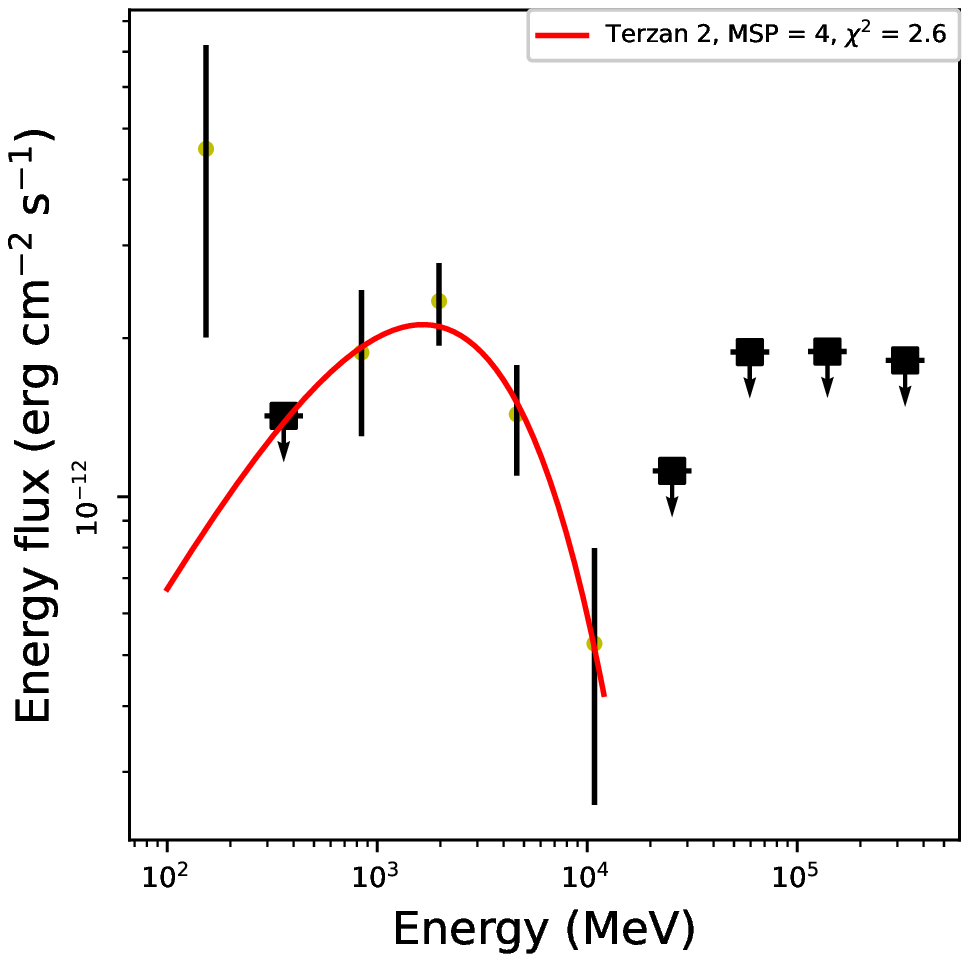}
\includegraphics[width=0.41\linewidth]{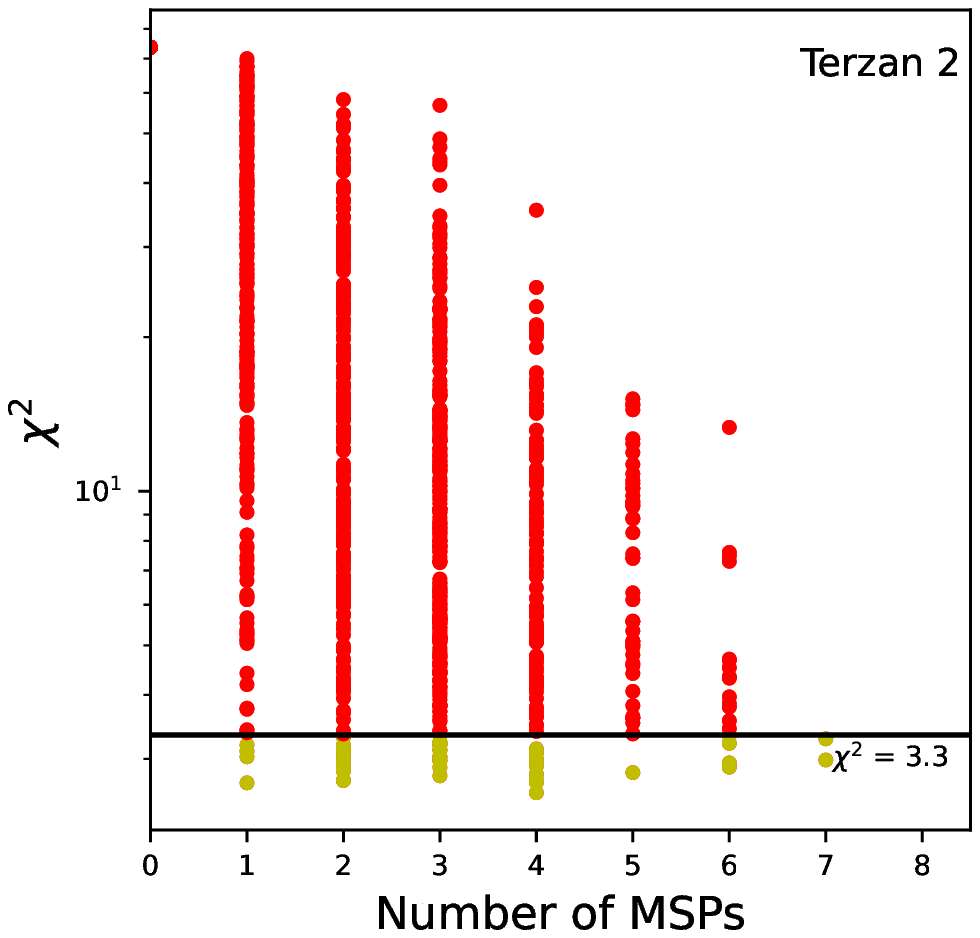}
        \includegraphics[width=0.41\linewidth]{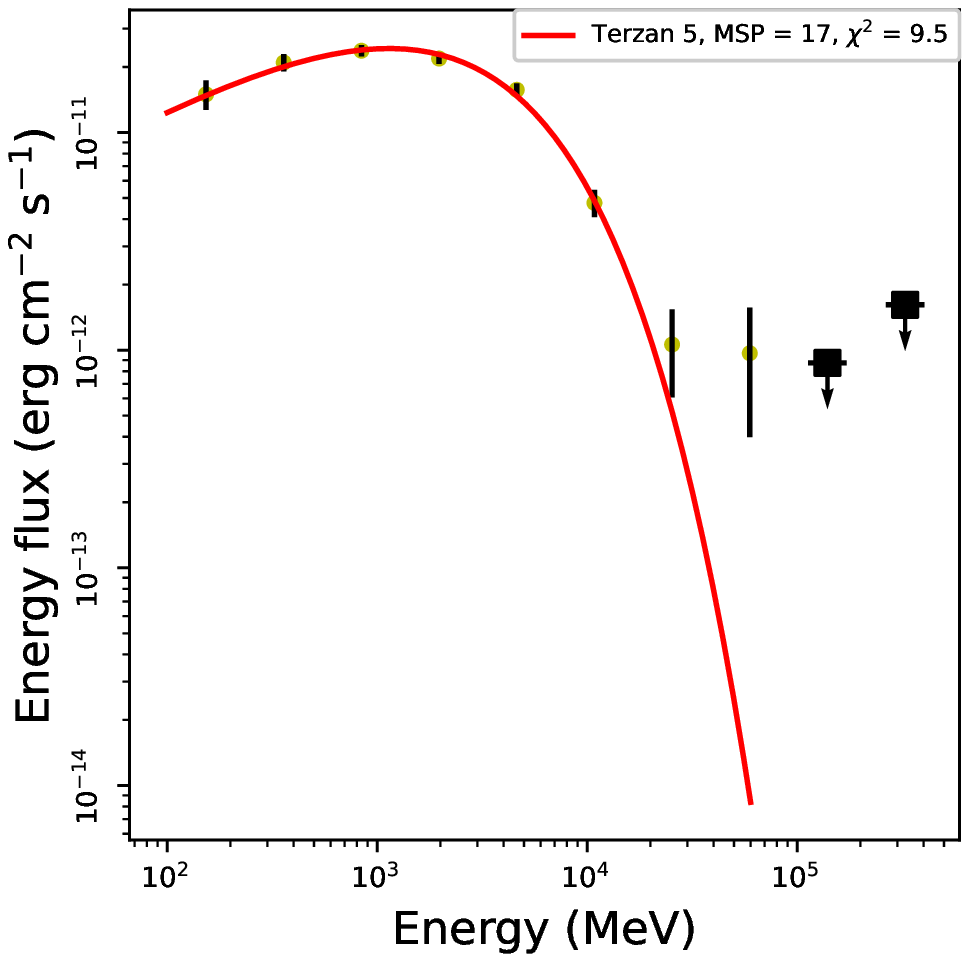}
\includegraphics[width=0.41\linewidth]{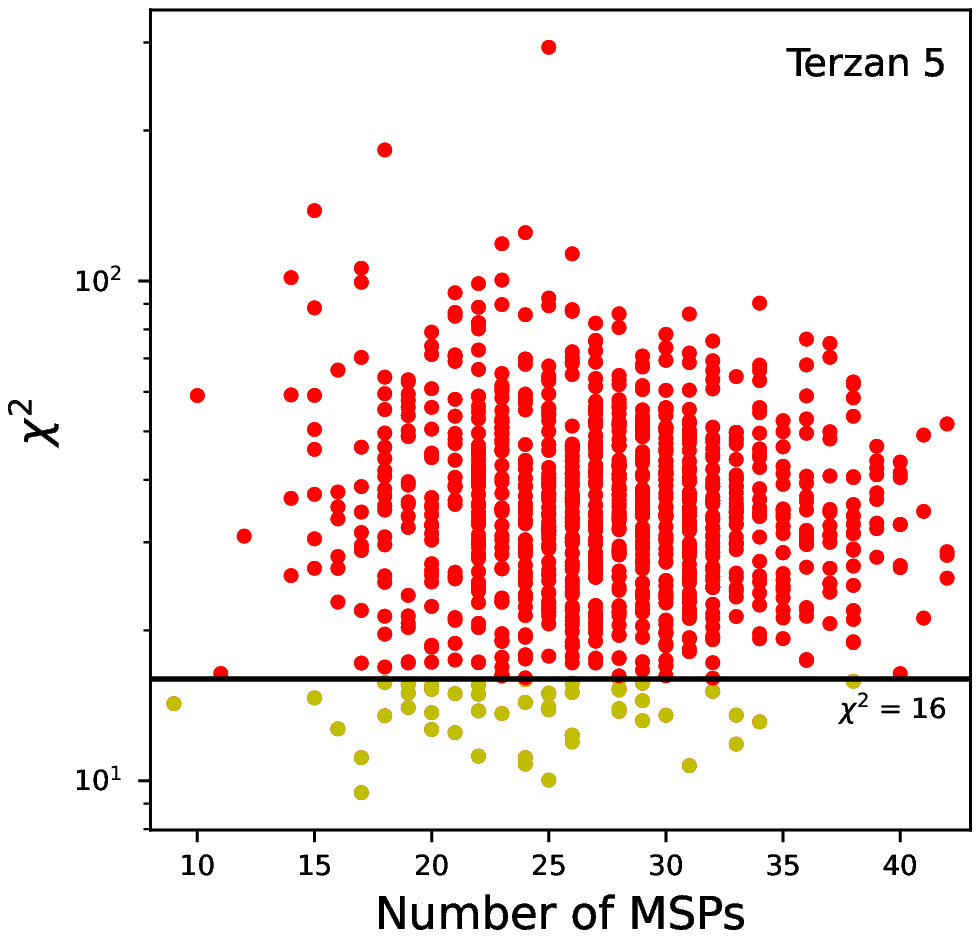}
\end{center}
        \caption{Same as Figure~\ref{fig:tuc}.}
\end{figure*}

\begin{figure*}
\begin{center}

	\includegraphics[width=0.41\linewidth]{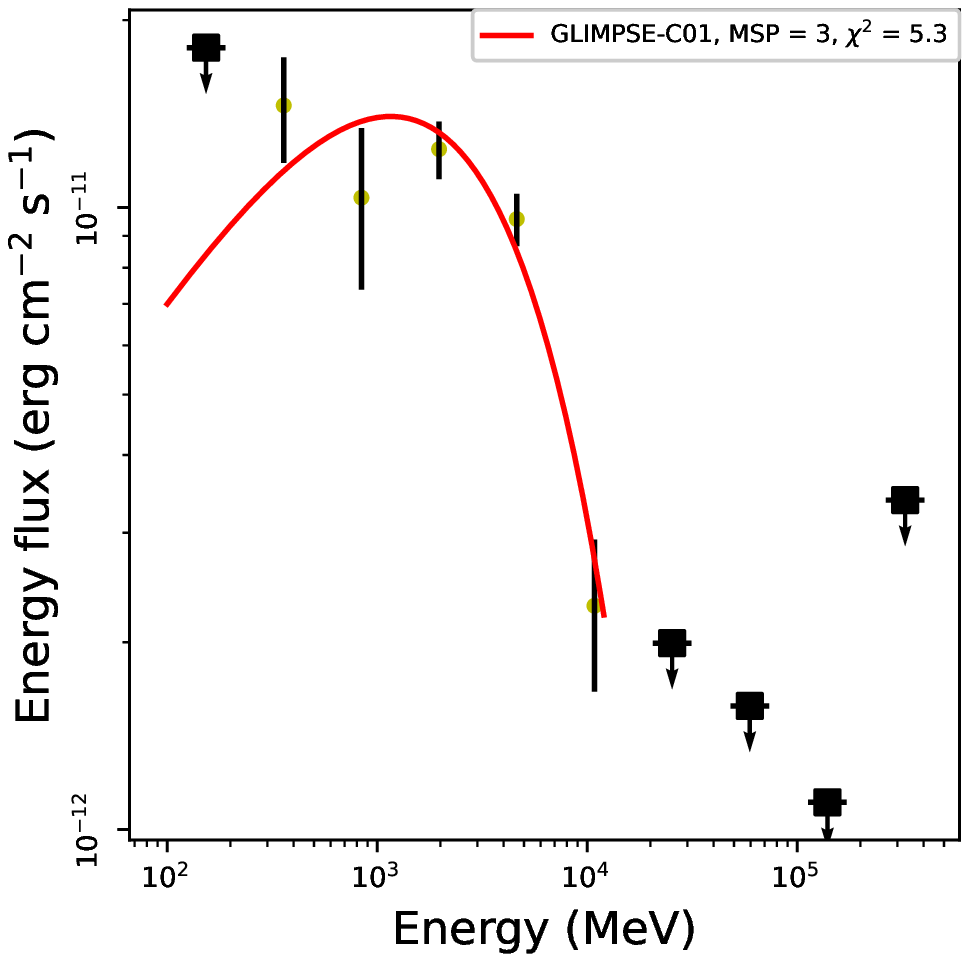}
\includegraphics[width=0.41\linewidth]{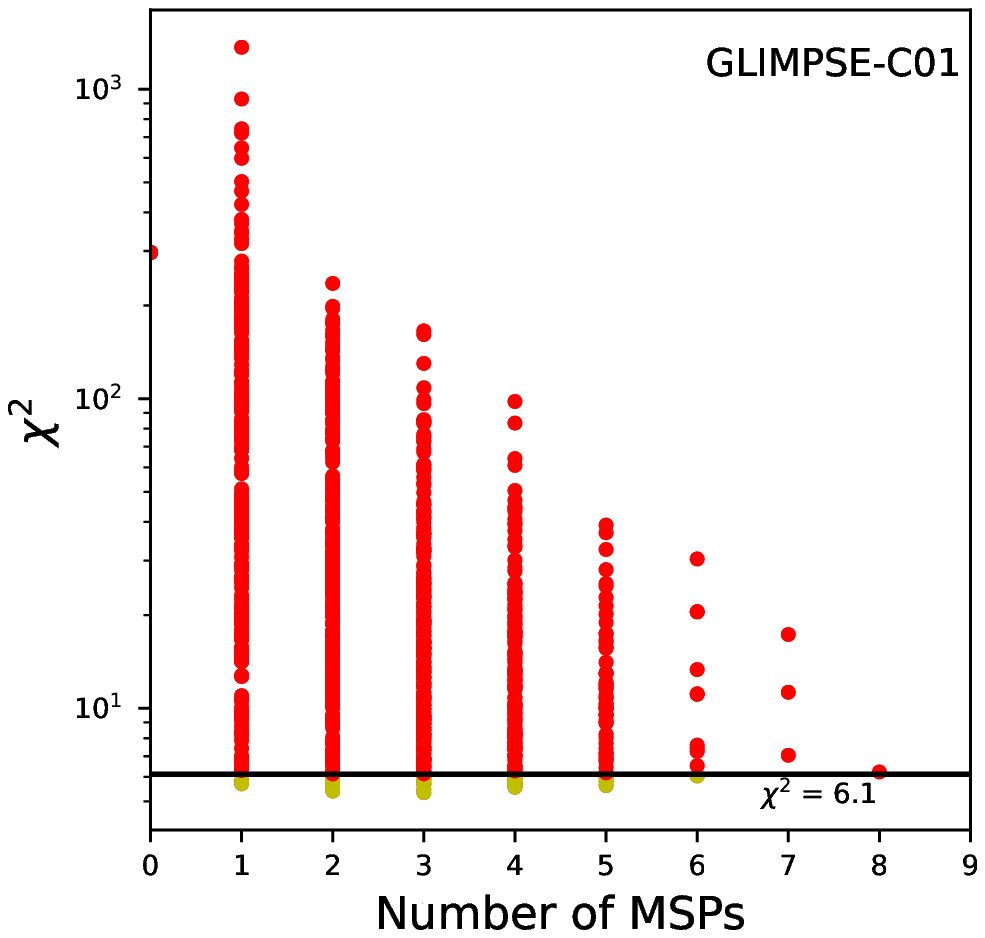}

	\includegraphics[width=0.41\linewidth]{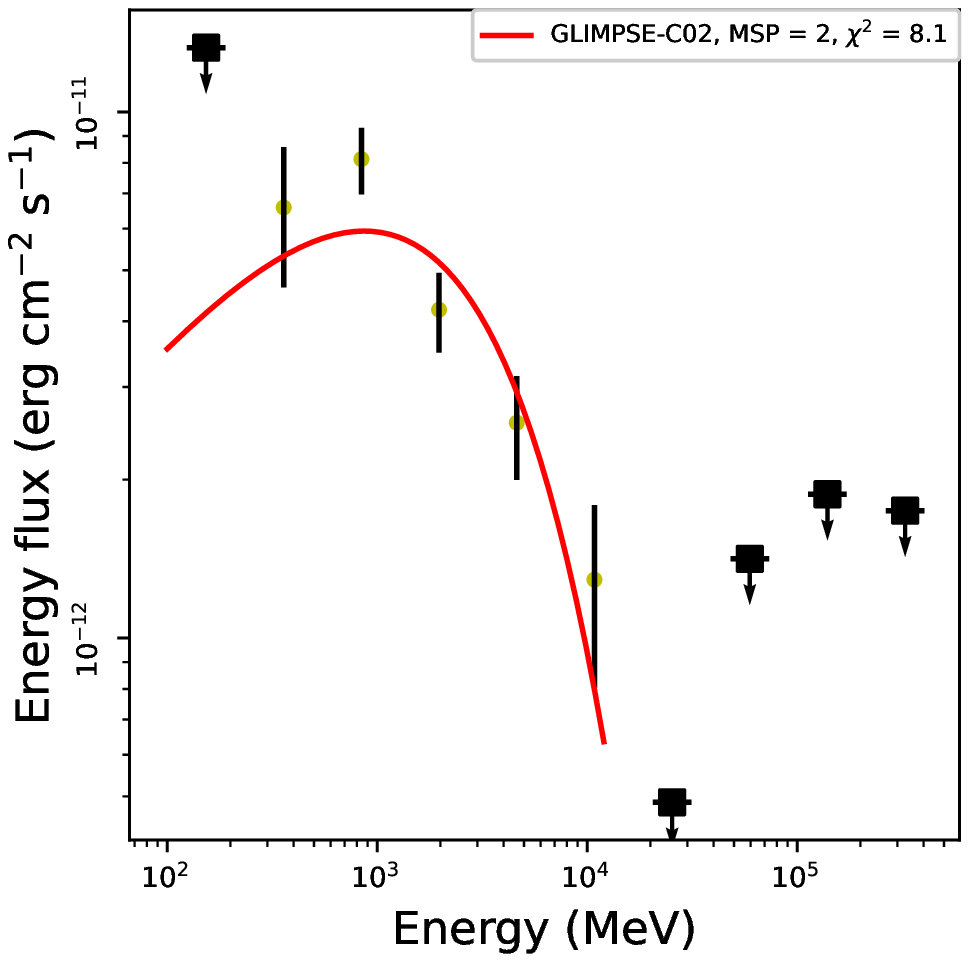}
\includegraphics[width=0.41\linewidth]{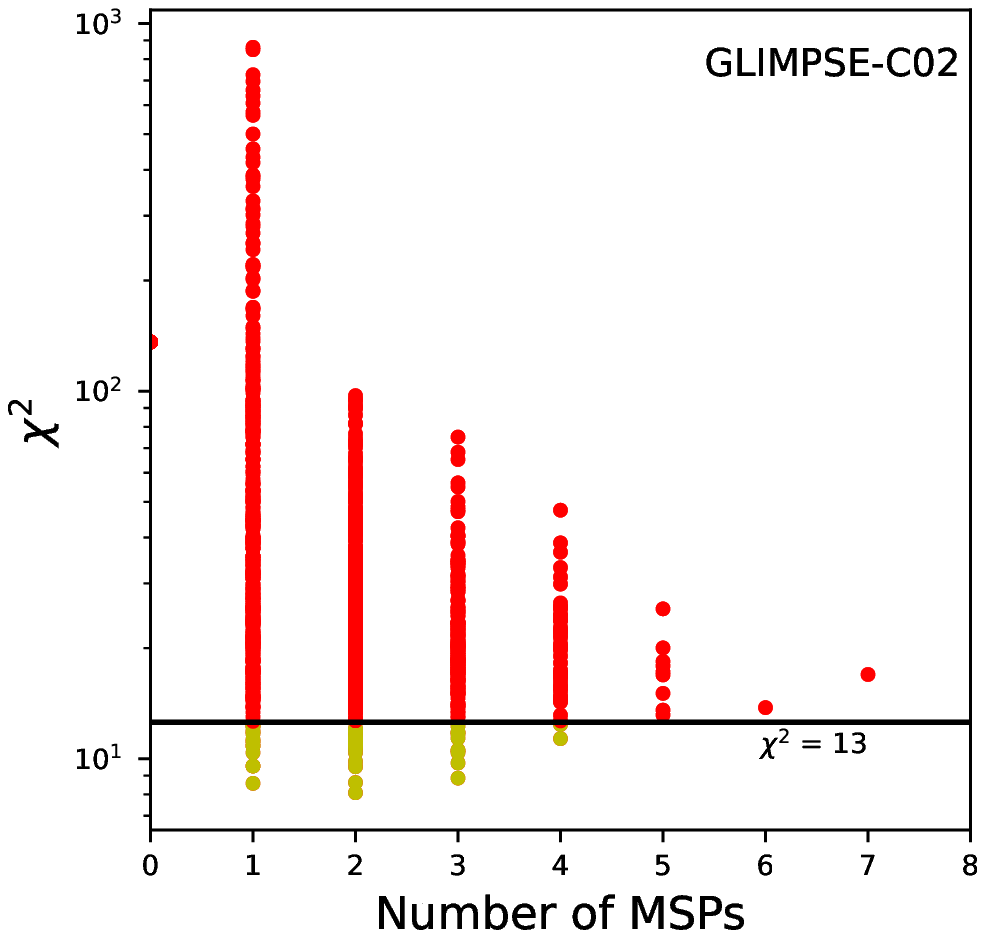}
\end{center}
	\caption{Same as Figure~\ref{fig:tuc}.}
\end{figure*}

\section{Spectral fitting for five nearby GCs}
\label{sec:ul}

Spectral upper limits derived for five nearby GCs are shown in the following
figure. 

\begin{figure*}
\begin{center}
        \includegraphics[width=0.41\linewidth]{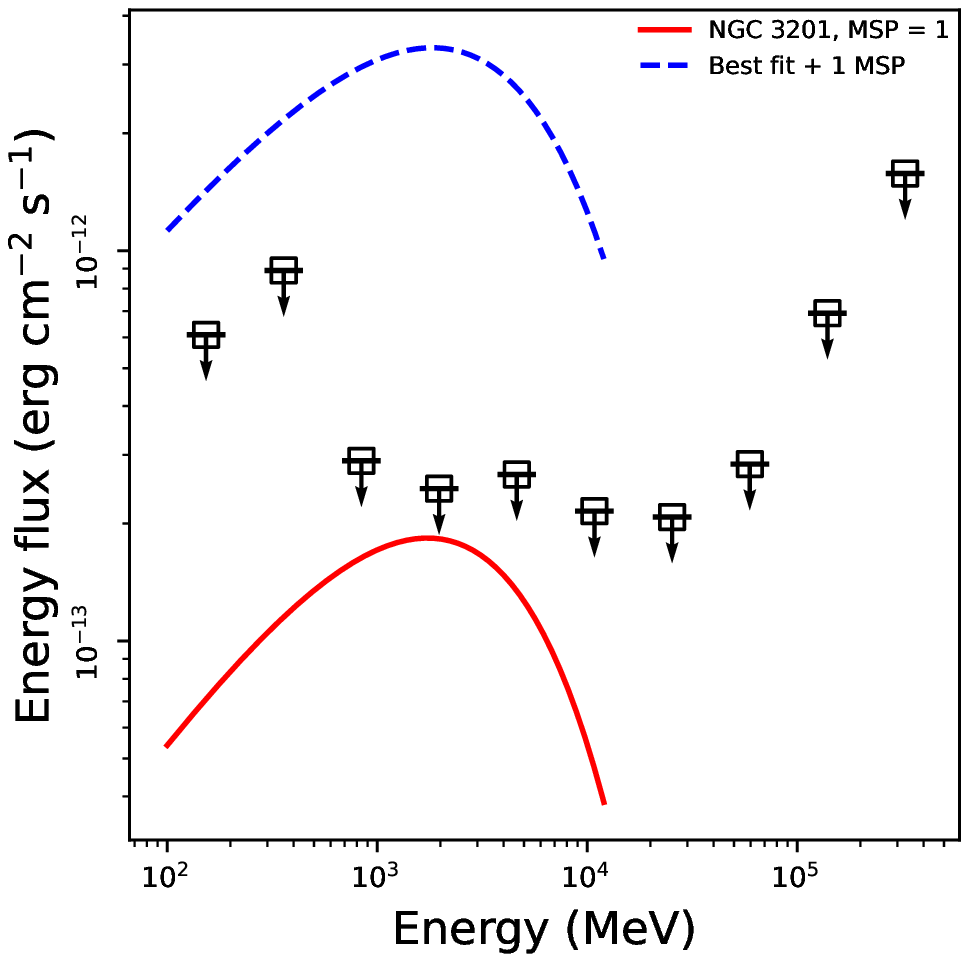}
\includegraphics[width=0.41\linewidth]{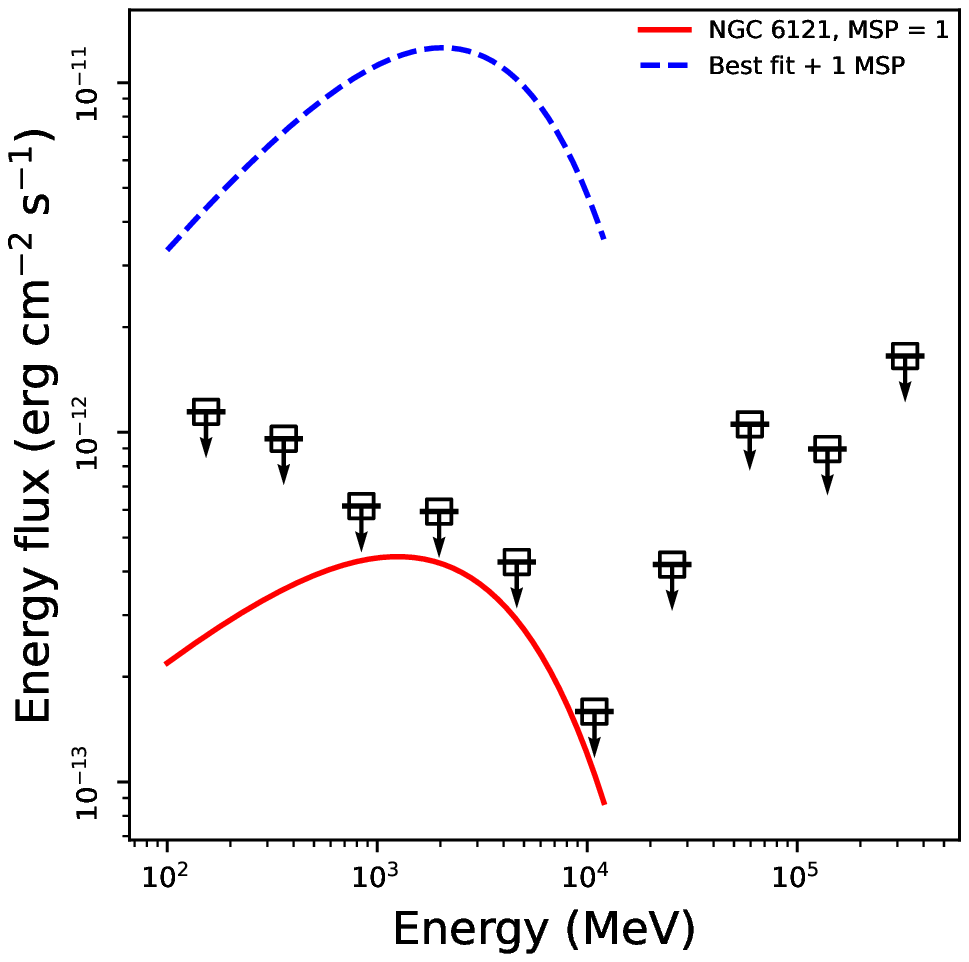}

	\includegraphics[width=0.41\linewidth]{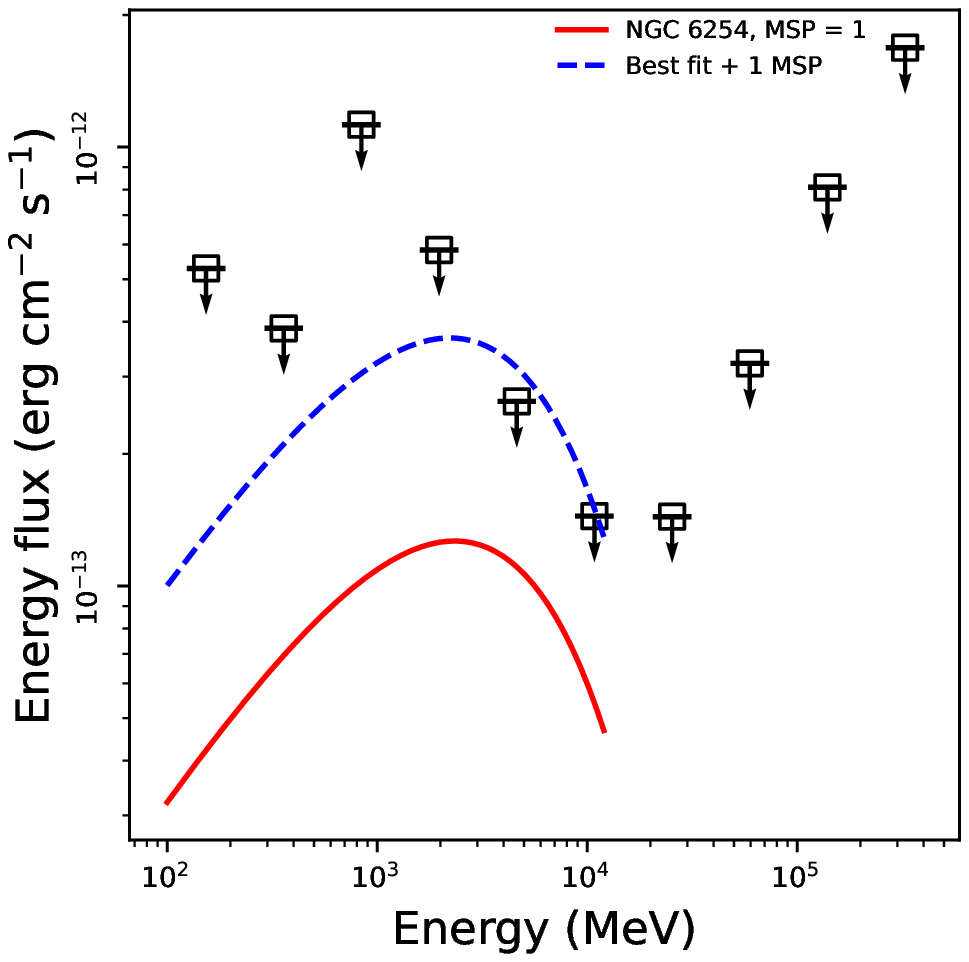}
\includegraphics[width=0.41\linewidth]{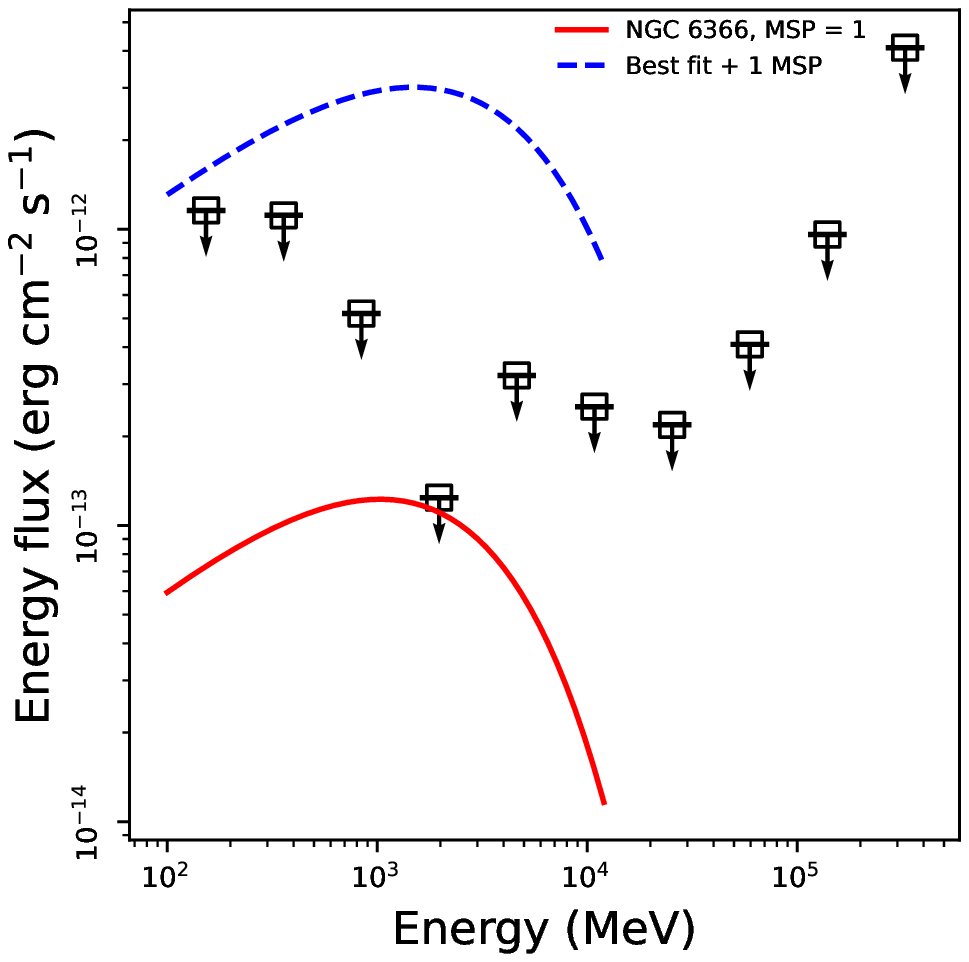}

	\includegraphics[width=0.41\linewidth]{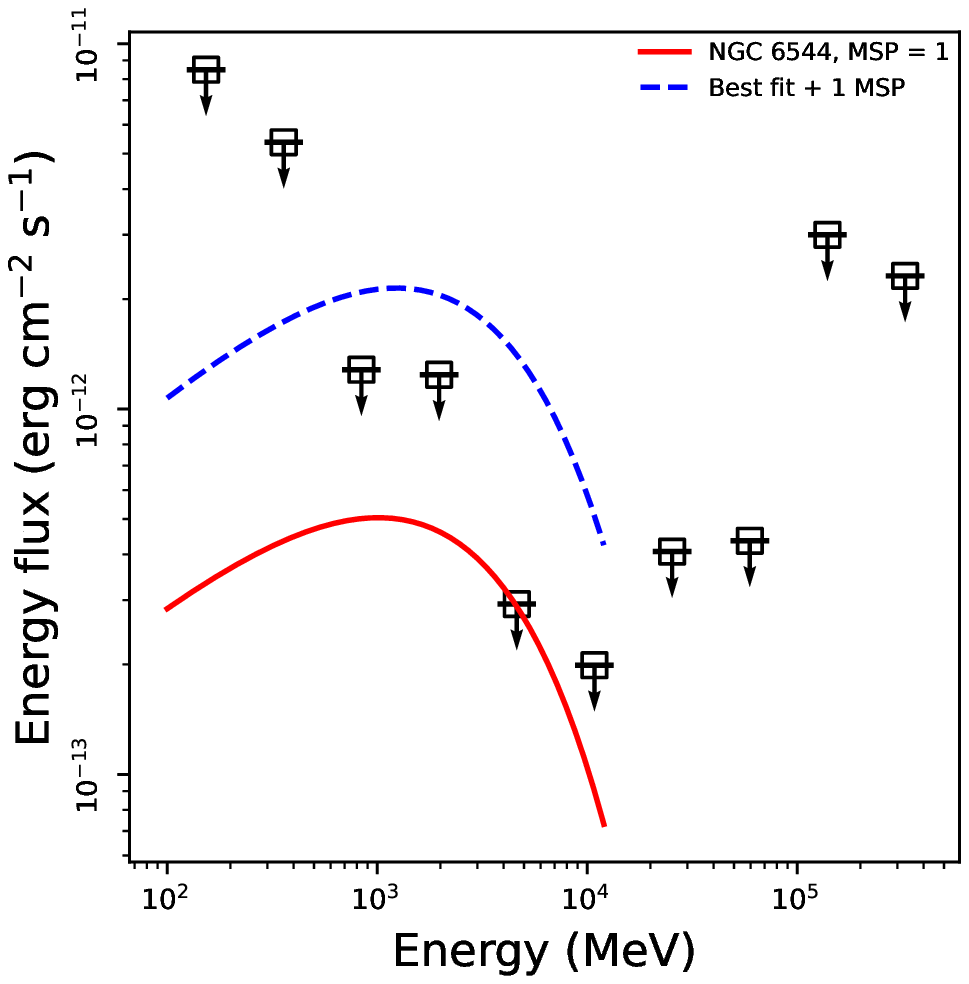}
\end{center}
	\caption{Spectral upper limits on \gr\ emission from NGC~3201, NGC~6121,
	NGC~6254, NGC~6366, and NGC~6544. Our spectral fitting to the upper 
	limits indicates that
	all of them are limited to have $\leq 1$ MSP.}
\label{fig:ul}
\end{figure*}

\end{document}